\documentclass{aa}

\usepackage{natbib}
\usepackage{txfonts}
\usepackage[dvips]{graphicx}
\usepackage{subfigure}
\usepackage{color}

\bibpunct{(}{)}{;}{a}{}{,}

\begin{document}

\def\kms{km\,s$^{-1}$}
\def\vsini{v$\sin{i}$}
\def\Vbroad{$V_{\rm broad}$}
\def\teff{$T_{\rm eff}$}
\def\logg{$\log{g}$}
\def\loggf{$\log{gf}$}
\def\C2{${\rm C}_2$}
\def\bv{$B-V$}

\titlerunning{Stellar photospheric parameters and C abundances}
\authorrunning{R. Da Silva et al.}

\title{Homogeneous photospheric parameters and C abundances \\
in G and K nearby stars with and without planets
\thanks{Based on public data from the ELODIE archive
(Moultaka et al. 2004, online access: http://atlas.obs-hp.fr/elodie/)}$^,$
\thanks{Tables 2, 3, and 7 are only available in electronic form
at the CDS via anonymous ftp to cdsarc.u-strasbg.fr (130.79.128.5)
or via http://cdsweb.u-strasbg.fr/cgi-bin/qcat?J/A+A/}}

\author{Ronaldo Da Silva\inst{1} \and Andr\'e C. Milone\inst{1} \and
Bacham E. Reddy\inst{2}}

\offprints{R. Da Silva,\\
\email{dasilvr2@gmail.com}}

\institute{
Astrophysics Division, Instituto Nacional de Pesquisas Espaciais,
12227-010, S\~ao Jos\'e dos Campos, Brazil
\and
Indian Institute of Astrophysics, Bengaluru, 560034, India}

\date{Received / accepted}

%
%
\abstract{}
{We present a determination of photospheric parameters and carbon abundances
for a sample of 172 G and K dwarf, subgiant, and giant stars with and
without detected planets in the solar neighbourhood. The analysis was based
on high signal-to-noise ratio and high resolution spectra observed with the
ELODIE spectrograph (Haute Provence Observatory, France) and for which the
observational data was publicly available. We intend to contribute precise
and homogeneous C abundances in studies that compare the behaviour of light
elements in stars, hosting planets or not. This will bring new arguments to
the discussion of possible anomalies that have been suggested and will
contribute to a better understanding of different planetary formation
process.}
{The photospheric parameters were computed through the excitation potential,
equivalent widths, and ionisation equilibrium of iron lines selected in the
spectra. Carbon abundances were derived from spectral synthesis applied to
prominent molecular head bands of \C2\ Swan ($\lambda$5128 and
$\lambda$5165) and to a C atomic line ($\lambda$5380.3). Careful attention
was drawn to carry out such a homogeneous procedure and to compute the
internal uncertainties.
}
{The distribution of [C/Fe] as a function of [Fe/H] shows no difference in
the behaviour of planet-host stars in comparison with stars for which no
planet was detected, for both dwarf and giant subsamples. This result is in
agreement with the hypothesis of primordial origin for the chemical
abundances presently observed instead of self-enrichment during the
planetary system formation and evolution. Additionally, giant stars are
clearly depleted in [C/Fe] (by about 0.14~dex) when compared with dwarfs,
which is probably related to evolution-induced mixing of H-burning products
in the envelope of evolved stars. Subgiant stars, although in small number,
seems to follow the same C abundance distribution as dwarfs. We also
analysed the kinematics of the sample stars that, in majority, are members
of the Galaxy's thin disc. Finally, comparisons with other analogue studies
were performed and, within the uncertainties, showed good agreement.}
{}


\maketitle

%
%
\section{Introduction}
\label{intro}

The Sun was usually assumed to be formed from the material representative of
local physical conditions in the Galaxy at the time of its formation and,
therefore, to represent a standard chemical composition. However, this
homogeneity hypothesis has been often put in question as a consequence of
many improvements in the observations techniques and data analysis. With the
discovery of extrasolar planetary systems, the study of heterogeneity
sources (e.g. stellar formation process, stellar nucleosynthesis and
evolution, collisions with molecular clouds, radial migration of stars in
the Galactic disc) has gained a new perspective and brought new questions.

It is now a fact that dwarf stars hosting giant planets are, on average,
richer in metal content than stars in the solar neighbourhood for which no
planet has been detected \citep[see e.g.][]{FischerValenti2005,Gonzalez2006,
Santosetal2001,Santosetal2004}. Two hypotheses have been suggested trying
to explain the origin of this anomaly:
{\it i) primordial hypothesis:} the chemical abundances presently observed
would represent those of the protostellar cloud from which the star was
formed;
{\it ii) self-enrichment hypothesis:} a significant amount of material
enriched in metals would be accreted by the star during the planetary system
formation and evolution.

It has been speculated that abundance anomalies between dwarf stars with and
without planets may not only involve the metal content of heavy elements but
also the abundance of light elements, like carbon and oxygen, measured by an
overabundance in the ratio [X/Fe] of one stellar group in spite to the
other, in a given metallicity range. \citet{GonzalezLaws2000} found that
[Na/Fe] and [C/Fe] in stars with planets are, on average, smaller than in
stars with no detected planets, for the same metallicity. Numerical
simulations performed by \citet{Robinsonetal2006} predicted an overabundance
of [O/Fe] in planet-host stars. The same result for this element was
obtained by \citet{Ecuvillonetal2006}, although the authors noticed that it
is not clear if this difference is due to the presence of planets.

In other publications, however, no difference was found in the abundance
ratios of light elements when comparing stars with and without planets
\citep{Ecuvillonetal2004a,Ecuvillonetal2004b,LuckHeiter2006,
TakedaHonda2005}. In particular, in more recent studies,
\citet{GonzalezLaws2007} and \citet{Bondetal2008} do not confirm the
overabundance of [O/Fe] in planet-host stars obtained by
\citet{Ecuvillonetal2006} and \citet{Robinsonetal2006}, showing that a
solution for the problem is not simple.

\begin{figure*}[t!]
\centering
\begin{minipage}[t]{0.4\textwidth}
\centering
\resizebox{\hsize}{!}{\includegraphics{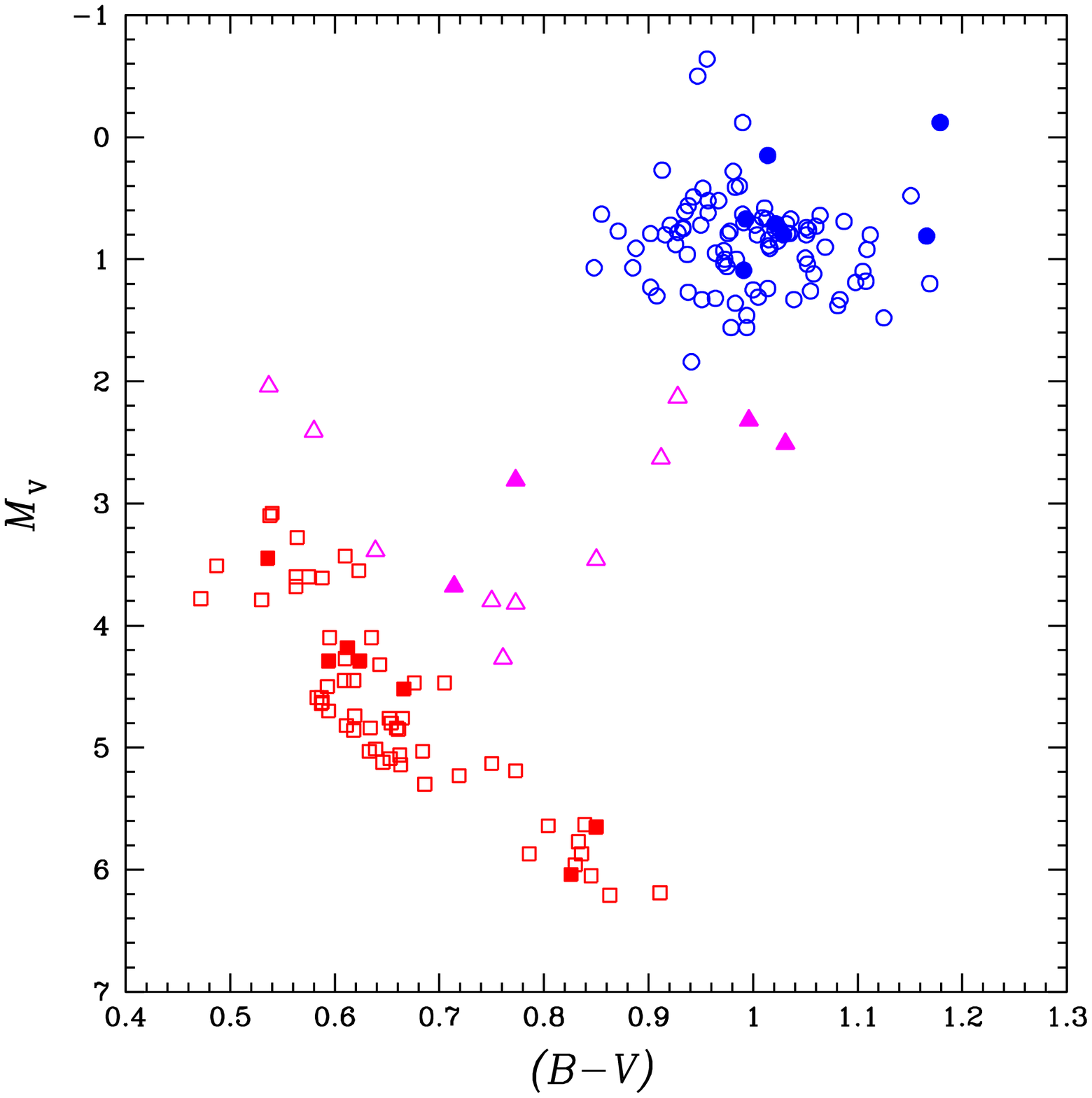}}
\end{minipage}
\begin{minipage}[t]{0.4\textwidth}
\centering
\resizebox{\hsize}{!}{\includegraphics{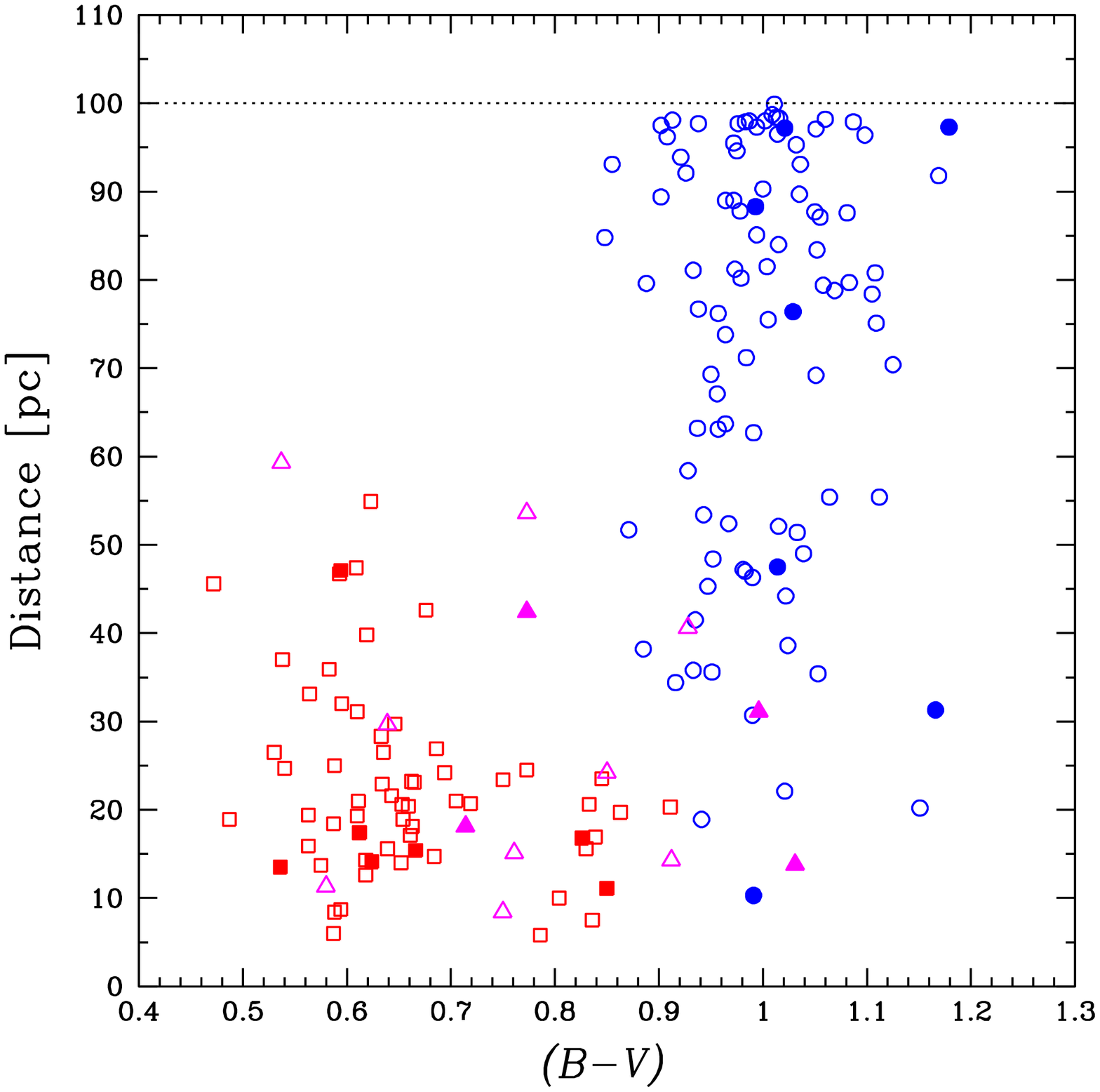}}
\end{minipage}
\caption{HR (left panel) and colour-distance (right panel) diagrams for 63
         dwarfs ({\color{red} $\square$}), 13 subgiants ({\color{magenta}
	 $\triangle$}), and 96 giants ({\color{blue} \Large $\circ$})
	 analysed in this work. Filled symbols represent stars with detected
	 planets. The distance limit of 100~pc is also shown (dotted line).}
\label{hr_diag}
\end{figure*}

Most of the studies above are not conclusive and the discussion about this
problem remains open. New tests are encouraged, using more precise and
homogeneous data, with a number of stars as large as possible. In the
present work we show the results of our analysis of high quality spectra, in
which we determined photospheric parameters and carbon abundances for 172 G
and K stars, including 18 planet hosts. Although this kind of investigation
is not new, the spectra analysed here offer the possibility to perform an
homogeneous and accurate study of the chemical anomalies that have been
proposed in the literature and will surely help to distinguish the different
stellar and planetary formation processes.

The abundance distribution of light elements in stars more evolved than the
Sun, hosting planets or not, have also been studied. \citet{Takedaetal2008}
analysed a large sample of late-G giants, including a few stars hosting
planets. For a solar metallicity, the authors found an underabundance of
[C/Fe] and [O/Fe] as well as an overabundance of [Na/Fe] in the atmosphere
of their sample stars compared to previous results for dwarf stars, which
they attributed to evolution-induced mixing of H-burning products in the
envelope of evolved stars.

Our sample comprises 63 dwarf stars (from which 7 have planets), 13
subgiants (4 with planets), and 96 giants (7 with planets). This allowed us
to investigate possible anomalies in the abundance ratios of stars more
evolved than the Sun. Excepting HD\,7924, whose planet has a minimum mass of
9 Earth masses (about half that of Neptune), the planet-host stars in
question come from systems that have at least one giant planet.

The data and the reduction process are presented in Sect.~\ref{obs}. The
determination of the photospheric parameters and their uncertainties are
described in Sect.~\ref{phot_par}. In Sect.~\ref{synth} we describe the
spectral synthesis method used to obtain the carbon abundances and their
uncertainties. Our results are presented and discussed in Sect.~\ref{res},
and final remarks and conclusions done in Sect.~\ref{concl}.

%
%
\section{Observation data and reduction}
\label{obs}

Our sample consists of 172 G and K dwarf, subgiant, and giant stars in the
solar neighbourhood (distance $<$ 100~pc) observed with the ELODIE
high-resolution \'echelle spectrograph \citep{Baranneetal1996} of the Haute
Provence Observatory (France). The analysis was done based on spectra that
were publicly available in the ELODIE archive \citep{Moultakaetal2004} when
the work started. The spectra have resolution $R = 42\,000$ and cover the
wavelength range 3895$-$6815~\AA. The resulting sample stars were selected
according to the following criteria:

\begin{itemize}

\item[\it i)] stars for which the averaged spectra have S/N $\geq$ 200;
among all individual spectra available in the database, only those having
S/N $\geq$ 30 and with an image type classified as {\it object fibre only}
(OBJO) were used;

\item[\it ii)] stars within a distance $\leq$ 100~pc (parallax $\pi \geq$
10~mas) and with spectral type between F8 and M1; earlier type stars have a
small number of spectral lines whereas dwarfs later than M1 are quite faint
to provide good quality spectra, and also quite cold (exhibiting a lot of
strong molecular features, such as the TiO bands), making difficult the
determination of precise abundances;

\item[\it iii)] stars for which no close binary companion is known, since
these objects may contaminate the observed spectra; we used the information
of angular separation between components ($rho$) available in the Hipparcos
catalogue \citep{ESA1997}, choosing only the cases with $rho >$ 10~arcsec.

\item[\it iv)] stars for which the determination of the photospheric
parameters (Sect.~\ref{phot_par}) is reliable;

\item[\it v)] stars with (\bv) values measured by Hipparcos and with
spectral cross-correlation parameters available in the ELODIE database; both
(\bv) and the width of the cross-correlation function are required in
the estimate of the stellar projected rotation velocity \vsini\
(see Sect.~\ref{synth}); and

\item[\it vi)] stars that passed the quality control of the spectral
synthesis (see Sect.~\ref{synth}).

\end{itemize}

The selected sample is plotted in the HR and colour-distance diagrams of
Fig.~\ref{hr_diag}, separating the subsamples of dwarfs, subgiants, and
giants. The transition boundaries between dwarfs and subgiants and also
between subgiants and giants are not clearly defined on an observational HR
plane. In this work, we chose to classify as subgiants those stars situated
1.5~mag above the lower limit of the main-sequence and having $M_{\rm v} >
2.0$~mag. Note that the distance of dwarfs and subgiants is not limited to
100~pc, but to about 60~pc. This is not an imposition of our selection
criteria, but a selection effect of ELODIE observation surveys instead.

For each sample star, the spectra available in the ELODIE database were
processed using IRAF\,\footnote{{\it Image Reduction and Analysis Facility},
distributed by the National Optical Astronomy Observatories (NOAO) in
Tucson, Arizona (EUA).} routines. First, they were normalised (a global
pre-normalisation) based on continuum windows selected in the wavelength
range. Then the normalised spectra were corrected from Doppler effect, i.e.,
transformed to a rest wavelength scale taking the solar spectrum as
reference, with a precision of better than 0.02~\AA\ in the correction.
After these two first steps, the spectra were averaged to reduce noise.
Finally, a more careful normalisation was done, this time only considering a
small wavelength region around the spectral features analysed in this work:
for the molecular bands the range 5100$-$5225~\AA\ were used, and for the C
atomic line the wavelength range was 5330$-$5430~\AA. At this point, the
stellar spectra were ready to be used by the synthesis method.

%
%
\section{Determination of photospheric parameters}
\label{phot_par}

A precise and homogeneous determination of chemical abundances in stars
depends on the calculation of realistic model atmospheres, which in turn
depends on accurate stellar photospheric parameters: the effective
temperature \teff, the metallicity [Fe/H], the surface gravity \logg, and
the micro-turbulence velocity $\xi$. We developed a code that uses these
four parameters as input and, iteratively changing their values, try to find
a solution for the model atmosphere and metal abundance that are physically
acceptable.

\begin{figure}[t!]
\centering
\begin{minipage}[t]{0.45\textwidth}
\centering
\resizebox{\hsize}{!}{\includegraphics[angle=-90]{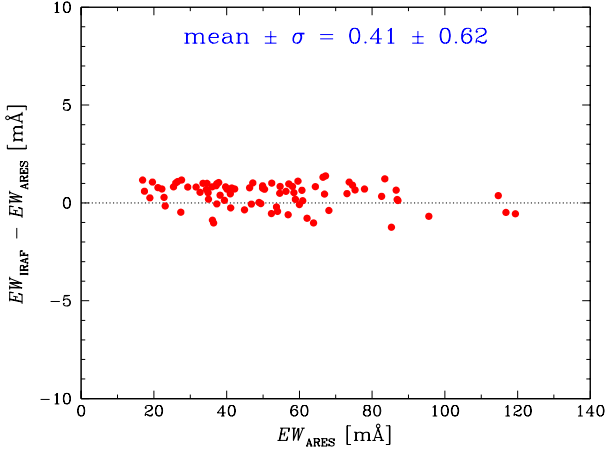}}
\end{minipage}
\caption{Equivalent widths of \ion{Fe}{i} and \ion{Fe}{ii} lines in the
         spectrum of the sunlight reflected by the Moon measured using ARES
	 in comparison to those measured using IRAF tasks.}
\label{ares_iraf}
\end{figure}

The abundance yielded by different spectral lines of the same element should
not depend on their excitation potential ($\chi$) or their equivalent width
($EW$). Also, neutral and ionised lines of the same element should provide
the same abundance as well. Therefore, in this work, the effective
temperature was computed through the excitation equilibrium of neutral iron
by removing any dependence in the [\ion{Fe}{i}/H] $versus$ $\chi$ diagram.
Additionally, by removing any dependence of [\ion{Fe}{i}/H] on $EW$, the
micro-turbulence velocity was estimated. The surface gravity was computed
through the ionisation equilibrium between \ion{Fe}{i} and \ion{Fe}{ii}, and
the metallicity was yielded by the $EW$ of \ion{Fe}{i} lines. In other
words, the photospheric parameters were determined following the conditions
below:
\\
\begin{tabular}{lr}
& \\
\big| slope([\ion{Fe}{i}/H] $versus$ $\chi$) \big| $< c_1$ &
(dependence on \teff) \\[0.2cm]
\big| slope([\ion{Fe}{i}/H] $versus$ $EW$) \big| $< c_2$ &
(dependence on $\xi$) \\[0.2cm]
\big| [\ion{Fe}{i}/H] $-$ [\ion{Fe}{ii}/H] \big| $< c_3$ &
(dependence on \logg) \\[0.2cm]
\big| [Fe/H] $-$ [\ion{Fe}{i}/H] \big| $< c_4$ & \\
& \\
\end{tabular} \\
where $c_1$, $c_2$, $c_3$, and $c_4$ are arbitrary constants as small as one
wishes. If at least one of the first three conditions is not satisfied, then
\teff, $\xi$, and/or \logg\ are changed by a given step. In the fourth
condition, the value of metallicity [Fe/H] used as input is compared to the
one provided by \ion{Fe}{i} lines and, if this condition is not satisfied,
the code defines [Fe/H] = [\ion{Fe}{i}/H]. Therefore, the code iteratively
executes several cycles until these four conditions are satisfied at the
same time.

Atomic line parameters (wavelength, oscillator strength $gf$, and
lower-level excitation potential $\chi$) for 72 \ion{Fe}{i} and 12
\ion{Fe}{ii} lines used in our analysis are listed in Table~\ref{linelist}.
They were all taken from the {\it Vienna Atomic Line Database} -- VALD
\citep{Kupkaetal2000,Kupkaetal1999,Piskunovetal1995,Ryabchikovaetal1997},
though the $gf$ values were revised to fit the $EW$ measured in the Kurucz
Solar Flux Atlas \citep{Kuruczetal1984}, along with a model atmosphere for
\teff\ = 5777~K, \logg\ = 4.44, $\xi$ = 1.0~\kms, and
$\log{\epsilon}_{\,\odot}$ = 7.47 (the solar Fe abundance). Concerning the
calculations of the van der Waals line damping parameters, we adopted the
Uns\"old approximation multiplied by 6.3.

The stellar model atmospheres used in the spectroscopic analysis are those
interpolated in a grid derived by \citet{Kurucz1993} for stars with \teff\
from 3500 to 50\,000~K, \logg\ from 0.0 to 5.0~dex, and [Fe/H] from $-$5.0
to 1.0~dex. These are plan-parallel and LTE models, computed over 72 layers.
For each layer, the quantities column density ($\rho$x), temperature ($T$),
gas pressure ($P_{\rm g}$), electronic density ($N_{\rm e}$), and Rosseland
mean opacity ($\kappa_{\rm Ross}$) are listed. The models also include the
micro-turbulence velocity, the elemental abundances in the format
$\log{\epsilon}$ (where $\log{\epsilon}_{\rm\,star} =
\log{\epsilon}_{\,\odot} +$ [Fe/H]), both assumed to be constant in all
layers, and a list of molecules used in the molecular equilibrium
computation. Although the Kurucz models used here were computed for a
micro-turbulence velocity $\xi$ = 2~\kms, in our iterative computation of
the photospheric parameters, $\xi$ was set as a free parameter instead. We
believe that this does not significantly affect the chemical analysis
performed here, since their uncertainties are dominated by the errors in the
other photospheric parameters.

The equivalent widths were measured using the {\it Automatic Routine for
line Equivalent widths in stellar Spectra} -- ARES \citep{Sousaetal2007}. In
order to test the reliability of the automatic measurements, we performed a
comparison between $EW$ measured in the solar spectrum (the sunlight
reflected by the Moon) using ARES and those measured one by one using IRAF
tasks (see Fig.~\ref{ares_iraf}). Notice that both procedures provide $EW$
that are consistent with each other within the uncertainties (the absolute
differences are smaller than 1.5~m\AA).

\begin{figure*}[t!]
\centering
\begin{minipage}[t]{0.33\textwidth}
\centering
\resizebox{\hsize}{!}{\includegraphics{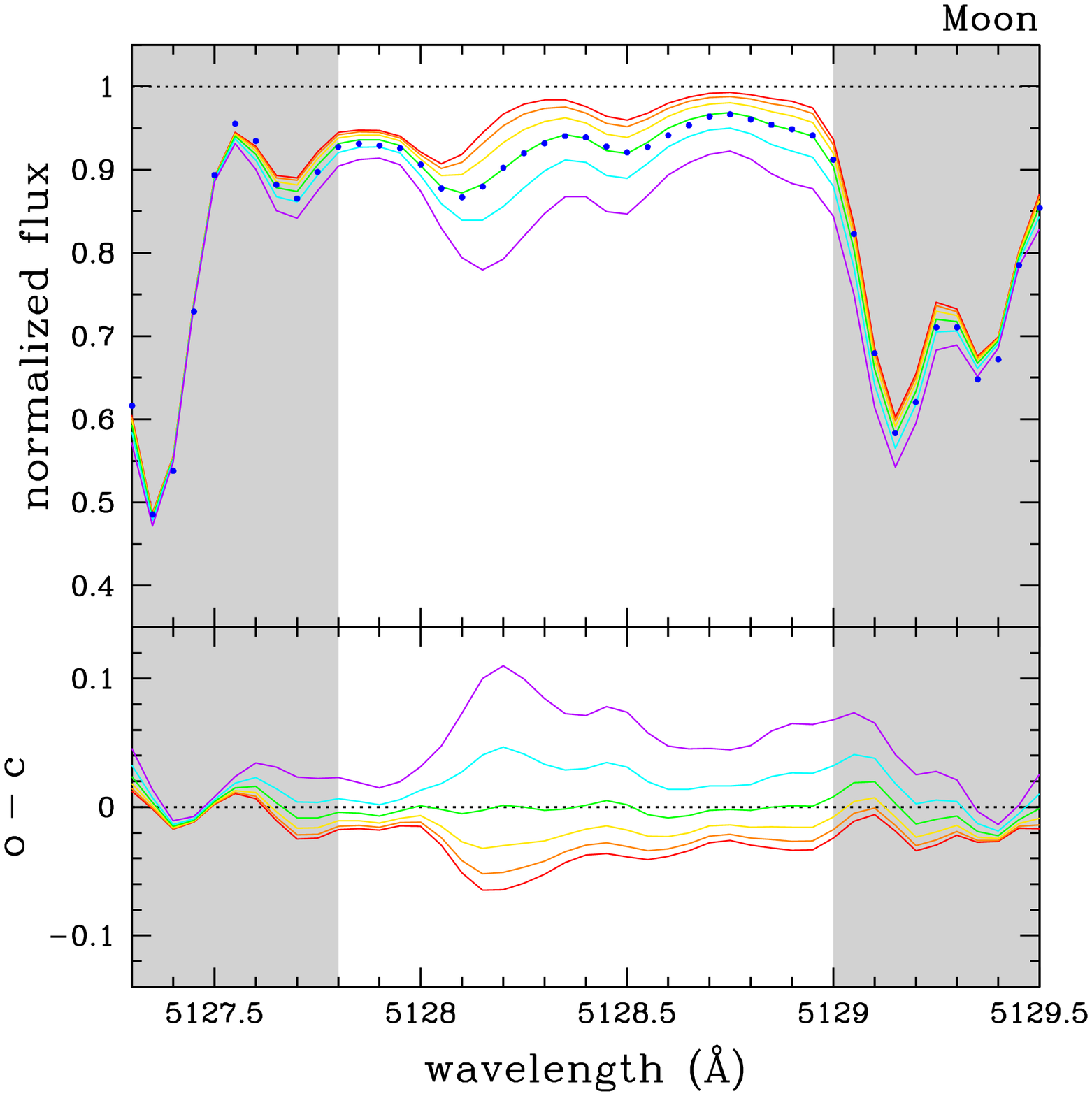}}
\end{minipage}
\begin{minipage}[t]{0.33\textwidth}
\centering
\resizebox{\hsize}{!}{\includegraphics{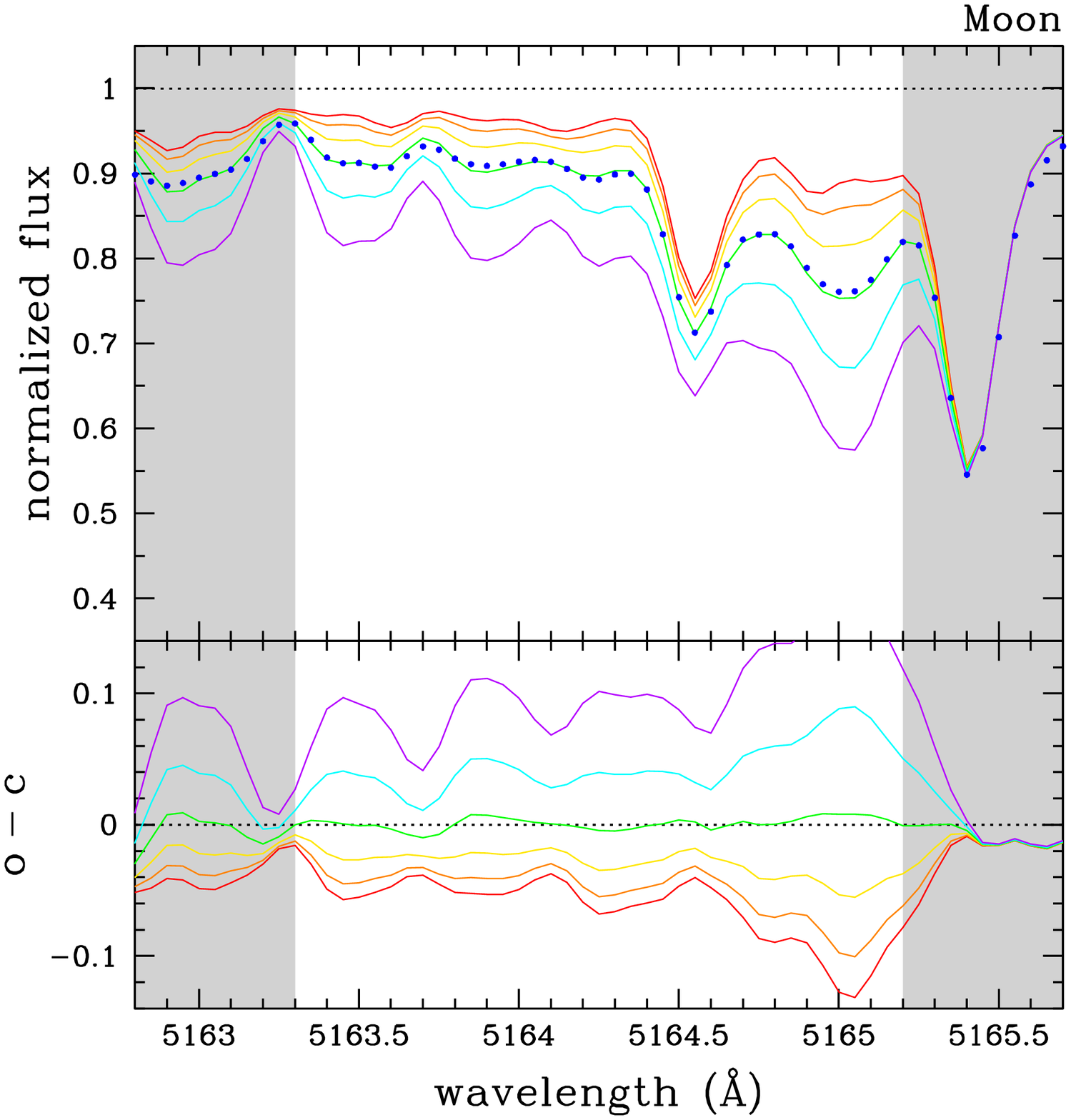}}
\end{minipage}
\begin{minipage}[t]{0.33\textwidth}
\centering
\resizebox{\hsize}{!}{\includegraphics{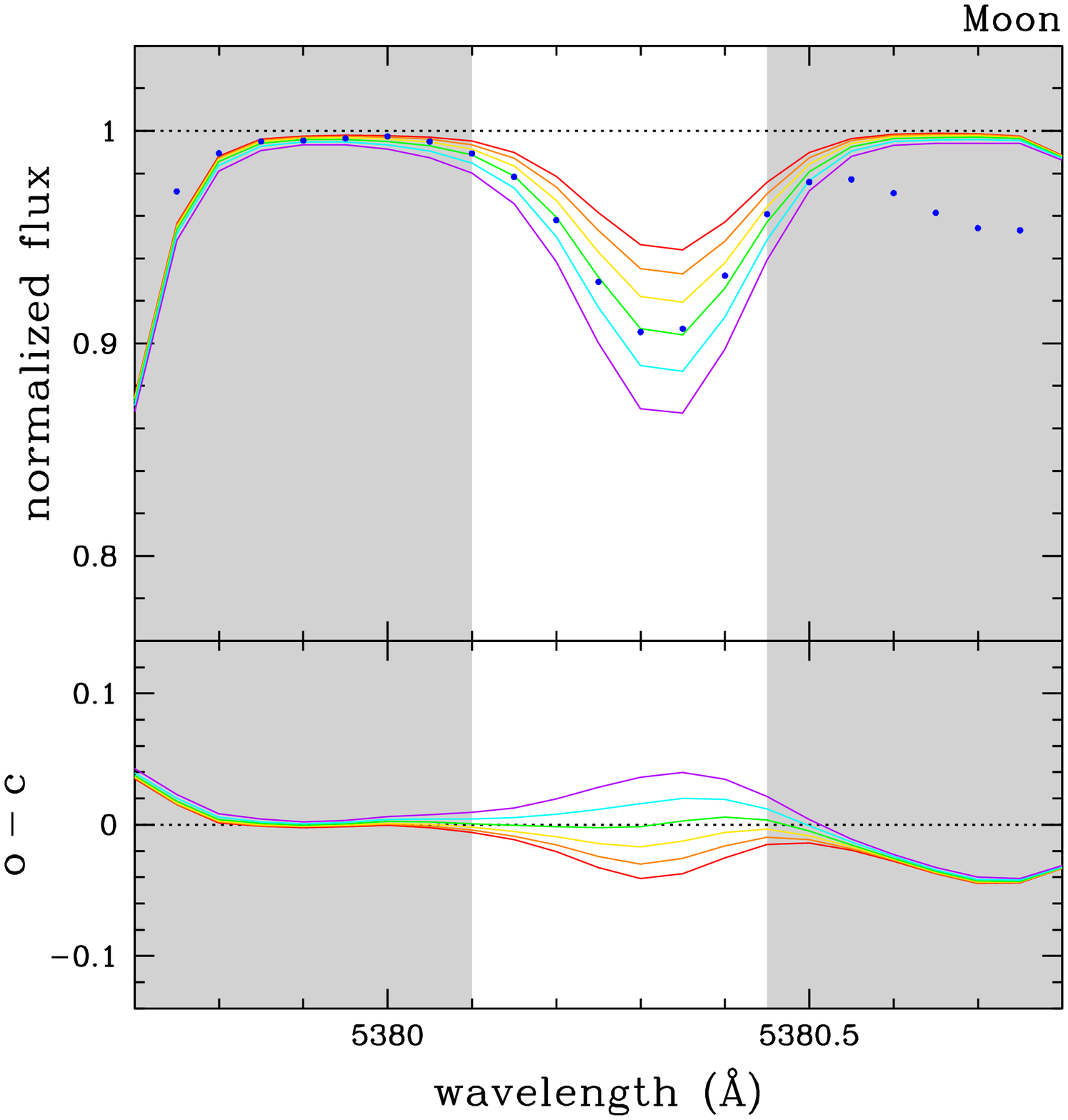}}
\end{minipage}
\caption{Example to illustrate how the spectral synthesis was applied to
         observed data (in this case the spectrum of the sunlight reflected
	 by the Moon). The three regions investigated in this work are shown
	 within the hatched areas: the molecular head bands around
	 $\lambda$5128 (left panel) and $\lambda$5165 (middle panel), and
	 the C atomic line at $\lambda$5380.3 (right panel). Six spectra
	 computed for different values of [C/Fe] and distinguished by
	 0.1~dex are shown. The differences between observed and computed
	 spectra (O$-$C) are also plotted.}
\label{moon_synth}
\end{figure*}

\subsection{Uncertainties in the photospheric parameters}
\label{phot_par_err}

We developed a routine that iteratively estimates the uncertainties in the
computed photospheric parameters of each star. The procedure is as follows:

\begin{itemize}

\item[\it i)] first, the micro-turbulence velocity is increased (decreased)
by a given step and new model atmospheres are computed; the change proceeds
iteratively until the angular coefficient of the linear regression in the
[\ion{Fe}{i}/H] $versus$ $EW$ diagram is of the same order of its standard
error; the absolute differences between increased (decreased) and best
values provide the $\xi$ upper (lower) limits; the uncertainty $\sigma(\xi)$
is an average of lower and upper values;

\item[\it ii)] next, a similar procedure is used for the effective
temperature, which is iteratively changed until the angular coefficient of
the linear regression in the [\ion{Fe}{i}/H] $versus$ $\chi$ diagram is of
the same order of its standard error; since micro-turbulence velocity and
effective temperature are not independent from each other, the uncertainty
$\sigma(\xi)$ estimated above is taken into account before changing \teff;
thus, the absolute differences between changed and best values of \teff\
provide the uncertainty $\sigma$(\teff), with the effect of $\sigma(\xi)$
properly removed;

\item[\it iii)] the uncertainty in the metallicity $\sigma$([Fe/H]) is the
standard deviation of the abundances yielded by individual \ion{Fe}{i}
lines;

\item[\it iv)] finally, the uncertainty in the surface gravity
$\sigma$(\logg) is estimated by iteratively changing its value until the
difference between the iron abundance provided by \ion{Fe}{i} and
\ion{Fe}{ii} is of the same order of $\sigma$([Fe/H]).

\end{itemize}

%
%
\section{Carbon abundance from spectral synthesis}
\label{synth}

\begin{figure*}[t!]
\centering
\begin{minipage}[t]{0.3\textwidth}
\centering
\resizebox{\hsize}{!}{\includegraphics[angle=-90]{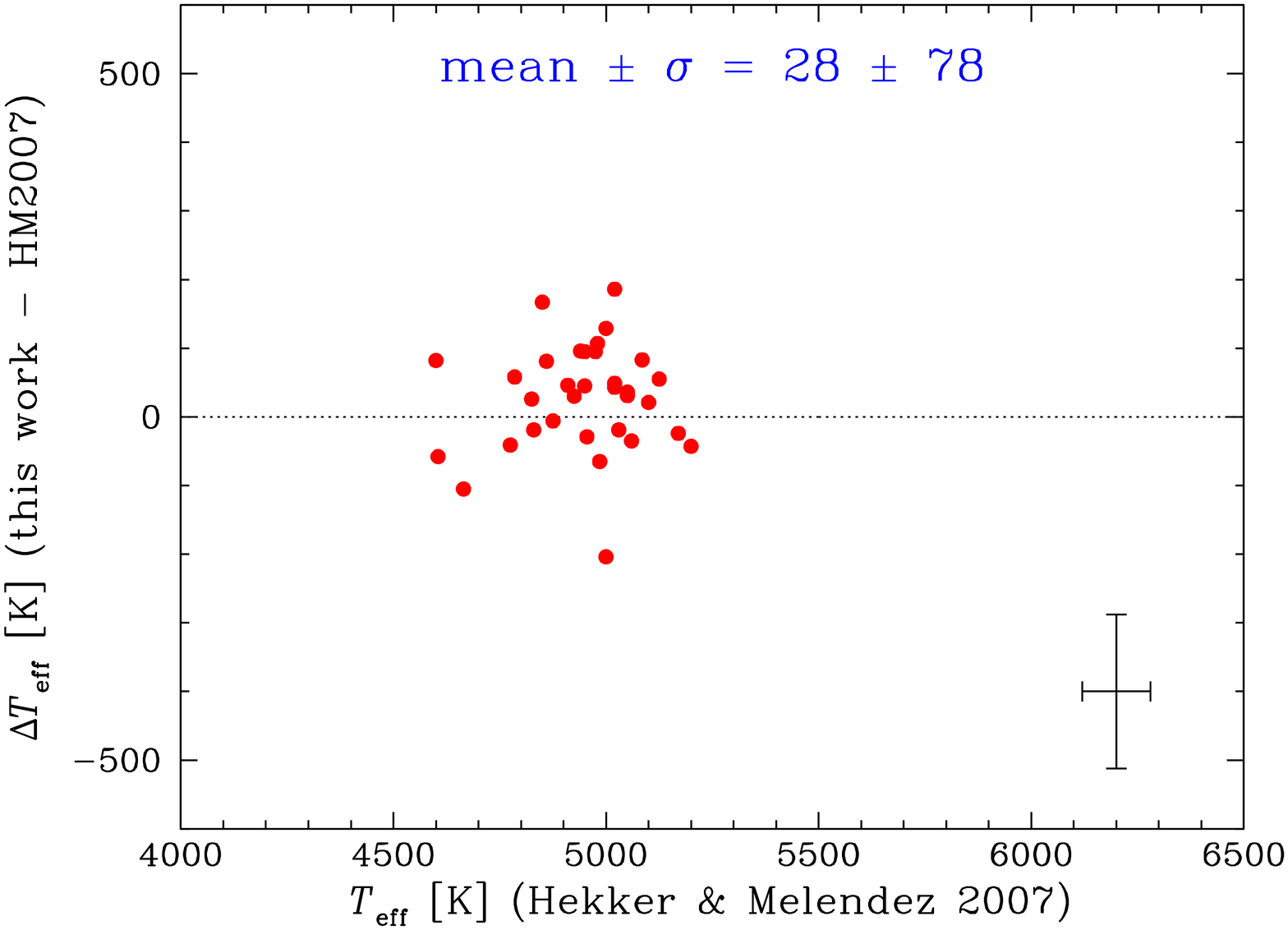}}
\end{minipage}
\begin{minipage}[t]{0.3\textwidth}
\centering
\resizebox{\hsize}{!}{\includegraphics[angle=-90]{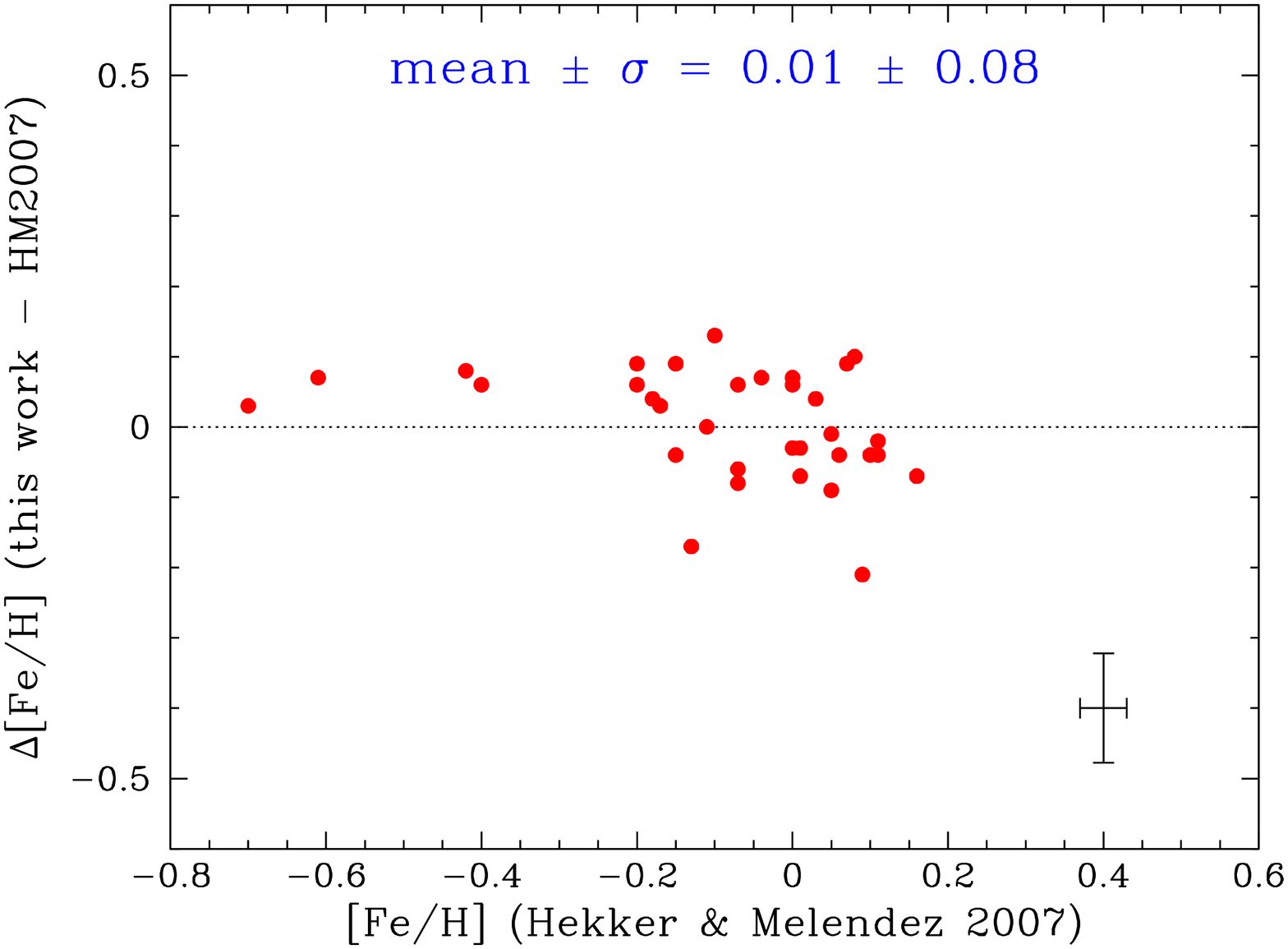}}
\end{minipage}
\begin{minipage}[t]{0.3\textwidth}
\centering
\resizebox{\hsize}{!}{\includegraphics[angle=-90]{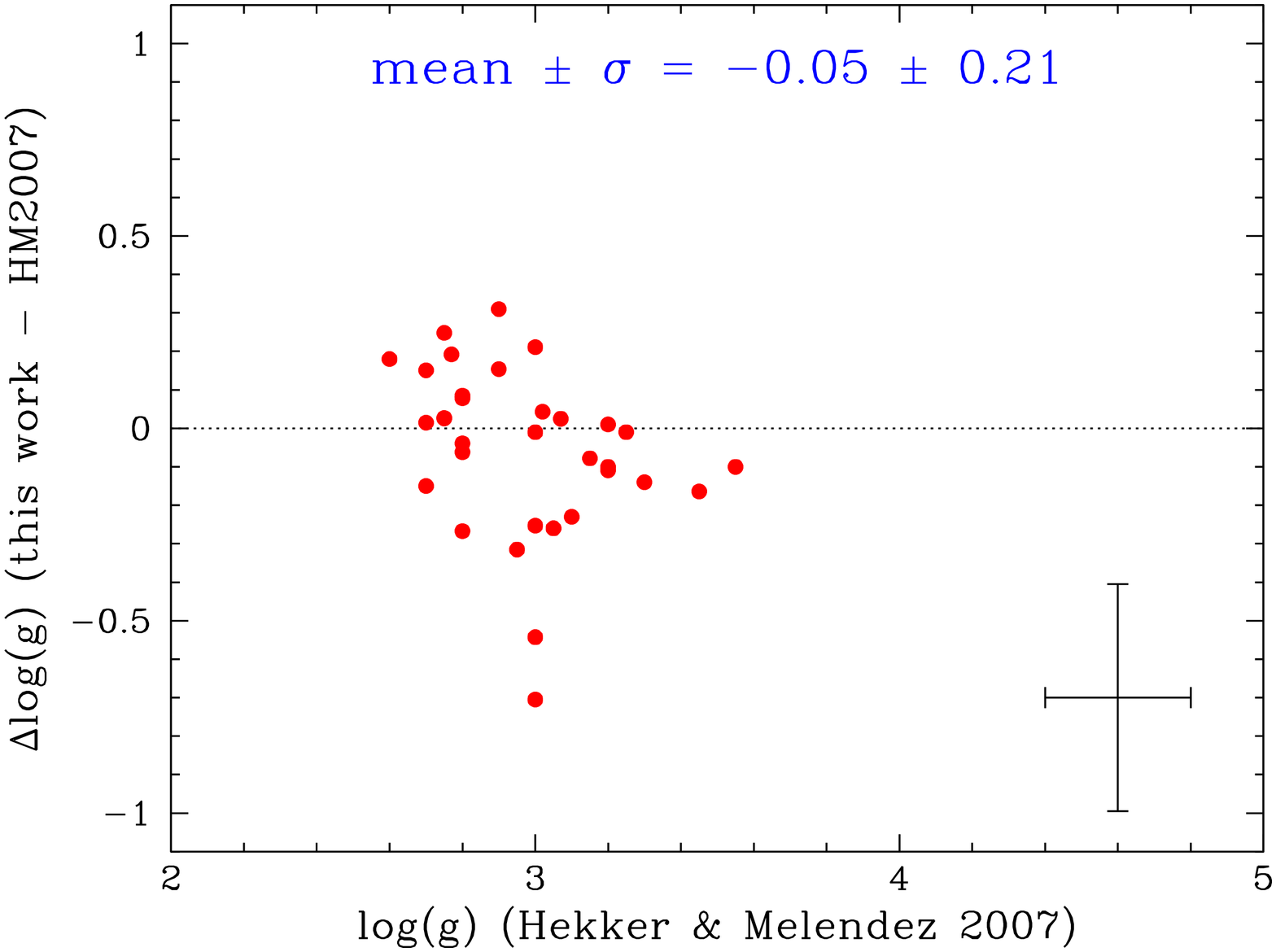}}
\end{minipage} \\
\begin{minipage}[t]{0.3\textwidth}
\centering
\resizebox{\hsize}{!}{\includegraphics[angle=-90]{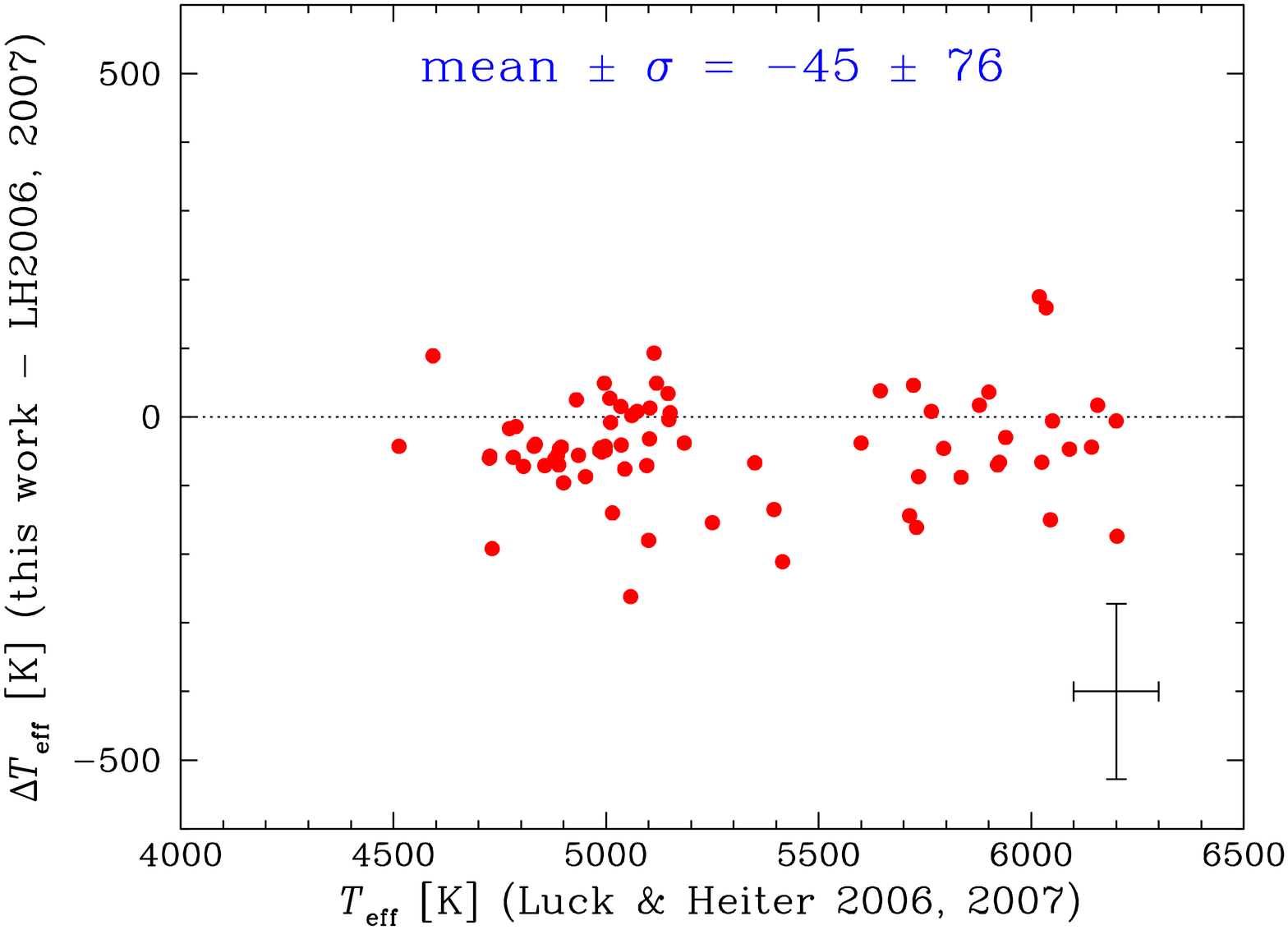}}
\end{minipage}
\begin{minipage}[t]{0.3\textwidth}
\centering
\resizebox{\hsize}{!}{\includegraphics[angle=-90]{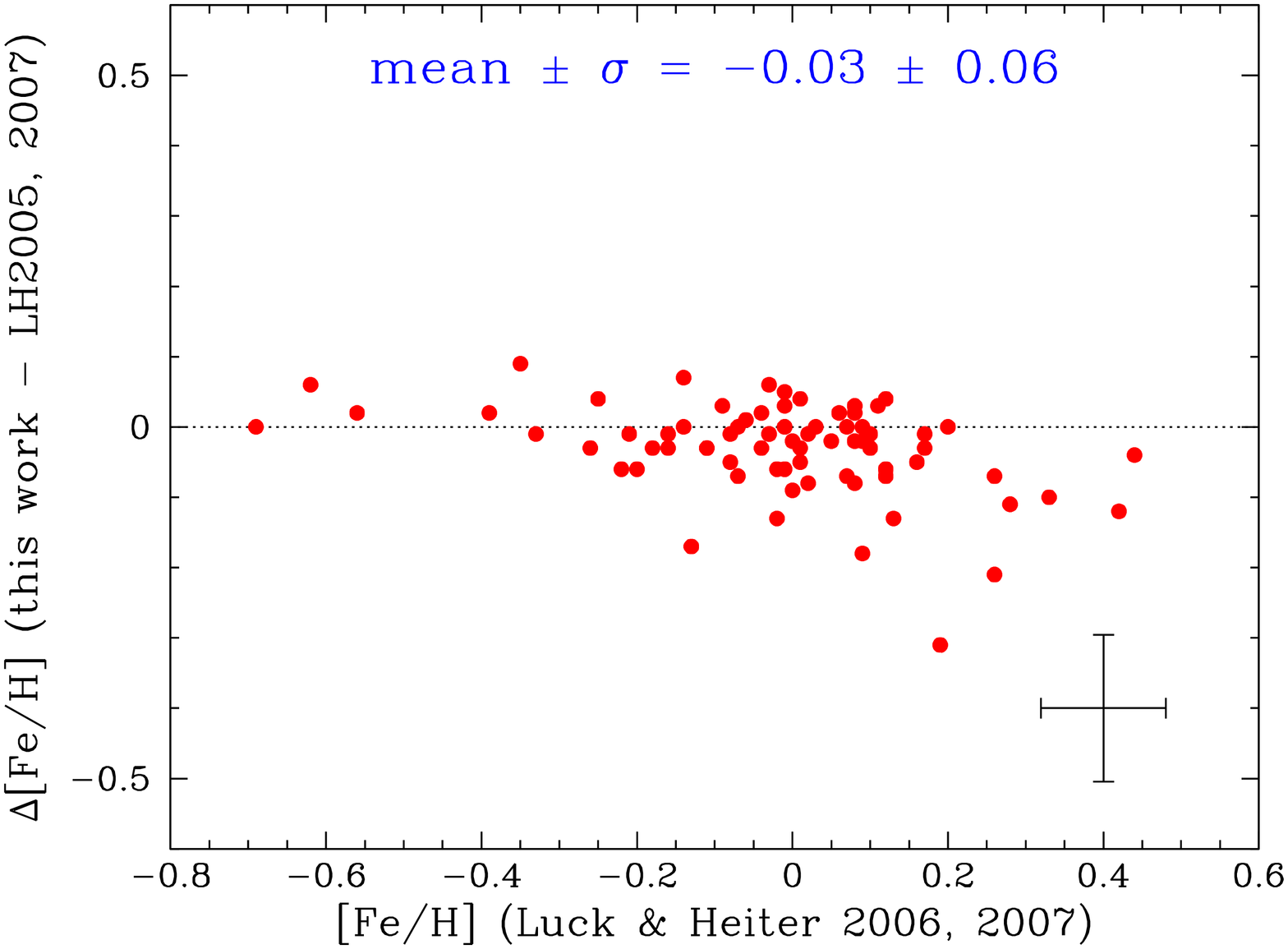}}
\end{minipage}
\begin{minipage}[t]{0.3\textwidth}
\centering
\resizebox{\hsize}{!}{\includegraphics[angle=-90]{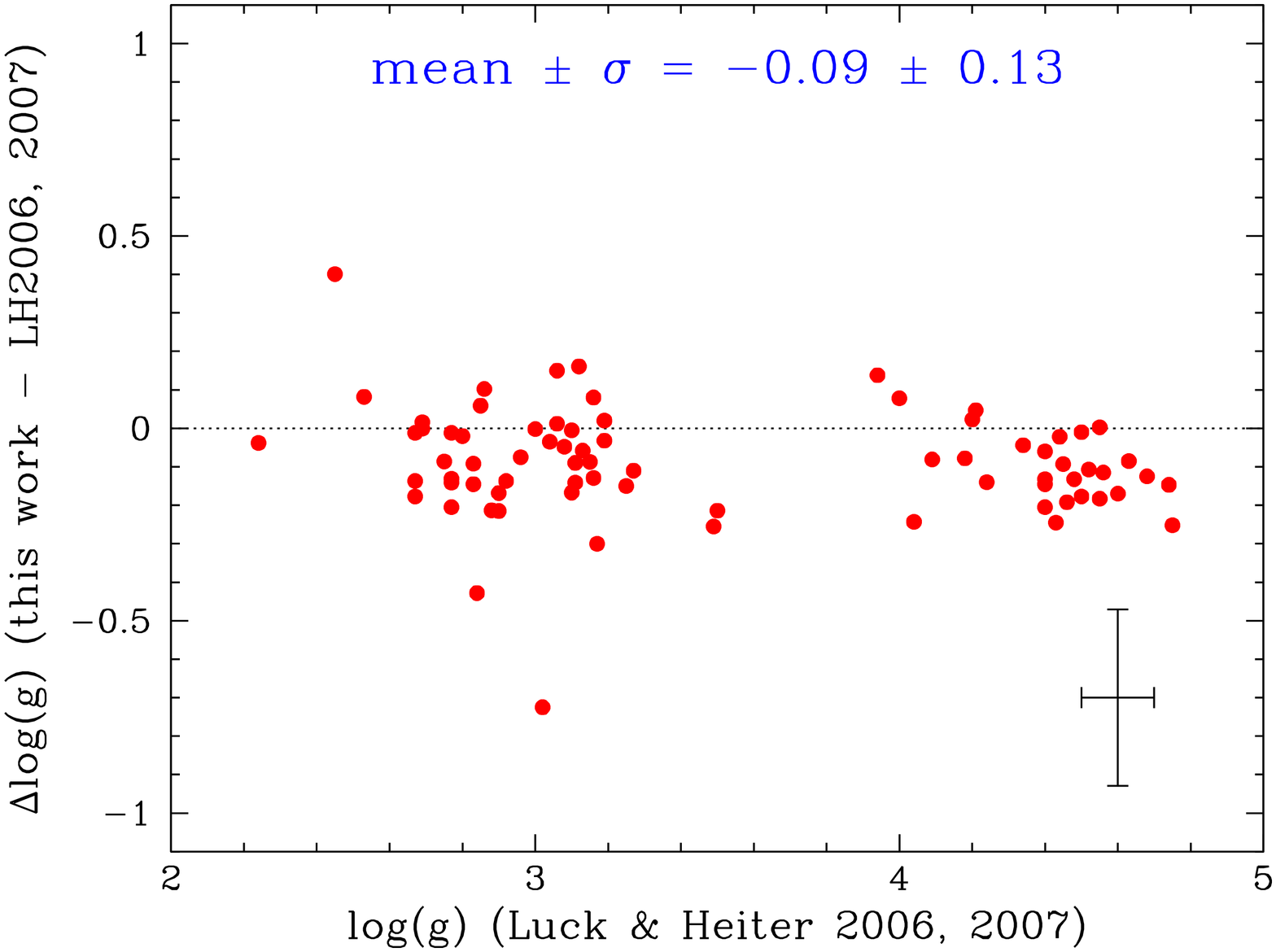}}
\end{minipage} \\
\begin{minipage}[t]{0.3\textwidth}
\centering
\resizebox{\hsize}{!}{\includegraphics[angle=-90]{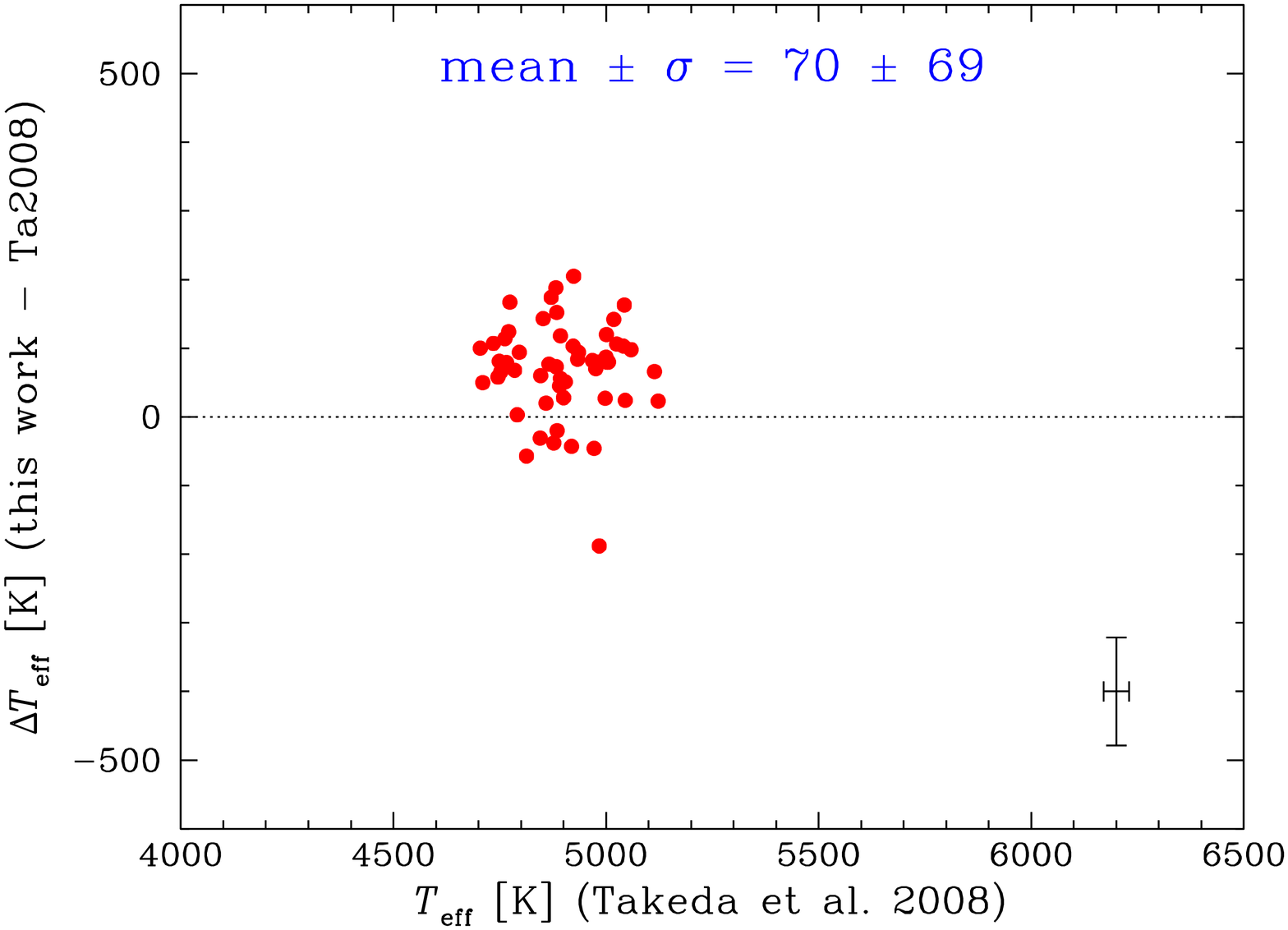}}
\end{minipage}
\begin{minipage}[t]{0.3\textwidth}
\centering
\resizebox{\hsize}{!}{\includegraphics[angle=-90]{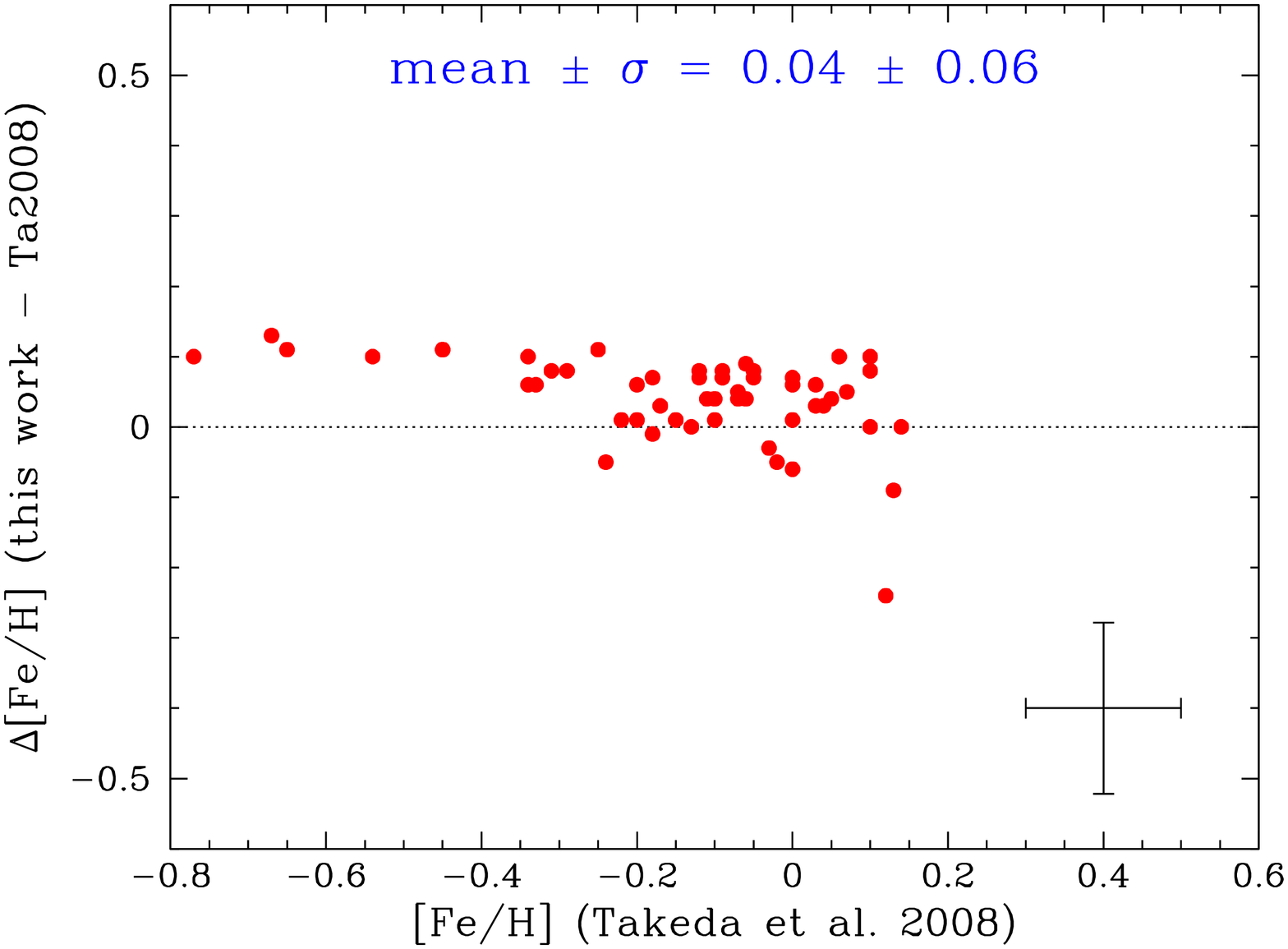}}
\end{minipage}
\begin{minipage}[t]{0.3\textwidth}
\centering
\resizebox{\hsize}{!}{\includegraphics[angle=-90]{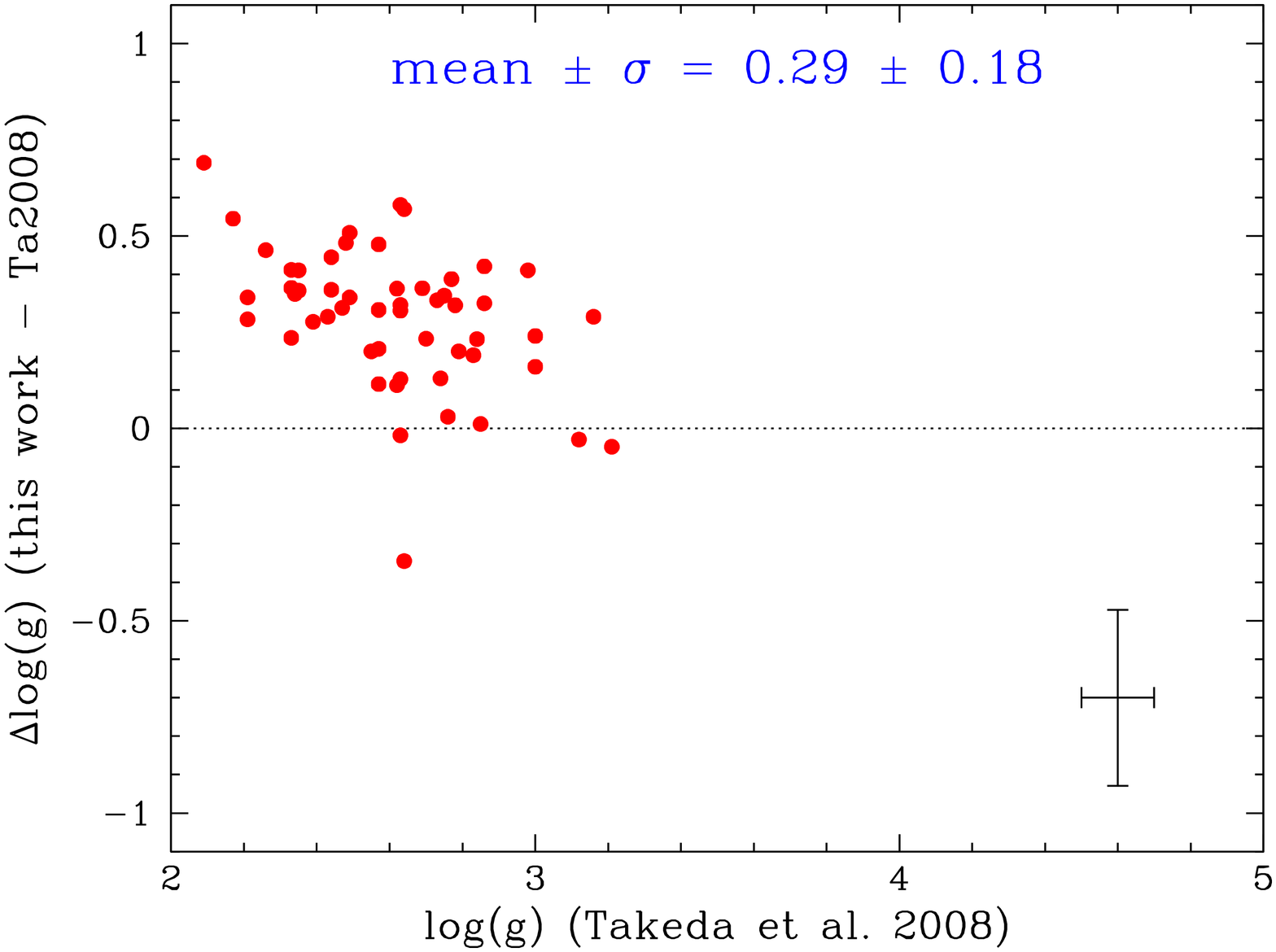}}
\end{minipage} \\
\begin{minipage}[t]{0.3\textwidth}
\centering
\resizebox{\hsize}{!}{\includegraphics[angle=-90]{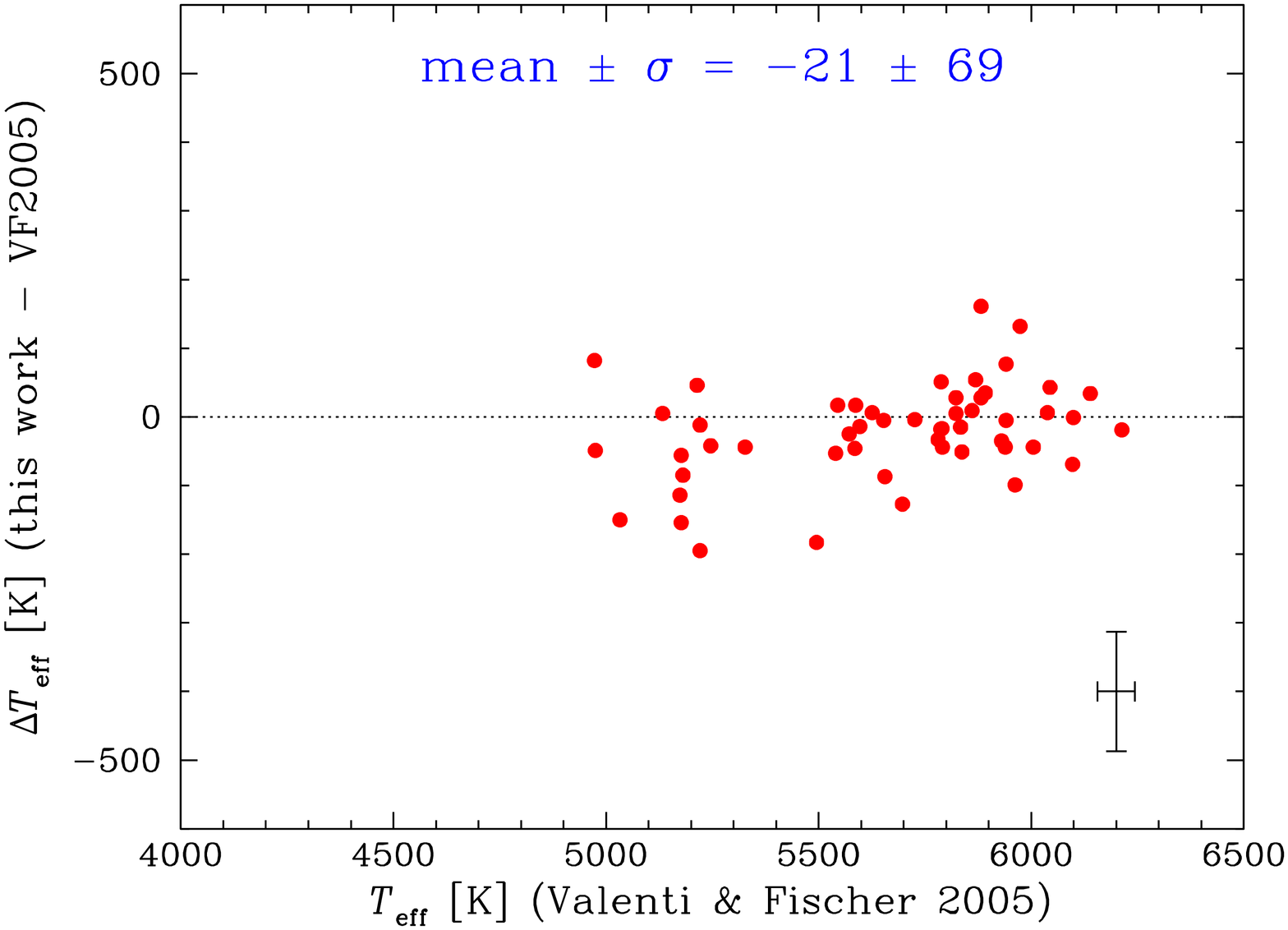}}
\end{minipage}
\begin{minipage}[t]{0.3\textwidth}
\centering
\resizebox{\hsize}{!}{\includegraphics[angle=-90]{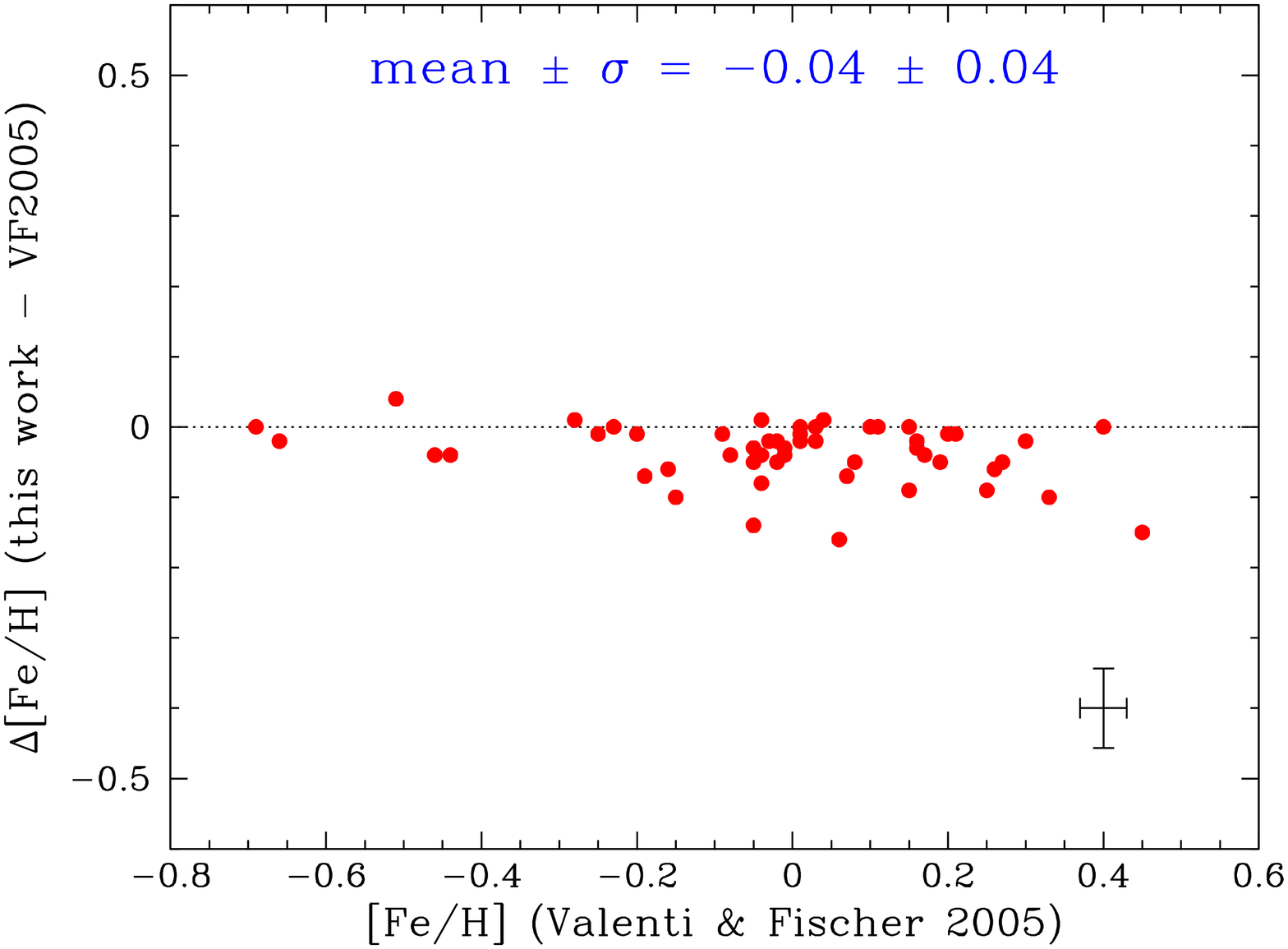}}
\end{minipage}
\begin{minipage}[t]{0.3\textwidth}
\centering
\resizebox{\hsize}{!}{\includegraphics[angle=-90]{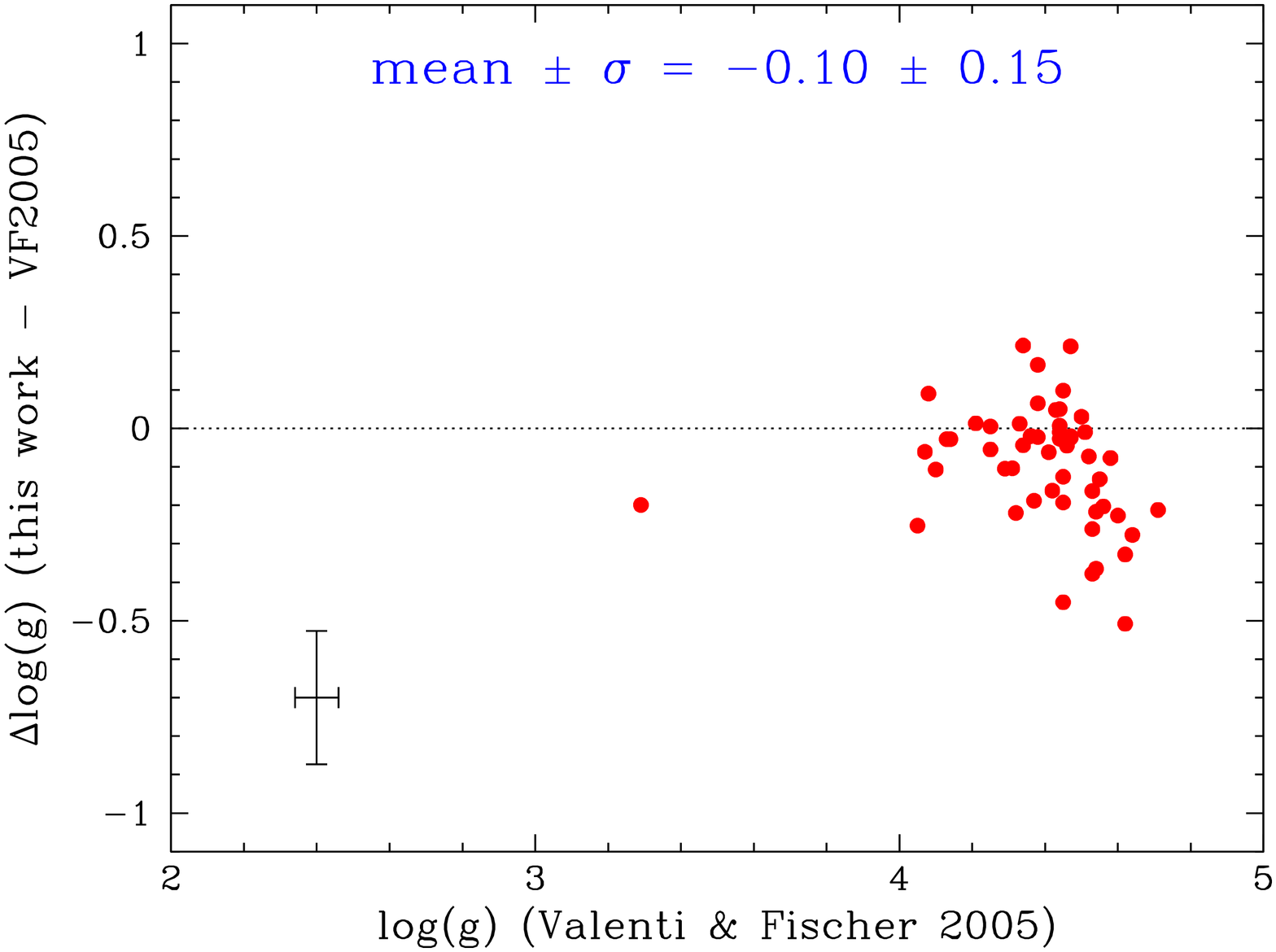}}
\end{minipage}
\caption{Our determination of effective temperature (left column panels),
         metallicity (middle column panels), and surface gravity (right
	 column panels) compared to the values published by other works
	 having stars in common. The error bars plotted represent typical
	 uncertainties (see description in the text). The mean value of the
	 differences ({\it this work} $-$ {\it comparison paper}) and their
	 standard deviations are also shown.}
\label{comp_paratm}
\end{figure*}

\begin{figure*}[t!]
\centering
\begin{minipage}[t]{0.3\textwidth}
\centering
\resizebox{\hsize}{!}{\includegraphics{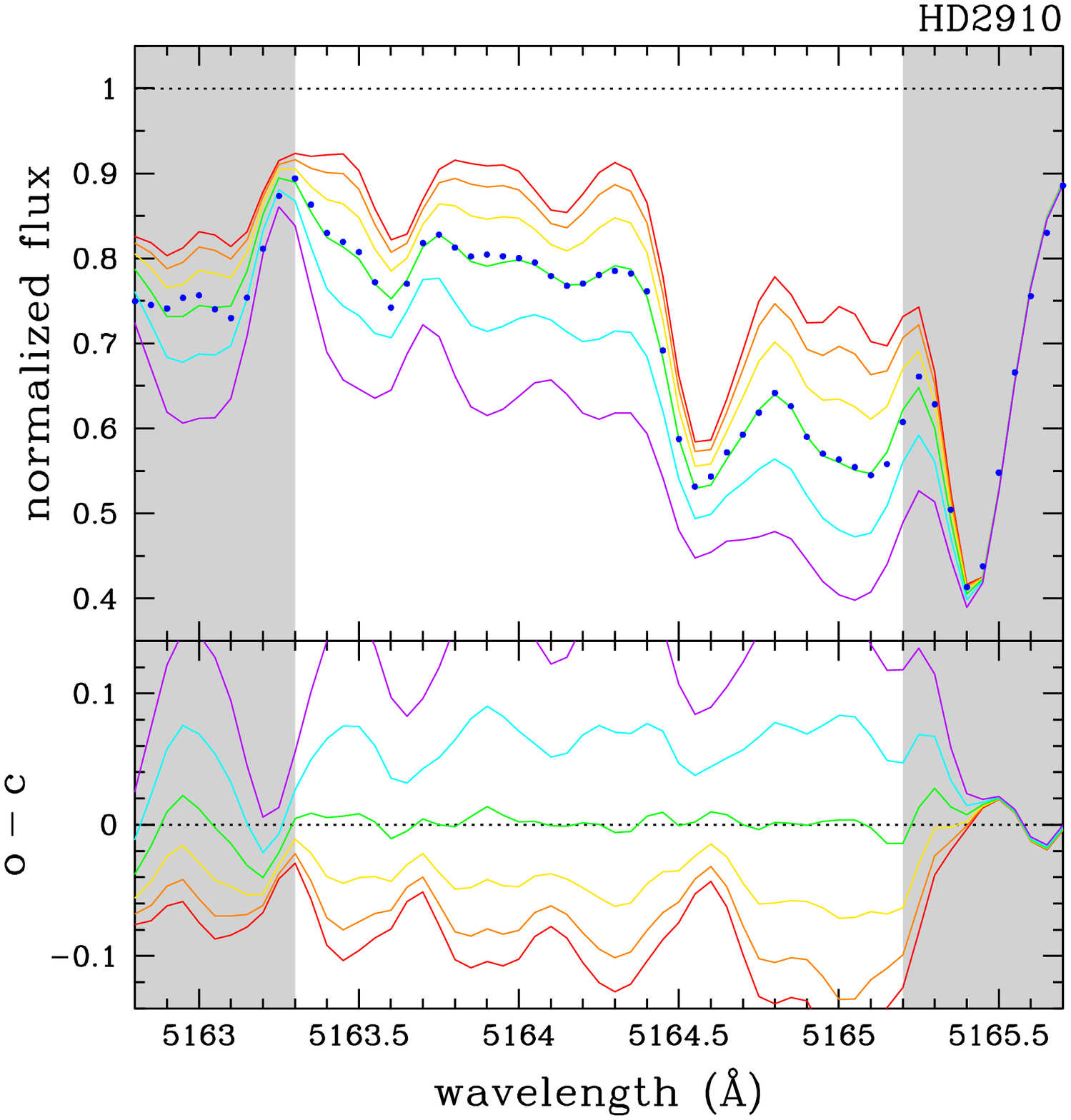}}
\end{minipage}
\begin{minipage}[t]{0.3\textwidth}
\centering
\resizebox{\hsize}{!}{\includegraphics{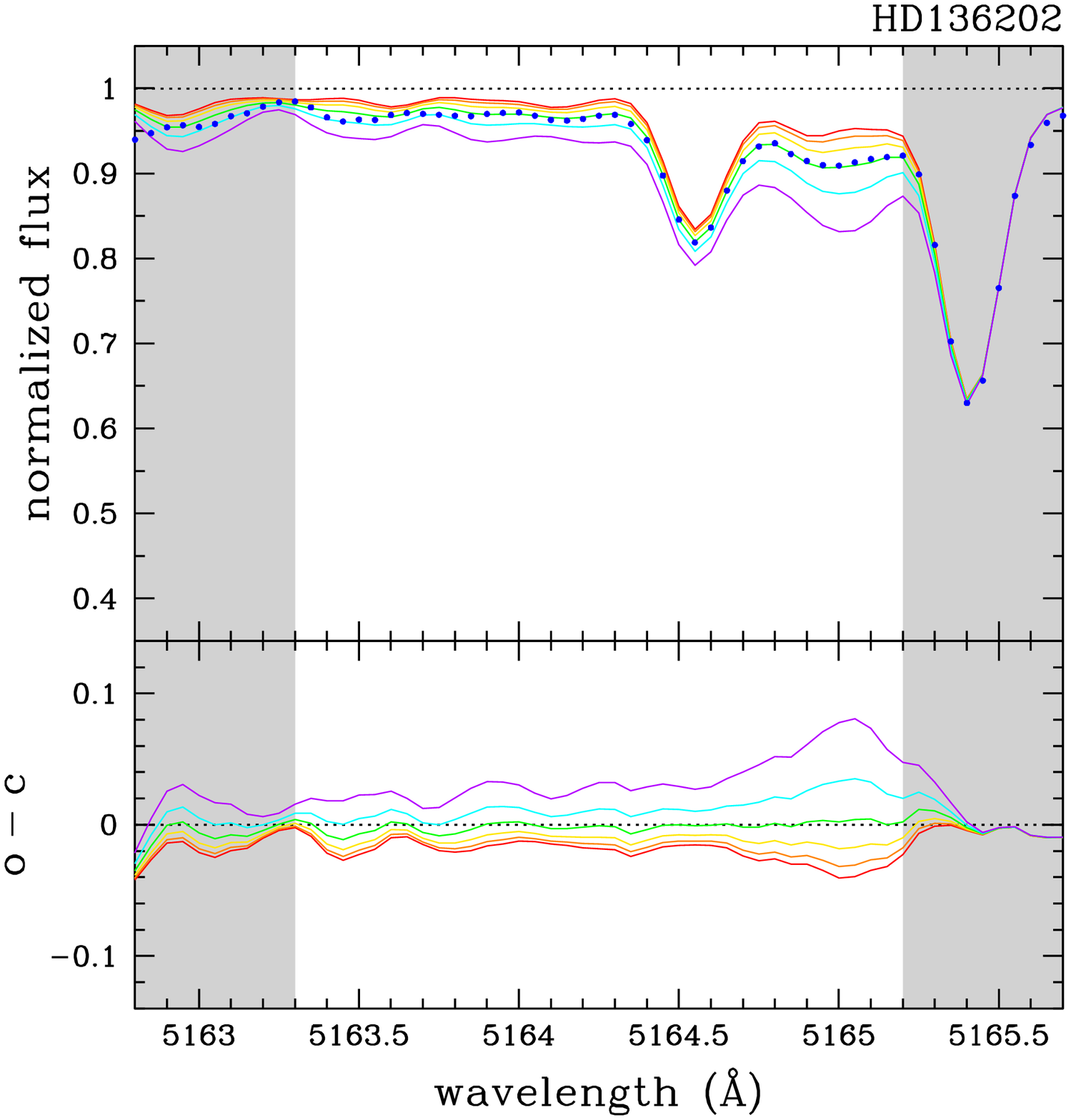}}
\end{minipage}
\begin{minipage}[t]{0.3\textwidth}
\centering
\resizebox{\hsize}{!}{\includegraphics{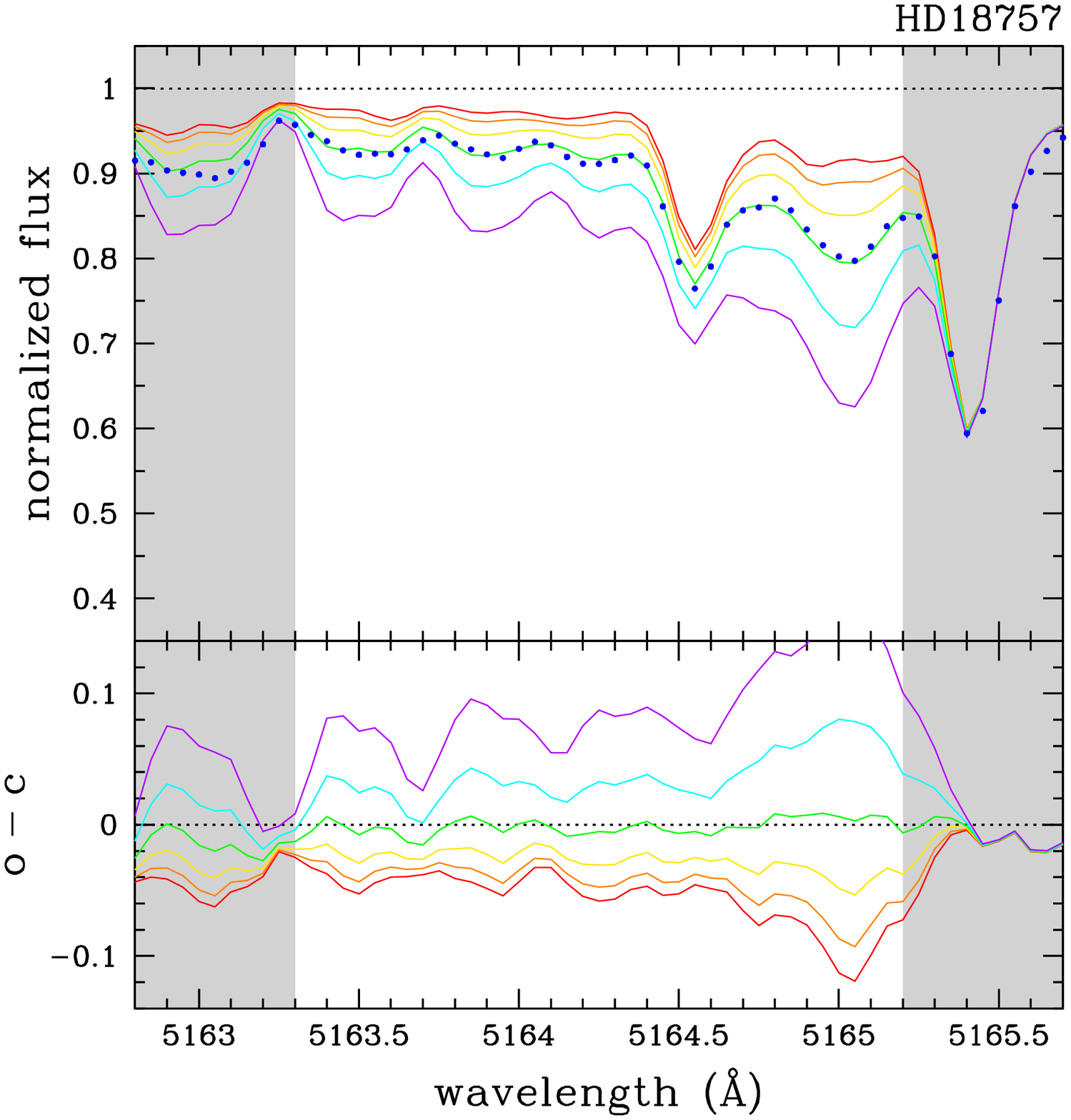}}
\end{minipage}
\caption{Examples of spectral synthesis applied to different types of stars.
         {\it Left panel:} a cold giant star (\teff\ = 4815~K), with [Fe/H]
	 and [C/Fe] close to the solar values;
	 {\it Middle panel:} a hot dwarf star (\teff\ = 6215~K), also with
	 solar values for Fe and C abundance; and
	 {\it Right panel:} a Fe-poor ([Fe/H] = $-$0.31) dwarf star, for
	 which both temperature and C abundance are close to solar.
	 Other parameters are listed in Tables~\ref{dwarfs_tab} and
	 \ref{giants_tab}.}
\label{ex_synth}
\end{figure*}

Spectral synthesis was performed to reproduce the observed spectra of the
sample stars and thus determine their carbon abundance. The technique was
applied to molecular lines of electronic-vibrational head bands of the \C2\
Swan System within spectral regions centred at $\lambda$5128 and
$\lambda$5165 as well as to a C atomic line at $\lambda$5380.3. The atomic
line at $\lambda$5052.2 is also commonly used as a C abundance indicator,
but we preferred not to use it since this line is blended with a strong Fe
line, which may affect the abundance determination, specially for C-poor
stars. Using the MOOG spectral synthesis code \citep{Sneden2002}, synthetic
spectra based on atomic and molecular lines were computed in wavelength
steps of 0.01~\AA, also considering the continuum opacity contribution in
ranges of 0.5~\AA\ and line-broadening corrections, and then fitted to the
observed spectra.

To compute a theoretical spectrum, the MOOG requires, besides a model
atmosphere for each star, some parameters of atomic and molecular spectral
lines, which come from the VALD online database and from \citet{Kurucz1995},
respectively, and some convolution parameters related to spectral line
profiles.

In addition to \C2, another molecule that contributes to the spectral line
formation in the studied wavelength regions is MgH, although its
contribution is relatively small. The $gf$ of \C2\ and MgH lines from the
Kurucz database were revised according to the normalisation of the
H\"onl-London factors \citep{WhitingNicholls1974}. The $gf$ values of atomic
and other molecular lines were also revised when needed to fit solar
spectrum, taken as a reference in our differential chemical analysis.

The parameters of all atomic and molecular lines used to compute the
synthetic spectra of the studied regions are listed in Tables~2 and 3, which
are only available in electronic form at the CDS. Table~2 contains the
following information: the wavelength of the spectral feature, the atomic
and molecular line identification, the lower-level excitation potential, the
oscillator strength, and the dissociation energy $D_0$ (only for molecular
features). Table~3 (strong atomic lines) contains the same information of
Table~2, excepting the dissociation energy parameter.

The convolution parameters responsible for spectral line broadening that are
important to our analysis are:
{\it i)} the spectroscopic instrumental broadening;
{\it ii)} a composite of velocity fields, such as rotation velocity and
macro-turbulence, that we named \Vbroad; and
{\it iii)} the limb darkening of the stellar disc.
We estimated the instrumental broadening by measuring the {\it Full Width
at Half Maximum} (FWHM) of thorium lines in a Th-Ar spectrum observed with
ELODIE. As a first estimate of \Vbroad, the projected rotation velocity
\vsini\ of the star was used, which was computed according to
\citet{Quelozetal1998}. Then, small corrections in \Vbroad\ based on an
eye-trained inspection of the spectral synthesis fit were applied when
needed. Concerning the stellar limb darkening, an estimate of the linear
coefficient ($u$) was performed by interpolating \teff\ and \logg\ in
Table~1 of \citet{DiazCordoves1995}, and it ranges from 0.63 to 0.83 for the
stars in our sample.

Figure~\ref{moon_synth} shows the spectral synthesis method applied to the
observed data. In this illustrative example, the spectrum of the sunlight
reflected by the Moon is plotted, showing how reliable is the reproduction
of the Sun's spectrum. Synthetic spectra were computed for the molecular
band heads around $\lambda$5128 and $\lambda$5165, and the C atomic line at
$\lambda$5380.3, in steps of 0.01~\AA, but resampled in steps of 0.05~\AA\
in order to consistently match the observed spectrum wavelength scale.

\subsection{Uncertainties in the C abundances}
\label{err_synth}

The three wavelength regions investigated in this work provide an
independent determination of C abundance and its respective uncertainty. To
estimate the uncertainties due to the errors in the photospheric parameters,
we developed a routine that takes into account the error propagation of
input parameters used by the MOOG spectral synthesis code, namely, \teff,
\logg, [Fe/H], $\xi$, and \Vbroad. Each one in turn, the MOOG input
parameters are iteratively changed by their errors, and new values of the
abundance ratio [C/Fe] are computed. The difference between new and best
determination provides the uncertainty due to each parameter. The
uncertainty $\sigma$([C/Fe]) is a quadratic sum of individual contributions.
The error in \Vbroad\ was estimated to be of the order of 1~\kms\ or
smaller. The error in the limb darkening coefficient was not considered
since its contribution can be neglected. In any case, the uncertainty in the
C abundances obtained here is dominated by the errors in \teff, \logg, and
[Fe/H].

%
%
\section{Results and discussion}
\label{res}

The photospheric parameters and C abundances obtained in the present work,
together with their uncertainties, are listed in Tables~\ref{dwarfs_tab},
\ref{subgiants_tab}, and \ref{giants_tab} for dwarfs, subgiants, and giants,
respectively. Table~7, only available in electronic form at the CDS,
contains the individual [C/Fe] determinations provided by the three
abundance indicators: the star name, the [C/Fe] abundance ratio and its
uncertainty yielded by the \C2\ molecular band indicator around
$\lambda$5128, [C/Fe] and its uncertainty computed from the $\lambda$5165
indicator, and [C/Fe] and its uncertainty computed from the $\lambda$5380.3
indicator.

Both molecular and atomic indicators agree quite well, providing a $rms$ of
0.06~dex in the C abundances. When comparing the two molecular indicators
$\lambda$5128 and $\lambda$5165, a $rms$ of only 0.02~dex is found. The fact
that the atomic indicator $\lambda$5380.3 is a quite weak line ($EW \sim
20$~m\AA\ in the Sun's spectrum) could explain the larger dispersion.
Nonetheless, we notice that the final C abundances listed in
Tables~\ref{dwarfs_tab}-\ref{giants_tab} are the result of a weighted
average of the abundances yielded by the three C abundance indicators, and
that the weights are inversely proportional to the individual uncertainties
in each determination.

Figure~\ref{comp_paratm} shows a comparison of the photospheric parameters
obtained in this work to those published by other works having stars in
common. Effective temperature, metallicity, and surface gravity of dwarf,
subgiant, and giant stars are compared with the results of
\citet{HekkerMelendez2007}, \citet{LuckHeiter2006,LuckHeiter2007},
\citet{Takedaetal2008}, and \citet{ValentiFischer2005}. The errors in the
difference {\it this work} $-$ {\it comparison paper} are a quadratic sum of
the errors in our photospheric parameters and those published by the papers.
Typical values are represented by error bars plotted in each panel of the
figure. The micro-turbulence velocity comparisons are not shown in
Fig.~\ref{comp_paratm}, but our determination is consistent, within the
uncertainties, with the publications above for which an estimate of this
parameter was also performed.

The photospheric parameters in the comparison papers are also based on an
homogeneous analysis of high signal-to-noise ratio and high resolution data.
We can observe in Fig.~\ref{comp_paratm} a very good agreement between our
estimates and different determinations. An exception is the surface gravity
values of \citet{Takedaetal2008}, which are systematically lower than in our
work. The authors also found that their \logg\ determination is
systematically lower than in previous studies, whose cause seems to be due
to different set of spectral lines used, as they suggested.

For the Sun, our estimate for the photospheric parameters is: \teff\ = 5724
$\pm$ 38~K, \logg\ = 4.37 $\pm$ 0.10, [Fe/H] = $-$0.03 $\pm$ 0.03, and $\xi$
= 0.87 $\pm$ 0.05~\kms. The broadening velocity was set to \Vbroad\ =
1.8~\kms, and the linear limb darkening coefficient, $u$ = 0.69, was
obtained in the same way as for the other stars. These values were used to
compute the solar model atmosphere, which in turn was used to obtain the
solar value of C abundance: [C/Fe] = 0.01 $\pm$ 0.01. Again, as for the
other stars, this is the result of a weighted average of the abundances
yielded by the three C abundance indicators, where the weights are
inversely proportional to the individual uncertainties in each
determination.

Figure~\ref{ex_synth} shows a few examples of spectral synthesis applied to
different stars: a cold giant star, a hot dwarf star, and a Fe-poor and
high-rotation giant star. The good fit of the synthetic spectra to the
observed data in these examples demonstrates that the spectral synthesis
method used here provides reliable results for the different spectral types
of our stellar sample.

\begin{figure*}[t!]
\centering
\begin{minipage}[t]{0.36\textwidth}
\centering
\resizebox{\hsize}{!}{\includegraphics[origin=br,angle=-90]{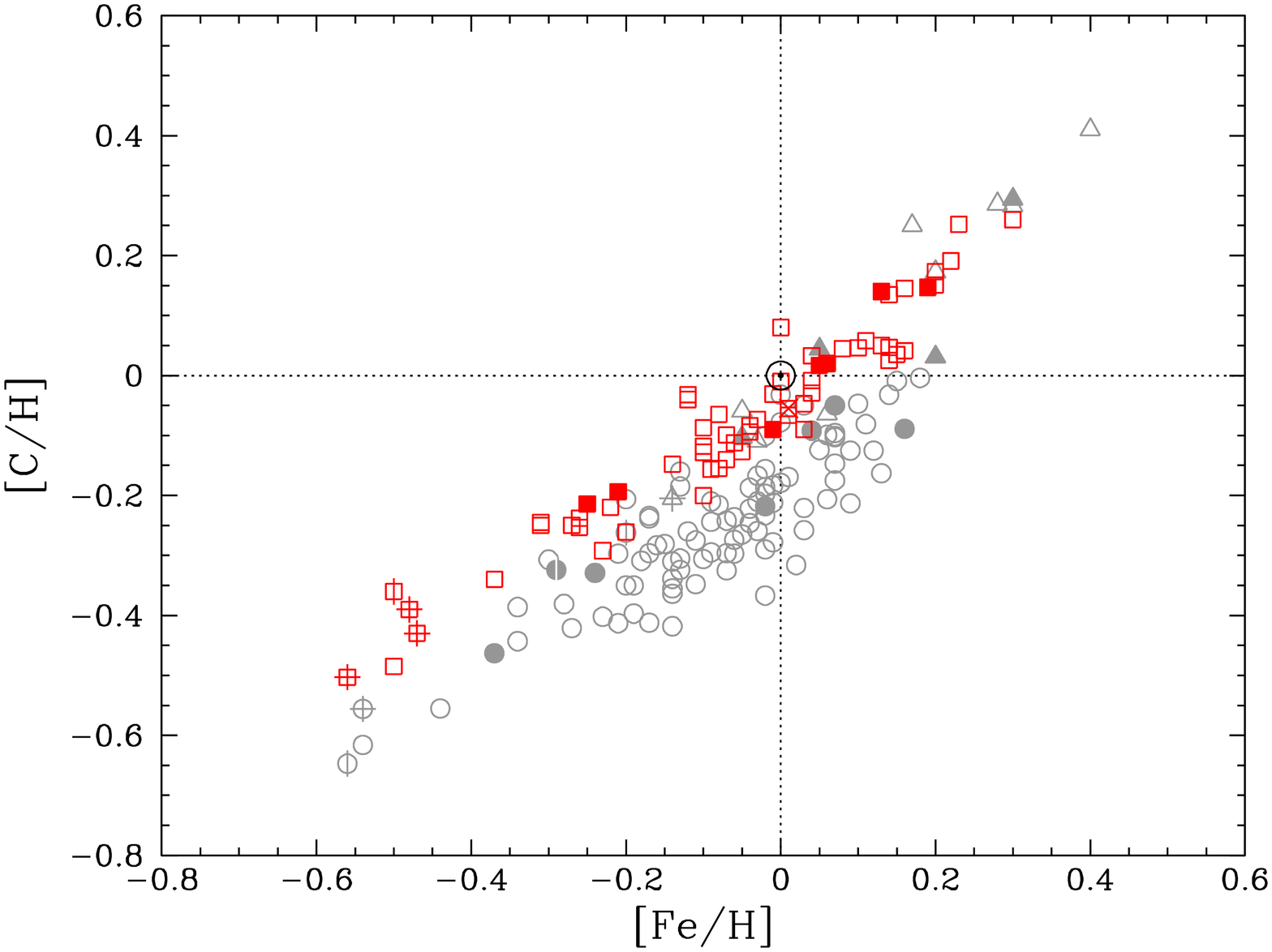}}
\end{minipage}
\begin{minipage}[t]{0.36\textwidth}
\centering
\resizebox{\hsize}{!}{\includegraphics[origin=br,angle=-90]{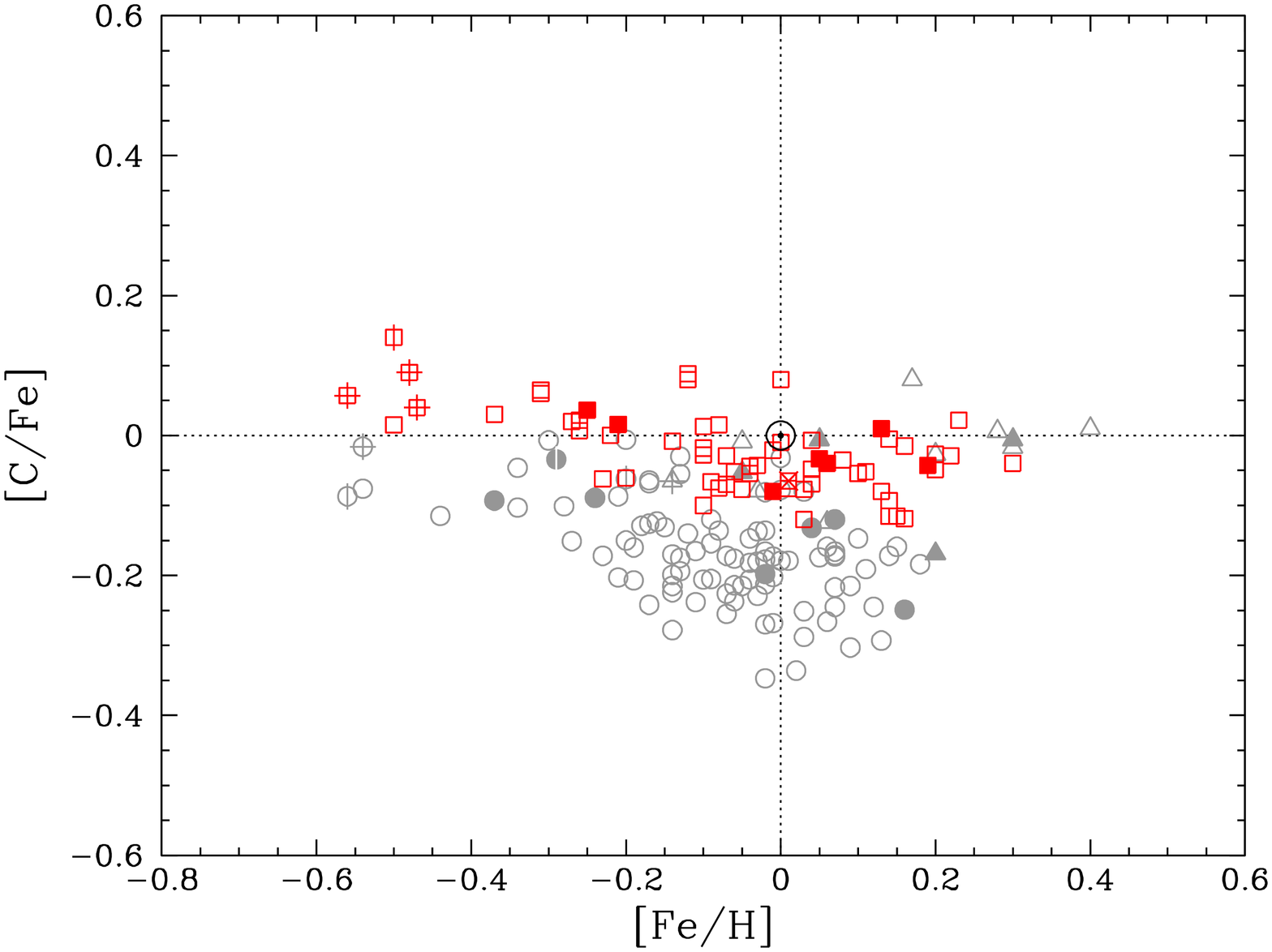}}
\end{minipage}
\begin{minipage}[t]{0.27\textwidth}
\centering
\resizebox{\hsize}{!}{\includegraphics[origin=br]{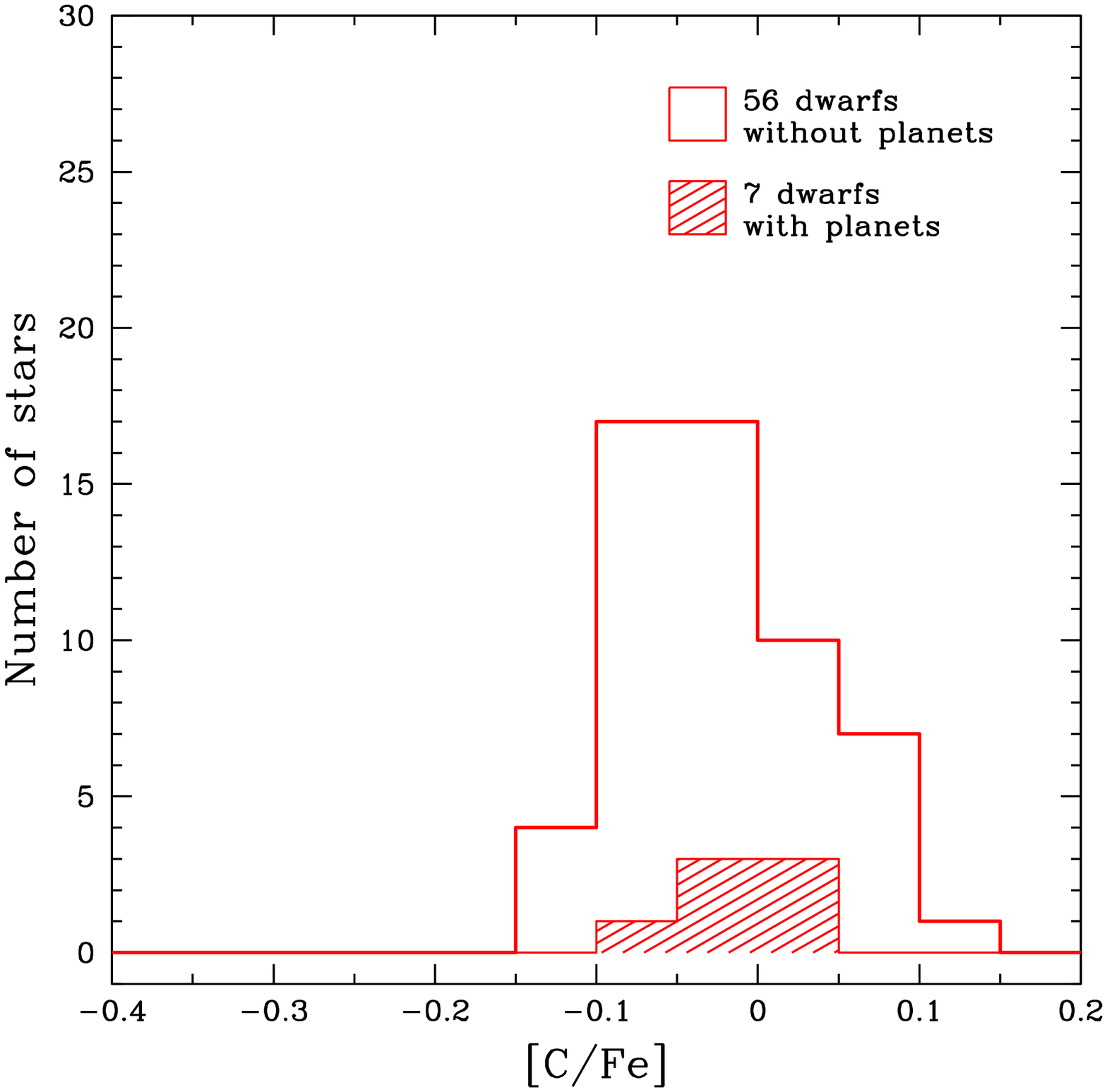}}
\end{minipage} \\
\begin{minipage}[t]{0.36\textwidth}
\centering
\resizebox{\hsize}{!}{\includegraphics[origin=br,angle=-90]{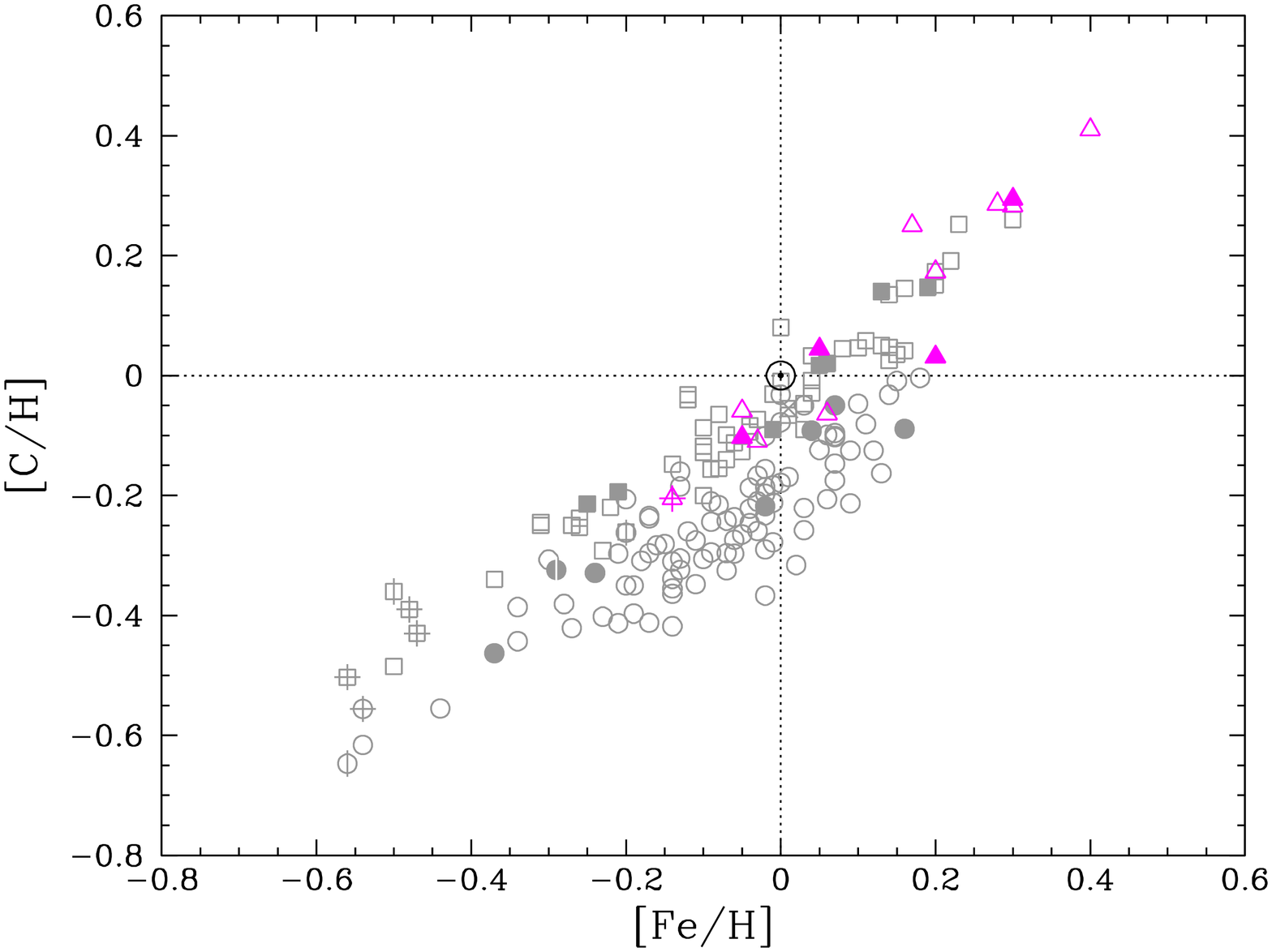}}
\end{minipage}
\begin{minipage}[t]{0.36\textwidth}
\centering
\resizebox{\hsize}{!}{\includegraphics[origin=br,angle=-90]{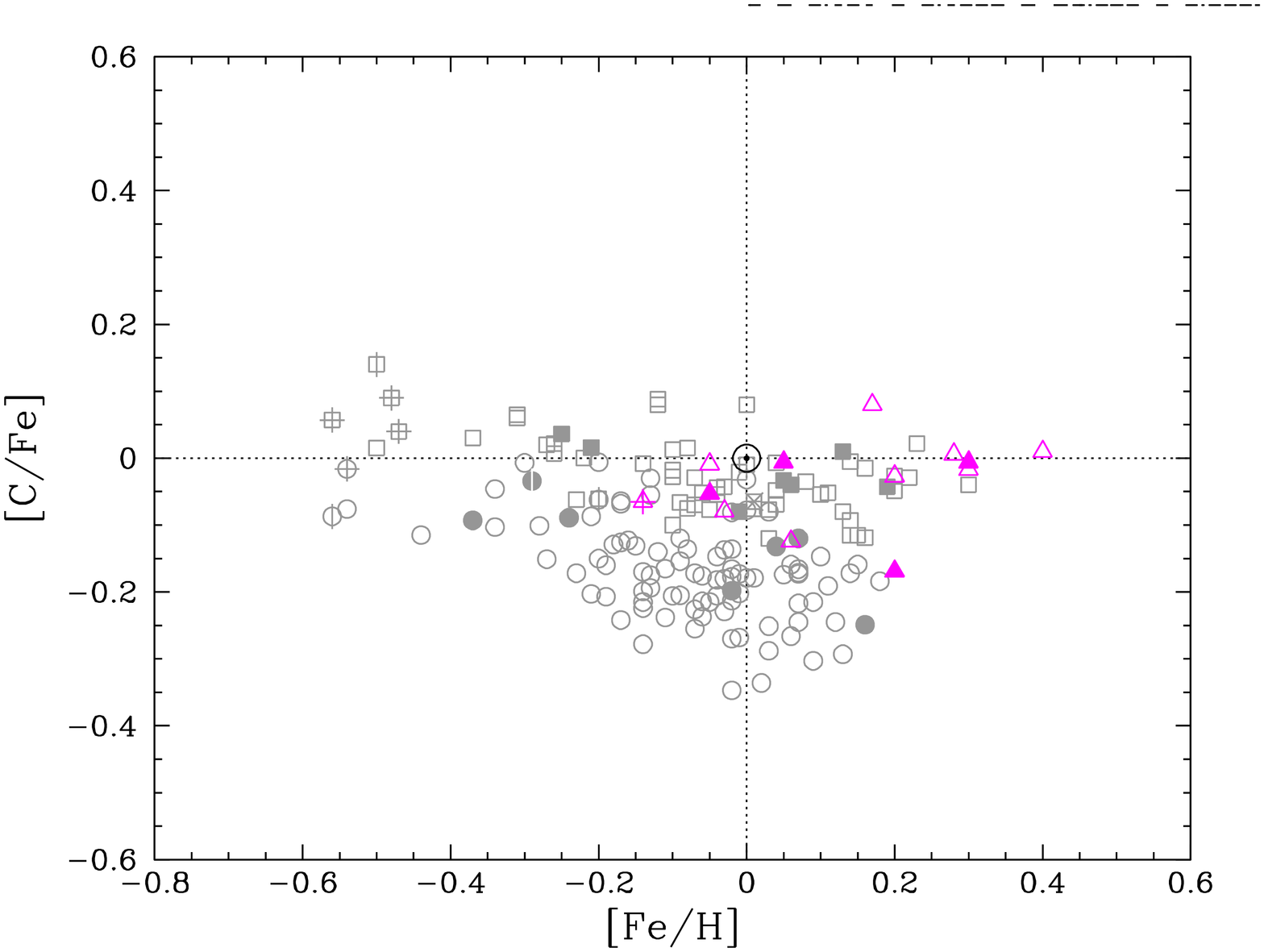}}
\end{minipage}
\begin{minipage}[t]{0.27\textwidth}
\centering
\resizebox{\hsize}{!}{\includegraphics[origin=br]{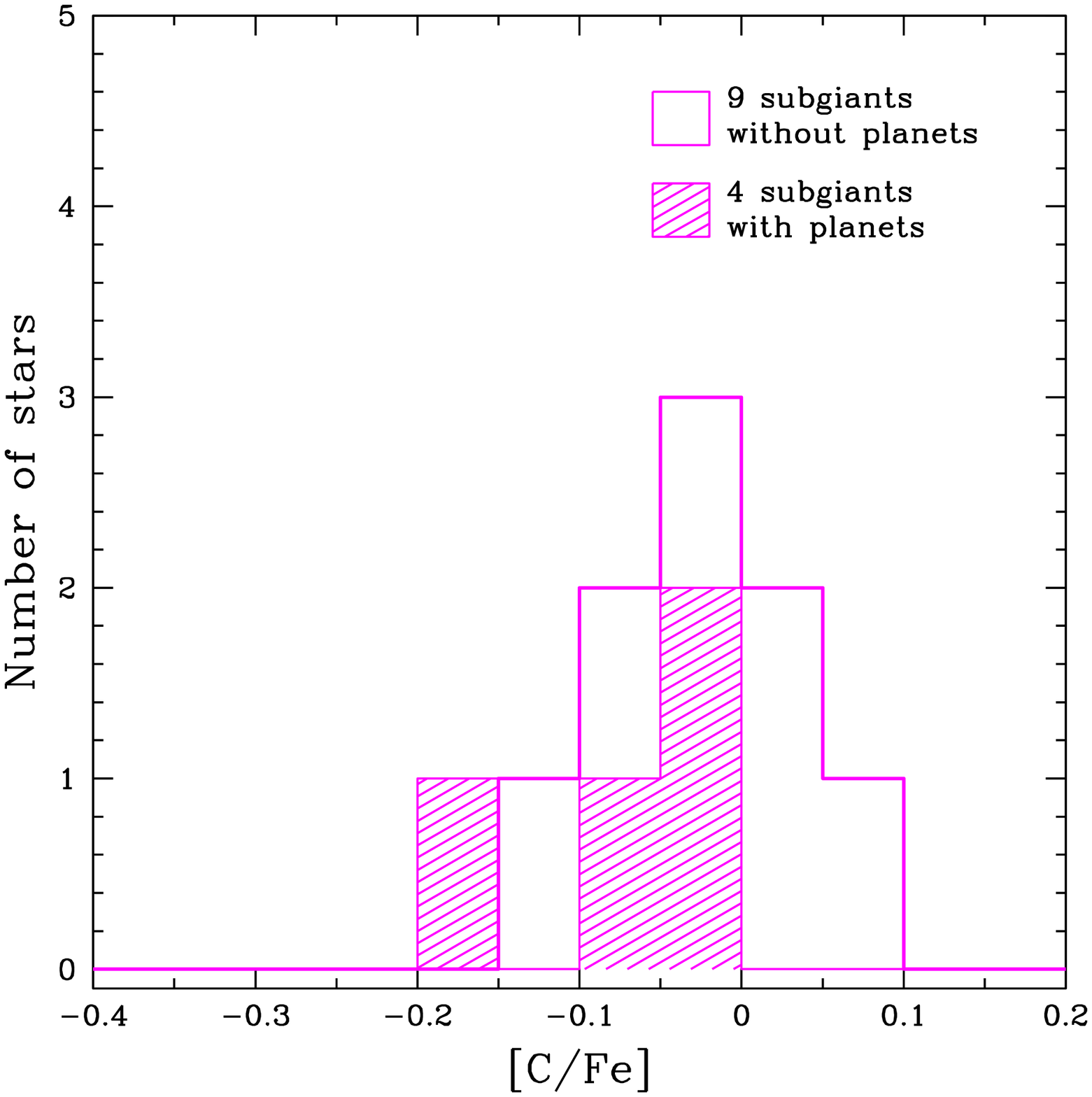}}
\end{minipage} \\
\begin{minipage}[t]{0.36\textwidth}
\centering
\resizebox{\hsize}{!}{\includegraphics[origin=br,angle=-90]{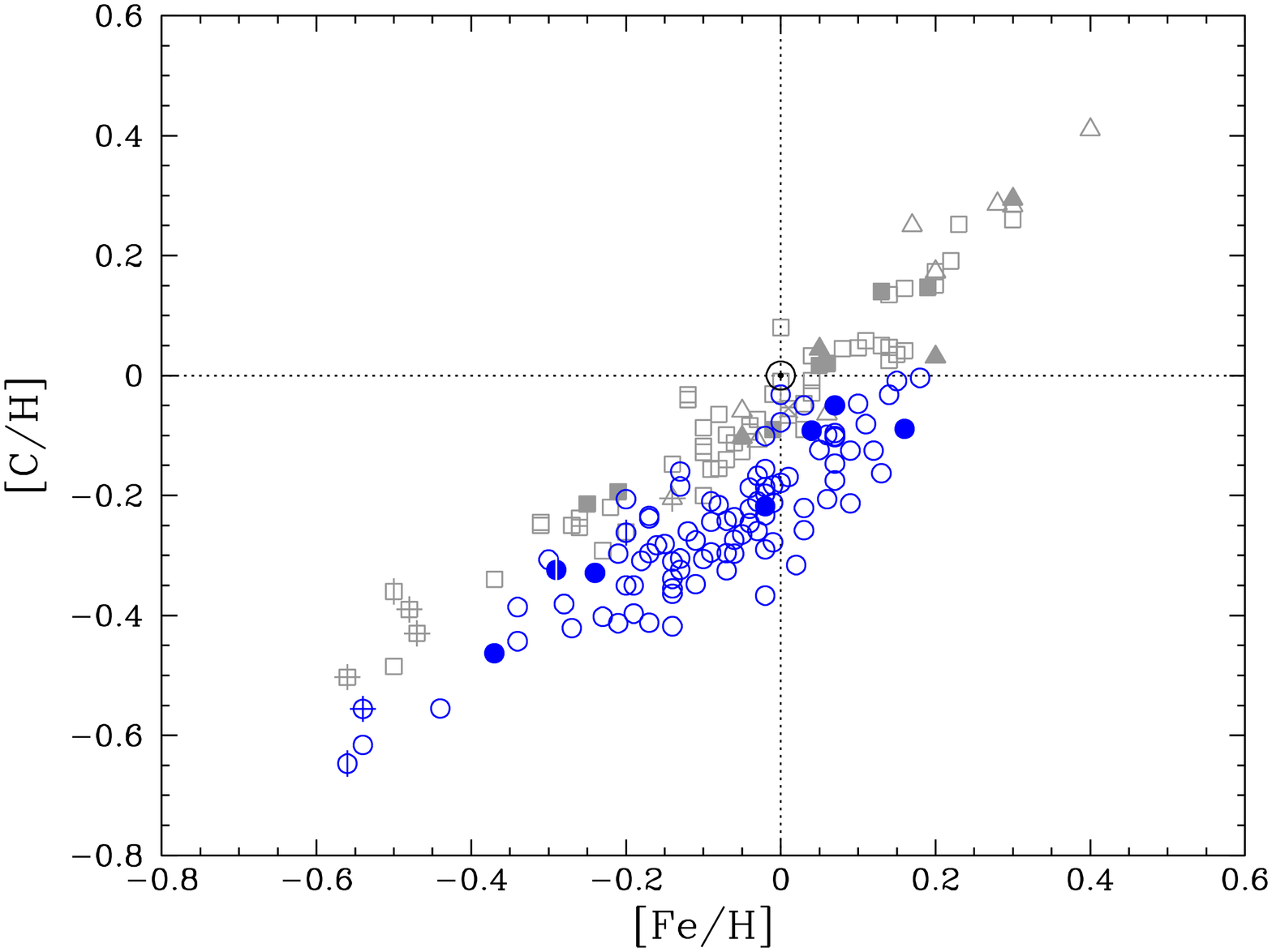}}
\end{minipage}
\begin{minipage}[t]{0.36\textwidth}
\centering
\resizebox{\hsize}{!}{\includegraphics[origin=br,angle=-90]{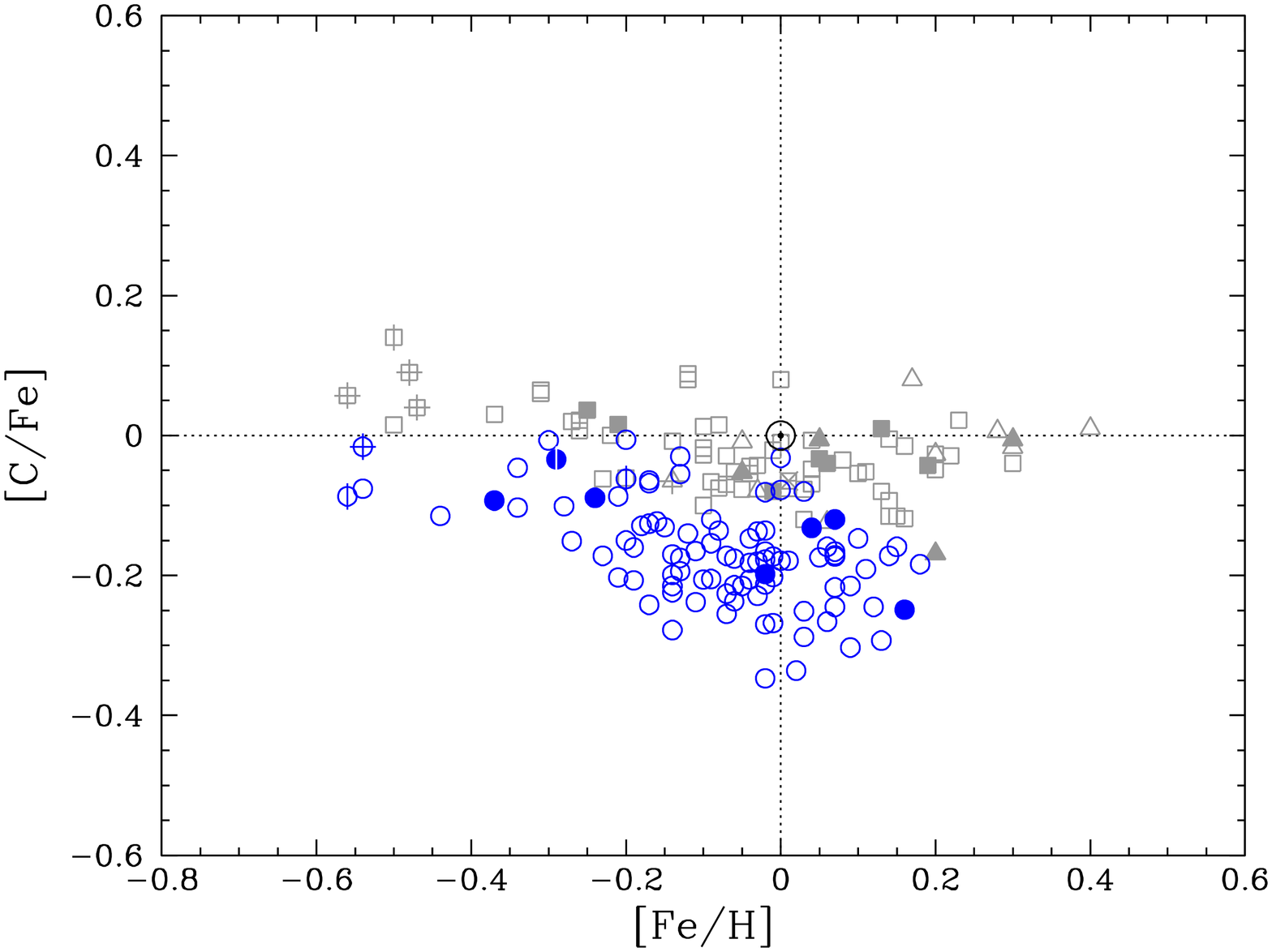}}
\end{minipage}
\begin{minipage}[t]{0.27\textwidth}
\centering
\resizebox{\hsize}{!}{\includegraphics[origin=br]{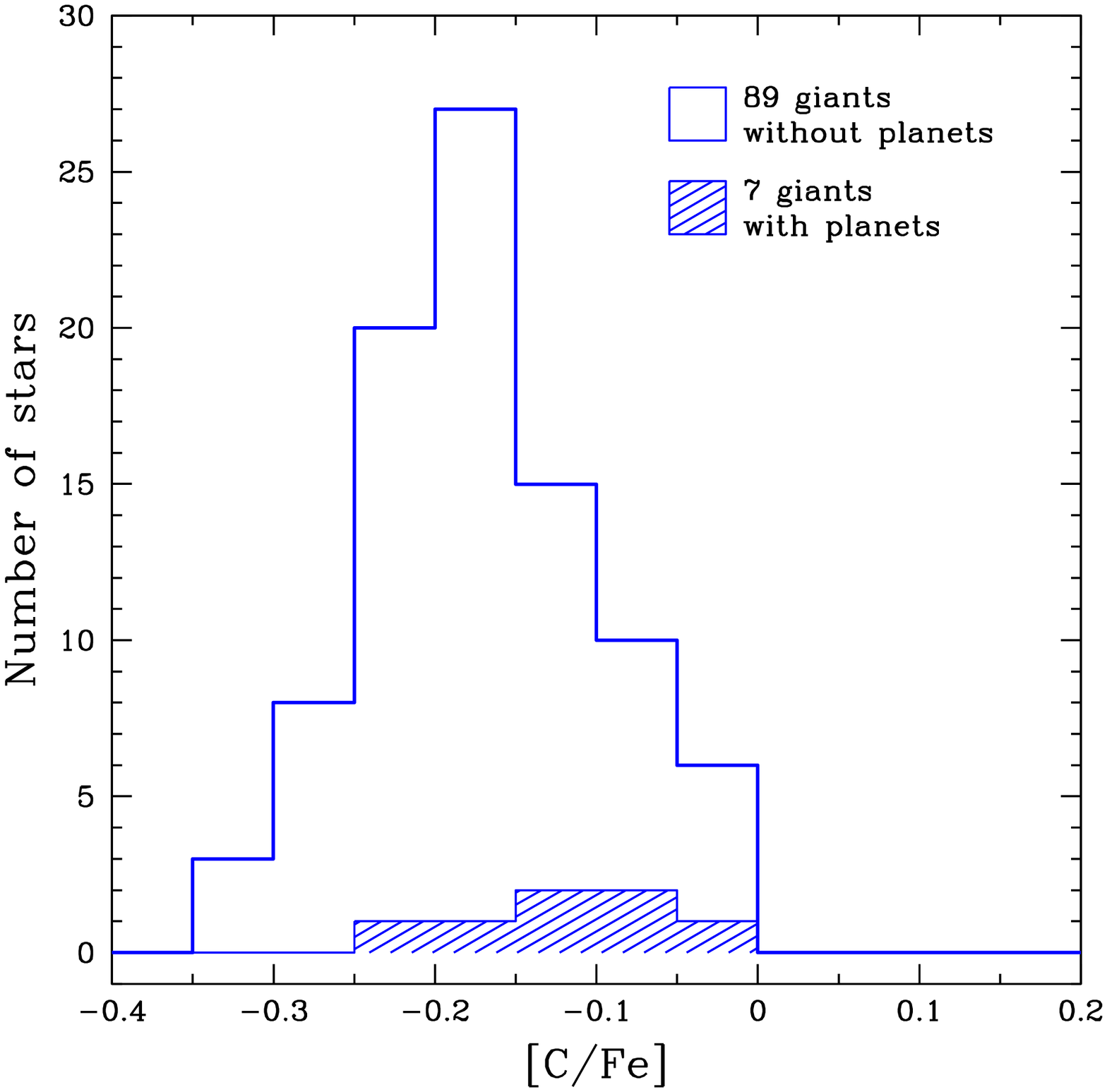}}
\end{minipage}
\caption{{\it Left and middle column panels:} [C/H] and [C/Fe] as a function
         of [Fe/H] for our sub-samples of dwarfs ({\color{red} $\square$}),
	 subgiants ({\color{magenta} $\triangle$}), and giants
	 ({\color{blue} \Large $\circ$}). Superposed symbols indicate the
	 population classification: thick disc members ($\shortmid$),
	 thin/thick disc (+), and one halo star ($\times$). The remaining
	 stars are thin disc members. Planet hosts are represented by filled
	 symbols. The Sun's position is also indicated. Each panel focus on
	 one sub-sample, and the two others are shown in light gray.
	 {\it Right column panels:} [C/Fe] distributions comparing stars
	 with and without planets.}
\label{cfe_feh}
\end{figure*}

Figure~\ref{cfe_feh} shows our results of carbon abundance plotted in the
form of abundance ratios: [C/H] and [C/Fe] in function of metallicity, and
[C/Fe] distributions. Both diagrams and histograms compare the C abundance
of planet-host stars with the abundance of stars for which no planet has
been detected. To clarify visualisation and simplify the discussion, the
three subsamples are presented separately: dwarfs on the top panels,
subgiants on the middle panels, and giants on the bottom panels. Choosing a
metallicity range in which both stars hosting planets or not are equally
represented ($-$0.4 $<$ [Fe/H] $<$ 0.4) and computing the [C/Fe] mean and
standard deviation, we have:
$i)$ $-0.02 \pm 0.04$ and $-0.03 \pm 0.05$ respectively for dwarfs with and
without planets;
$ii)$ $-0.06 \pm 0.07$ and $-0.03 \pm 0.05$ respectively for subgiants with
and without planets; and
$iii)$ $-0.13 \pm 0.07$ and $-0.17 \pm 0.07$ respectively for giants with
and without planets.
Although it seems that planet-host giants are, on average, richer in [C/Fe]
than giants without planets (especially regarding the histogram), according
to these values, there is no indication that, in all subsamples, stars with
and without planets share different C abundance ratios. In addition,
applying a t-test for unequal sample sizes and equal variance, we obtain
that the [C/Fe] distributions are indistinguishable with respect to the
presence or the absence of planets. These results support the
{\it primordial hypothesis} discussed in Sect.~\ref{intro} instead of
{\it self-enrichment}.

Figure~\ref{cfe_feh} also shows that [C/Fe] is clearly depleted (by about
0.14~dex) in the atmosphere of giants in comparison with dwarf stars. This
is in agreement with the results of \citet{Takedaetal2008} and
\citet{Liuetal2010}, which they attributed to evolution-induced mixing of
H-burning products in the envelope of evolved stars in the sense that
carbon-deficient material, produced by the CN-cycle, would be dredged up to
the stellar photosphere.

\begin{figure}[t!]
\centering
\begin{minipage}[t]{0.45\textwidth}
\centering
\resizebox{\hsize}{!}{\includegraphics{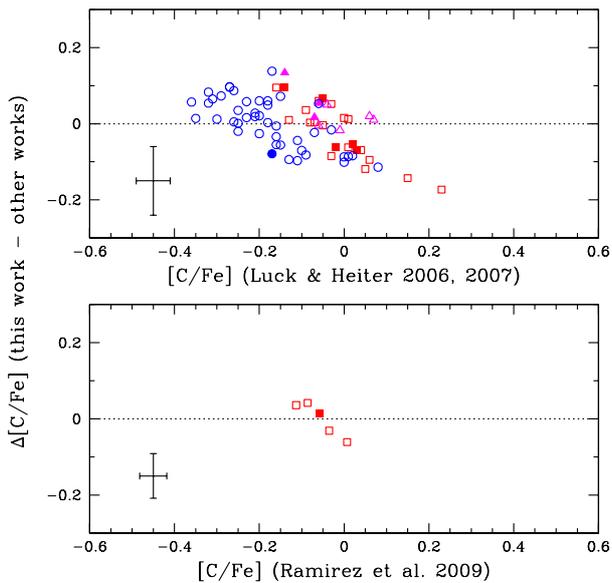}}
\end{minipage}
\caption{[C/Fe] determination form this study compared to the results of
         other works having stars in common. The error bars plotted
	 represent typical uncertainties. Symbols are the same as in
	 Fig.~\ref{cfe_feh}.}
\label{comp_cfe}
\end{figure}

Our [C/Fe] determinations for dwarfs are somewhat lower than in other works
\citep{Ecuvillonetal2004b,GonzalezLaws2000,Reddyetal2003}. We found that the
Sun is slightly overabundant in carbon than other dwarfs in the same
metallicity range, which is the opposite situation found by those authors.
\citet{LuckHeiter2006,LuckHeiter2007} analysed a large sample of stars also
using the spectral synthesis method in the determination of C abundances.
In our study we have several stars in common with their papers, and a
comparison is shown in Fig.~\ref{comp_cfe}. A systematic difference can be
observed: our C determination passes from overabundant to underabundant with
increasing [C/Fe], at least for dwarfs and giants. We notice, however, that
the differences stand mostly from $-0.1$ to $+0.1$~dex, and are compatible
with typical uncertainties.

On the other hand, other recent studies corroborate our results in the sense
that the Sun seems to be overabundant in carbon with respect to other solar
metallicity dwarfs \citep{Ramirezetal2009}. We have only 5 stars in common
with this work, and they are also shown in Fig.~\ref{comp_cfe}. In addition,
Fig.~1 in their paper for the [C/Fe] abundance ratio is very similar to
Fig.~\ref{cfe_feh} for the [C/Fe] distribution of dwarfs in this paper.

It is not unexpected if systematic differences are found between samples
analysed by different methods: different model atmospheres, or different set
of atomic and molecular lines could produce offsets and trends with respect
to other works. Here, stars with planets were compared with their analogues
without planets, and dwarfs and subgiants were compared with giants, and
they were all analysed using the same method. Therefore, any possible offset
that may exist among different works will not affect the analysis of our
differential comparison among these subsamples, either the conclusions that
we draw.

\subsection{Kinematics properties}
\label{kinem}

Kinematic properties of the entire sample was considered to investigate the
Galaxy population membership. Computation of the kinematics required
astrometry (parallaxes and proper motions) and radial velocities. The
astrometry was taken from the new reduction of the Hipparcos catalogue
\citep{vanLeeuwen2007} and the values of radial velocities were measured
from the spectra. The space velocities ($U$, $V$, and $W$) were computed
with respect to the local standard of rest (LSR), where the solar motion
($U$, $V$, $W$) = (10.0, 5.3, 7.2)~\kms\ was adopted
\citep[see][]{DehnenBinney1998}.

With the kinematic data, we have grouped the entire sample into three main
populations: thin disc, thick disc, and halo. The probability that a given
star in the sample belongs to one of the three populations is computed based
on the procedure outlined in \citet{Reddyetal2006} and references therein.
A star whose probability $P_{\rm thin}$, $P_{\rm thick}$, or $P_{\rm halo}$
is greater than or equal to 75\% is considered as thin, thick, or halo star,
respectively. If the probabilities are in-between, they are considered as
either thin/thick disc or thick/halo stars. Out of 172 stars in the sample,
vast majority (162) are of the thin disc population and a few metal-poor
stars are of the thick disc (just 4). HD\,10780 is the only halo star in the
sample. With the exception of one thick disc giant, all planet hosts in our
sample are thin disc members. Population groups are indicated in
Tables~\ref{dwarfs_tab}-\ref{giants_tab} and also in Fig.~\ref{cfe_feh}.
The C abundance results seem to be indistinguishable among the different
populations.

%
%
\section{Conclusions}
\label{concl}

The results presented here represent an homogeneous determination of
photospheric parameters and carbon abundances for a large number of G and K
stars, comprising 63 dwarfs, 13 subgiants, and 96 giants, from which 18 have
already had at least one detected planetary companion at the time of
developing this work (mostly giant planets indeed). Our analysis used high
signal-to-noise ratio and high-resolution spectra that are public available
in the ELODIE online database. We derived the photospheric parameters
through the excitation potential, equivalent widths, and ionisation
equilibrium of iron lines selected in the spectra. In order to compute the C
abundances, we performed spectral synthesis applied to two \C2\ molecular
head bands ($\lambda$5128 and $\lambda$5165) and one C atomic line
($\lambda$5380.3).

The photospheric parameters here estimated (\teff, \logg, [Fe/H], and $\xi$)
are in very good agreement with several works that have stars in common with
our sample. These comparison samples were also analysed based on high
signal-to-noise ratio and high-resolution data. Our estimates are the result
of a precise and homogeneous study, both required conditions to compute
reliable model atmospheres used in abundance determinations. Concerning the
C abundances, our results point out that:

\begin{itemize}

\item[\it i)] regarding the subsamples of dwarfs, subgiants, and giants,
there is no clear indication that stars with and without planets have
different [C/Fe] or [C/H] abundance distributions;

\item[\it ii)] [C/Fe] is clearly underabundant (by about 0.14~dex) in the
atmosphere of giants in comparison with dwarf stars, which is probably the
result of carbon-deficient material, produced by the CN-cycle, dredged-up to
the envelope of evolved stars; subgiant stars, although in small number,
seem to follow the same behaviour of dwarfs; and

\item[\it iii)] the Sun is slightly overabundant in carbon in comparison to
other dwarf stars with the same metallicity.

\end{itemize}

The first of the results above are based on small-number statistics. In
order to draw more reliable conclusions, a larger number of planet-host
stars is required, covering a metallicity range as large as possible.
Adding more elements to the study, e.g. nitrogen, oxygen, and some
refractory metals, would also expand the analysis to a larger context. In
fact, the investigation of volatile and refractory elements with respect to
the distribution of their abundances in function of the condensation
temperature ($T_{\rm C}$) will shed light on recent controversies aroused by
\citet{Chaveroetal2010}. The flat distribution found by these authors should
be confirmed with more precise abundance determination for
$T_{\rm C} \lesssim 300$~K (which includes C, N, and O).

The systematic differences in [C/Fe] found between this and other works are
probably related to different analysis method employed to compute the
abundances: model atmospheres, the atomic and molecular lines used, etc.
Nevertheless, this will not affect our main results since they were based on
differential comparisons among subsamples analysed with the same approach.

In this work, we also considered the kinematic properties of our sample to
investigate C abundances among different population groups. The stars were
separated according to their Galaxy population membership: thin disc, thick
disc, or halo stars. We found that most of these stars are members of the
thin disc. Moreover, excepting one thick disc star, all planet-host stars
are thin disc members. This is probably related either to the fact that
giant planets are normally not much detected in less metal-rich stars (the
thick disc members indeed) or to the fact that the observation samples are
usually limited in distance, which naturally selects thin disc stars.

\begin{acknowledgements}
R. Da Silva thanks the Instituto Nacional de Pesquisas Espaciais (INPE) for
its support.
\end{acknowledgements}

\bibliographystyle{aa}
\bibliography{daSilvaetal2010}

\begin{table*}[p]
\centering
  \caption[]{Atomic parameters and solar equivalent widths for 72
             \ion{Fe}{i} and 12 \ion{Fe}{ii} lines used. The equivalent
	     widths listed are those measured (using ARES) in the solar
	     spectrum of the sunlight reflected by the Moon.}
  \label{linelist}
\begin{tabular}{cccccccccccc}
\noalign{\smallskip}
\cline{1-5}\cline{8-12}
\noalign{\smallskip}
$\lambda$ [\AA] & ident. & $\chi$ [eV] & \loggf & $EW_{\odot}$ [m\AA] &
& &
$\lambda$ [\AA] & ident. & $\chi$ [eV] & \loggf & $EW_{\odot}$ [m\AA] \\
\noalign{\smallskip}
\cline{1-5}\cline{8-12}
\noalign{\smallskip}
4080.88 & \ion{Fe}{i}  & 3.65 & $-$1.543 &  58.5 & & & 5983.69 & \ion{Fe}{i}  & 4.55 & $-$0.719 &  66.5 \\
5247.06 & \ion{Fe}{i}  & 0.09 & $-$4.932 &  68.1 & & & 5984.82 & \ion{Fe}{i}  & 4.73 & $-$0.335 &  83.5 \\
5322.05 & \ion{Fe}{i}  & 2.28 & $-$2.896 &  62.1 & & & 6024.06 & \ion{Fe}{i}  & 4.55 & $-$0.124 & 114.7 \\
5501.48 & \ion{Fe}{i}  & 0.96 & $-$3.053 & 116.8 & & & 6027.06 & \ion{Fe}{i}  & 4.08 & $-$1.180 &  64.4 \\
5522.45 & \ion{Fe}{i}  & 4.21 & $-$1.419 &  44.9 & & & 6056.01 & \ion{Fe}{i}  & 4.73 & $-$0.498 &  73.1 \\
5543.94 & \ion{Fe}{i}  & 4.22 & $-$1.070 &  63.9 & & & 6065.49 & \ion{Fe}{i}  & 2.61 & $-$1.616 & 119.4 \\
5546.51 & \ion{Fe}{i}  & 4.37 & $-$1.124 &  54.0 & & & 6079.01 & \ion{Fe}{i}  & 4.65 & $-$1.009 &  46.3 \\
5560.22 & \ion{Fe}{i}  & 4.43 & $-$1.064 &  52.3 & & & 6082.72 & \ion{Fe}{i}  & 2.22 & $-$3.566 &  34.6 \\
5587.58 & \ion{Fe}{i}  & 4.14 & $-$1.656 &  36.4 & & & 6089.57 & \ion{Fe}{i}  & 5.02 & $-$0.883 &  35.0 \\
5618.64 & \ion{Fe}{i}  & 4.21 & $-$1.298 &  49.4 & & & 6094.38 & \ion{Fe}{i}  & 4.65 & $-$1.566 &  19.6 \\
5619.60 & \ion{Fe}{i}  & 4.39 & $-$1.435 &  34.4 & & & 6096.67 & \ion{Fe}{i}  & 3.98 & $-$1.776 &  37.3 \\
5633.95 & \ion{Fe}{i}  & 4.99 & $-$0.385 &  67.2 & & & 6151.62 & \ion{Fe}{i}  & 2.18 & $-$3.296 &  49.9 \\
5635.83 & \ion{Fe}{i}  & 4.26 & $-$1.556 &  32.7 & & & 6157.73 & \ion{Fe}{i}  & 4.07 & $-$1.240 &  60.7 \\
5638.27 & \ion{Fe}{i}  & 4.22 & $-$0.809 &  77.9 & & & 6165.36 & \ion{Fe}{i}  & 4.14 & $-$1.503 &  41.4 \\
5641.44 & \ion{Fe}{i}  & 4.26 & $-$0.969 &  66.9 & & & 6180.21 & \ion{Fe}{i}  & 2.73 & $-$2.636 &  60.0 \\
5649.99 & \ion{Fe}{i}  & 5.10 & $-$0.785 &  36.1 & & & 6188.00 & \ion{Fe}{i}  & 3.94 & $-$1.631 &  47.2 \\
5651.47 & \ion{Fe}{i}  & 4.47 & $-$1.763 &  18.9 & & & 6200.32 & \ion{Fe}{i}  & 2.61 & $-$2.395 &  73.7 \\
5652.32 & \ion{Fe}{i}  & 4.26 & $-$1.751 &  27.4 & & & 6226.74 & \ion{Fe}{i}  & 3.88 & $-$2.066 &  26.5 \\
5653.87 & \ion{Fe}{i}  & 4.39 & $-$1.402 &  37.8 & & & 6229.24 & \ion{Fe}{i}  & 2.84 & $-$2.893 &  39.7 \\
5661.35 & \ion{Fe}{i}  & 4.28 & $-$1.828 &  23.1 & & & 6240.65 & \ion{Fe}{i}  & 2.22 & $-$3.294 &  49.9 \\
5662.52 & \ion{Fe}{i}  & 4.18 & $-$0.601 &  95.6 & & & 6265.14 & \ion{Fe}{i}  & 2.18 & $-$2.559 &  87.1 \\
5667.52 & \ion{Fe}{i}  & 4.18 & $-$1.292 &  52.4 & & & 6380.75 & \ion{Fe}{i}  & 4.19 & $-$1.321 &  53.7 \\
5679.03 & \ion{Fe}{i}  & 4.65 & $-$0.756 &  59.6 & & & 6498.94 & \ion{Fe}{i}  & 0.96 & $-$4.631 &  48.9 \\
5701.55 & \ion{Fe}{i}  & 2.56 & $-$2.162 &  82.6 & & & 6608.03 & \ion{Fe}{i}  & 2.28 & $-$3.959 &  16.9 \\
5731.77 & \ion{Fe}{i}  & 4.26 & $-$1.124 &  56.3 & & & 6627.55 & \ion{Fe}{i}  & 4.55 & $-$1.481 &  27.6 \\
5741.85 & \ion{Fe}{i}  & 4.26 & $-$1.626 &  31.6 & & & 6703.57 & \ion{Fe}{i}  & 2.76 & $-$3.022 &  37.3 \\
5752.04 & \ion{Fe}{i}  & 4.55 & $-$0.917 &  56.9 & & & 6726.67 & \ion{Fe}{i}  & 4.61 & $-$1.053 &  46.8 \\
5775.08 & \ion{Fe}{i}  & 4.22 & $-$1.124 &  58.9 & & & 6733.16 & \ion{Fe}{i}  & 4.64 & $-$1.429 &  26.0 \\
5793.92 & \ion{Fe}{i}  & 4.22 & $-$1.622 &  33.5 & & & 6750.16 & \ion{Fe}{i}  & 2.42 & $-$2.614 &  74.6 \\
5806.73 & \ion{Fe}{i}  & 4.61 & $-$0.893 &  54.6 & & & 6752.71 & \ion{Fe}{i}  & 4.64 & $-$1.233 &  37.1 \\
5809.22 & \ion{Fe}{i}  & 3.88 & $-$1.614 &  50.4 & & & 5234.63 & \ion{Fe}{ii} & 3.22 & $-$2.233 &  85.3 \\
5814.81 & \ion{Fe}{i}  & 4.28 & $-$1.820 &  22.2 & & & 5325.56 & \ion{Fe}{ii} & 3.22 & $-$3.203 &  39.4 \\
5852.22 & \ion{Fe}{i}  & 4.55 & $-$1.187 &  41.0 & & & 5414.07 & \ion{Fe}{ii} & 3.22 & $-$3.569 &  25.4 \\
5855.08 & \ion{Fe}{i}  & 4.61 & $-$1.529 &  22.8 & & & 5425.25 & \ion{Fe}{ii} & 3.20 & $-$3.228 &  40.1 \\
5856.09 & \ion{Fe}{i}  & 4.29 & $-$1.564 &  34.9 & & & 5991.38 & \ion{Fe}{ii} & 3.15 & $-$3.533 &  29.3 \\
5862.36 & \ion{Fe}{i}  & 4.55 & $-$0.404 &  86.9 & & & 6084.11 & \ion{Fe}{ii} & 3.20 & $-$3.777 &  21.1 \\
5905.68 & \ion{Fe}{i}  & 4.65 & $-$0.775 &  58.1 & & & 6149.25 & \ion{Fe}{ii} & 3.89 & $-$2.719 &  36.0 \\
5916.26 & \ion{Fe}{i}  & 2.45 & $-$2.920 &  57.1 & & & 6247.56 & \ion{Fe}{ii} & 3.89 & $-$2.349 &  54.7 \\
5927.79 & \ion{Fe}{i}  & 4.65 & $-$1.057 &  42.2 & & & 6369.46 & \ion{Fe}{ii} & 2.89 & $-$4.127 &  17.4 \\
5929.68 & \ion{Fe}{i}  & 4.55 & $-$1.211 &  38.2 & & & 6416.93 & \ion{Fe}{ii} & 3.89 & $-$2.635 &  41.1 \\
5930.19 & \ion{Fe}{i}  & 4.65 & $-$0.326 &  86.6 & & & 6432.69 & \ion{Fe}{ii} & 2.89 & $-$3.564 &  41.0 \\
5934.66 & \ion{Fe}{i}  & 3.93 & $-$1.091 &  75.3 & & & 6456.39 & \ion{Fe}{ii} & 3.90 & $-$2.114 &  60.9 \\
\cline{1-5}\cline{8-12}
\end{tabular}
\end{table*}

\addtocounter{table}{2}
\begin{table*}[p]
\centering
  \caption[]{Photospheric parameters and [C/Fe] abundance ratios for 63
             dwarf stars. The stars with planets are listed first, and then
	     the stars for which no planet has been detected. The broadening
	     velocity \Vbroad\ and the Galaxy population group (thin disc,
	     thick disc, or halo) are also shown.}
  \label{dwarfs_tab}
\begin{tabular}{lccc r@{ }l cc r@{}l r@{}l}
\noalign{\smallskip}\hline\hline\noalign{\smallskip}
Star &
\parbox[c]{1.1cm}{\centering Spectral type} &
\parbox[c]{1.4cm}{\centering Population group} &
\parbox[c]{1.1cm}{\centering \Vbroad\ [\kms]} &
\multicolumn{2}{c}{\parbox[c]{1.0cm}{\centering \teff\ $ \pm\ \sigma$ [K]}} &
\parbox[c]{1.2cm}{\centering \logg\ $\pm\ \sigma$} &
\parbox[c]{1.1cm}{\centering $\xi$ $\pm\ \sigma$ [\kms]} &
\multicolumn{2}{c}{[Fe/H] $\pm\ \sigma$} &
\multicolumn{2}{c}{[C/Fe] $\pm\ \sigma$} \\
\noalign{\smallskip}\hline\noalign{\smallskip}
HD\,143761  & G0\,Va      & thin       & 3.1  & 5851 & $\pm$ 45  & 4.34 $\pm$ 0.13 & 1.04 $\pm$ 0.06 & $-$0&.21 $\pm$ 0.04 &    0&.02 $\pm$ 0.03 \\
HD\,209458  & G0\,V       & thin       & 2.5  & 6098 & $\pm$ 50  & 4.45 $\pm$ 0.14 & 1.23 $\pm$ 0.07 & $-$0&.01 $\pm$ 0.04 & $-$0&.08 $\pm$ 0.07 \\
HD\,217014  & G2.5\,IVa   & thin       & 2.4  & 5769 & $\pm$ 50  & 4.26 $\pm$ 0.11 & 0.88 $\pm$ 0.06 &    0&.19 $\pm$ 0.03 & $-$0&.04 $\pm$ 0.02 \\
HD\,3651    & K0\,V       & thin       & 0.0  & 5026 & $\pm$ 154 & 4.00 $\pm$ 0.17 & 0.14 $\pm$ 0.55 &    0&.13 $\pm$ 0.05 &    0&.01 $\pm$ 0.02 \\
HD\,7924    & K0          & thin       & 1.2  & 5121 & $\pm$ 51  & 4.50 $\pm$ 0.14 & 0.26 $\pm$ 0.10 & $-$0&.25 $\pm$ 0.05 &    0&.04 $\pm$ 0.02 \\
HD\,95128   & G1\,V       & thin       & 2.2  & 5910 & $\pm$ 56  & 4.36 $\pm$ 0.12 & 1.00 $\pm$ 0.08 &    0&.05 $\pm$ 0.04 & $-$0&.03 $\pm$ 0.03 \\
HD\,9826    & F8\,V       & thin       & 10.1 & 6194 & $\pm$ 68  & 4.20 $\pm$ 0.21 & 1.64 $\pm$ 0.10 &    0&.06 $\pm$ 0.07 & $-$0&.04 $\pm$ 0.08 \\
\hline\noalign{\smallskip}
HD\,10307   & G1.5\,V     & thin       & 2.7  & 5859 & $\pm$ 54  & 4.27 $\pm$ 0.16 & 1.01 $\pm$ 0.07 &    0&.04 $\pm$ 0.04 & $-$0&.07 $\pm$ 0.03 \\
HD\,10476   & K1\,V       & thin       & 0.0  & 5096 & $\pm$ 83  & 4.27 $\pm$ 0.17 & 0.26 $\pm$ 0.26 & $-$0&.08 $\pm$ 0.04 &    0&.01 $\pm$ 0.01 \\
HD\,10780   & K0\,V       & halo       & 0.0  & 5283 & $\pm$ 87  & 4.32 $\pm$ 0.13 & 0.41 $\pm$ 0.18 &    0&.01 $\pm$ 0.04 & $-$0&.07 $\pm$ 0.01 \\
HD\,109358  & G0\,V       & thin       & 2.4  & 5895 & $\pm$ 46  & 4.43 $\pm$ 0.13 & 1.12 $\pm$ 0.07 & $-$0&.22 $\pm$ 0.04 &    0&.00 $\pm$ 0.03 \\
HD\,12051   & G5          & thin       & 0.0  & 5312 & $\pm$ 108 & 4.11 $\pm$ 0.19 & 0.41 $\pm$ 0.21 &    0&.20 $\pm$ 0.05 & $-$0&.03 $\pm$ 0.02 \\
HD\,12235   & G2\,IV      & thin       & 5.6  & 6028 & $\pm$ 56  & 4.18 $\pm$ 0.18 & 1.36 $\pm$ 0.06 &    0&.23 $\pm$ 0.05 &    0&.02 $\pm$ 0.03 \\
HD\,12846   & G2\,V       & thin       & 0.4  & 5632 & $\pm$ 70  & 4.26 $\pm$ 0.20 & 0.96 $\pm$ 0.11 & $-$0&.27 $\pm$ 0.06 &    0&.02 $\pm$ 0.06 \\
HD\,135599  & K0          & thin       & 3.1  & 5209 & $\pm$ 101 & 4.42 $\pm$ 0.14 & 0.72 $\pm$ 0.18 & $-$0&.10 $\pm$ 0.05 & $-$0&.03 $\pm$ 0.02 \\
HD\,136202  & F8\,III-IV  & thin       & 4.7  & 6215 & $\pm$ 43  & 4.13 $\pm$ 0.15 & 1.53 $\pm$ 0.06 &    0&.04 $\pm$ 0.04 & $-$0&.05 $\pm$ 0.04 \\
HD\,140538  & G2.5\,V     & thin       & 1.3  & 5648 & $\pm$ 72  & 4.41 $\pm$ 0.16 & 0.73 $\pm$ 0.11 &    0&.03 $\pm$ 0.05 & $-$0&.12 $\pm$ 0.03 \\
HD\,14214   & G0.5\,IV    & thin       & 3.8  & 6114 & $\pm$ 46  & 4.26 $\pm$ 0.17 & 1.33 $\pm$ 0.06 &    0&.16 $\pm$ 0.04 & $-$0&.01 $\pm$ 0.03 \\
HD\,142373  & F8\,Ve      & thin/thick & 1.6  & 5870 & $\pm$ 48  & 4.11 $\pm$ 0.20 & 1.39 $\pm$ 0.08 & $-$0&.48 $\pm$ 0.04 &    0&.09 $\pm$ 0.09 \\
HD\,146233  & G2\,Va      & thin       & 1.8  & 5747 & $\pm$ 50  & 4.35 $\pm$ 0.12 & 0.76 $\pm$ 0.07 &    0&.03 $\pm$ 0.04 & $-$0&.08 $\pm$ 0.03 \\
HD\,154931  & G0          & thin       & 2.9  & 5927 & $\pm$ 48  & 4.14 $\pm$ 0.17 & 1.33 $\pm$ 0.07 & $-$0&.10 $\pm$ 0.05 & $-$0&.02 $\pm$ 0.04 \\
HD\,163183  & G0          & thin       & 4.4  & 6014 & $\pm$ 110 & 4.65 $\pm$ 0.27 & 1.46 $\pm$ 0.14 & $-$0&.07 $\pm$ 0.07 & $-$0&.07 $\pm$ 0.14 \\
HD\,16397   & G0\,V       & thin/thick & 0.7  & 5839 & $\pm$ 52  & 4.53 $\pm$ 0.12 & 0.94 $\pm$ 0.10 & $-$0&.47 $\pm$ 0.04 &    0&.04 $\pm$ 0.04 \\
HD\,176841  & G5          & thin       & 2.9  & 5857 & $\pm$ 80  & 4.33 $\pm$ 0.17 & 0.76 $\pm$ 0.12 &    0&.30 $\pm$ 0.05 & $-$0&.04 $\pm$ 0.03 \\
HD\,178428  & G5\,V       & thin       & 1.4  & 5629 & $\pm$ 67  & 4.15 $\pm$ 0.17 & 0.76 $\pm$ 0.09 &    0&.14 $\pm$ 0.04 & $-$0&.01 $\pm$ 0.02 \\
HD\,1835    & G3\,V       & thin       & 6.6  & 5786 & $\pm$ 61  & 4.45 $\pm$ 0.18 & 1.06 $\pm$ 0.09 &    0&.16 $\pm$ 0.05 & $-$0&.12 $\pm$ 0.03 \\
HD\,184499  & G0\,V       & thick      & 0.7  & 5775 & $\pm$ 61  & 4.21 $\pm$ 0.17 & 1.12 $\pm$ 0.11 & $-$0&.50 $\pm$ 0.06 &    0&.14 $\pm$ 0.10 \\
HD\,185144  & K0\,V       & thin       & 0.0  & 5204 & $\pm$ 63  & 4.37 $\pm$ 0.17 & 0.22 $\pm$ 0.23 & $-$0&.26 $\pm$ 0.05 &    0&.02 $\pm$ 0.02 \\
HD\,186408  & G1.5\,Vb    & thin       & 2.0  & 5748 & $\pm$ 103 & 4.30 $\pm$ 0.23 & 1.03 $\pm$ 0.12 &    0&.10 $\pm$ 0.07 & $-$0&.05 $\pm$ 0.05 \\
HD\,18757   & G4\,V       & thin       & 0.0  & 5640 & $\pm$ 44  & 4.38 $\pm$ 0.10 & 0.68 $\pm$ 0.07 & $-$0&.31 $\pm$ 0.04 &    0&.07 $\pm$ 0.02 \\
HD\,187691  & F8\,V       & thin       & 4.1  & 6173 & $\pm$ 45  & 4.25 $\pm$ 0.20 & 1.30 $\pm$ 0.05 &    0&.14 $\pm$ 0.04 & $-$0&.12 $\pm$ 0.05 \\
HD\,190771  & G5\,IV      & thin       & 4.1  & 5819 & $\pm$ 56  & 4.45 $\pm$ 0.16 & 1.02 $\pm$ 0.08 &    0&.13 $\pm$ 0.04 & $-$0&.08 $\pm$ 0.03 \\
HD\,197076A & G5\,V       & thin       & 3.2  & 5828 & $\pm$ 44  & 4.45 $\pm$ 0.12 & 0.81 $\pm$ 0.07 & $-$0&.10 $\pm$ 0.04 & $-$0&.10 $\pm$ 0.03 \\
HD\,199960  & G1\,V       & thin       & 3.8  & 5863 & $\pm$ 42  & 4.21 $\pm$ 0.12 & 1.02 $\pm$ 0.05 &    0&.22 $\pm$ 0.03 & $-$0&.03 $\pm$ 0.02 \\
HD\,200790  & F8\,V       & thin       & 6.7  & 6182 & $\pm$ 55  & 4.08 $\pm$ 0.18 & 1.53 $\pm$ 0.07 & $-$0&.01 $\pm$ 0.05 & $-$0&.02 $\pm$ 0.06 \\
HD\,206374  & G8\,V       & thin       & 1.1  & 5604 & $\pm$ 60  & 4.45 $\pm$ 0.14 & 0.69 $\pm$ 0.10 & $-$0&.08 $\pm$ 0.04 & $-$0&.07 $\pm$ 0.02 \\
HD\,206860  & G0\,V       & thin       & 9.4  & 6106 & $\pm$ 69  & 4.68 $\pm$ 0.23 & 1.37 $\pm$ 0.10 & $-$0&.04 $\pm$ 0.05 & $-$0&.04 $\pm$ 0.07 \\
HD\,208313  & K0\,V       & thin       & 0.6  & 4883 & $\pm$ 132 & 4.17 $\pm$ 0.21 & 0.23 $\pm$ 0.44 & $-$0&.12 $\pm$ 0.06 &    0&.09 $\pm$ 0.02 \\
HD\,218059  & F8          & thin       & 3.2  & 6343 & $\pm$ 72  & 4.43 $\pm$ 0.23 & 1.70 $\pm$ 0.15 & $-$0&.31 $\pm$ 0.06 &    0&.06 $\pm$ 0.11 \\
HD\,218209  & G6\,V       & thin       & 0.0  & 5539 & $\pm$ 45  & 4.37 $\pm$ 0.21 & 0.59 $\pm$ 0.10 & $-$0&.50 $\pm$ 0.04 &    0&.01 $\pm$ 0.02 \\
HD\,218868  & K0          & thin       & 1.6  & 5487 & $\pm$ 141 & 4.32 $\pm$ 0.12 & 0.56 $\pm$ 0.32 &    0&.20 $\pm$ 0.05 & $-$0&.05 $\pm$ 0.02 \\
HD\,221354  & K2\,V       & thin       & 0.0  & 5138 & $\pm$ 116 & 4.18 $\pm$ 0.12 & 0.40 $\pm$ 0.28 &    0&.00 $\pm$ 0.05 &    0&.08 $\pm$ 0.01 \\
HD\,221851  & G5\,V       & thin       & 1.7  & 5088 & $\pm$ 103 & 4.33 $\pm$ 0.15 & 0.48 $\pm$ 0.26 & $-$0&.14 $\pm$ 0.04 & $-$0&.01 $\pm$ 0.02 \\
HD\,222143  & G3/4\,V     & thin       & 3.0  & 5923 & $\pm$ 67  & 4.55 $\pm$ 0.15 & 1.06 $\pm$ 0.09 &    0&.15 $\pm$ 0.04 & $-$0&.12 $\pm$ 0.03 \\
HD\,224465  & G5          & thin       & 1.6  & 5688 & $\pm$ 40  & 4.29 $\pm$ 0.10 & 0.75 $\pm$ 0.07 &    0&.04 $\pm$ 0.03 & $-$0&.01 $\pm$ 0.01 \\
HD\,22484   & F9\,IV-V    & thin       & 4.0  & 6044 & $\pm$ 53  & 4.22 $\pm$ 0.15 & 1.21 $\pm$ 0.07 & $-$0&.07 $\pm$ 0.05 & $-$0&.03 $\pm$ 0.04 \\
HD\,24496   & G0          & thin       & 1.5  & 5547 & $\pm$ 142 & 4.36 $\pm$ 0.20 & 0.62 $\pm$ 0.27 &    0&.01 $\pm$ 0.08 & $-$0&.08 $\pm$ 0.03 \\
HD\,25825   & G0          & thin       & 3.6  & 6018 & $\pm$ 88  & 4.54 $\pm$ 0.23 & 0.94 $\pm$ 0.13 &    0&.00 $\pm$ 0.06 & $-$0&.01 $\pm$ 0.12 \\
HD\,28344   & G2\,V       & thin       & 6.6  & 5961 & $\pm$ 60  & 4.48 $\pm$ 0.15 & 1.26 $\pm$ 0.07 &    0&.14 $\pm$ 0.05 & $-$0&.09 $\pm$ 0.04 \\
HD\,29587   & G2\,V       & thin/thick & 0.0  & 5683 & $\pm$ 67  & 4.55 $\pm$ 0.20 & 0.95 $\pm$ 0.16 & $-$0&.56 $\pm$ 0.05 &    0&.06 $\pm$ 0.04 \\
HD\,38230   & K0\,V       & thin       & 0.0  & 5060 & $\pm$ 115 & 4.15 $\pm$ 0.15 & 0.32 $\pm$ 0.31 & $-$0&.12 $\pm$ 0.06 &    0&.08 $\pm$ 0.02 \\
HD\,38858   & G4\,V       & thin       & 0.4  & 5722 & $\pm$ 47  & 4.50 $\pm$ 0.12 & 0.73 $\pm$ 0.08 & $-$0&.23 $\pm$ 0.04 & $-$0&.06 $\pm$ 0.03 \\
HD\,39587   & G0\,V       & thin       & 8.7  & 6043 & $\pm$ 71  & 4.55 $\pm$ 0.23 & 1.40 $\pm$ 0.10 & $-$0&.04 $\pm$ 0.05 & $-$0&.05 $\pm$ 0.04 \\
HD\,42807   & G2\,V       & thin       & 4.5  & 5705 & $\pm$ 42  & 4.49 $\pm$ 0.12 & 0.94 $\pm$ 0.06 & $-$0&.06 $\pm$ 0.03 & $-$0&.05 $\pm$ 0.02 \\
HD\,43587   & F9\,V       & thin       & 1.8  & 5927 & $\pm$ 39  & 4.34 $\pm$ 0.14 & 1.10 $\pm$ 0.06 & $-$0&.03 $\pm$ 0.04 & $-$0&.04 $\pm$ 0.04 \\
HD\,45067   & F8\,V       & thin       & 5.9  & 6087 & $\pm$ 48  & 4.17 $\pm$ 0.17 & 1.39 $\pm$ 0.07 & $-$0&.05 $\pm$ 0.04 & $-$0&.08 $\pm$ 0.05 \\
HD\,4614    & G3\,V       & thin       & 2.3  & 5936 & $\pm$ 46  & 4.49 $\pm$ 0.12 & 0.95 $\pm$ 0.09 & $-$0&.26 $\pm$ 0.04 &    0&.01 $\pm$ 0.04 \\
HD\,59747   & G5\,V       & thin       & 2.0  & 5023 & $\pm$ 131 & 4.29 $\pm$ 0.12 & 0.83 $\pm$ 0.20 & $-$0&.10 $\pm$ 0.05 &    0&.01 $\pm$ 0.02 \\
HD\,693     & F5\,V       & thin       & 1.7  & 6220 & $\pm$ 64  & 4.22 $\pm$ 0.27 & 2.24 $\pm$ 0.26 & $-$0&.37 $\pm$ 0.06 &    0&.03 $\pm$ 0.12 \\
HD\,72905   & G1.5\,Vb    & thin       & 8.8  & 5959 & $\pm$ 93  & 4.59 $\pm$ 0.19 & 1.55 $\pm$ 0.13 & $-$0&.09 $\pm$ 0.07 & $-$0&.07 $\pm$ 0.08 \\
HD\,72945   & F8\,V       & thin       & 3.1  & 5977 & $\pm$ 55  & 4.54 $\pm$ 0.11 & 1.07 $\pm$ 0.07 &    0&.08 $\pm$ 0.04 & $-$0&.04 $\pm$ 0.04 \\
HD\,76151   & G2\,V       & thin       & 1.5  & 5773 & $\pm$ 59  & 4.42 $\pm$ 0.17 & 0.77 $\pm$ 0.09 &    0&.11 $\pm$ 0.04 & $-$0&.05 $\pm$ 0.03 \\
HD\,89269   & G5          & thin       & 0.0  & 5577 & $\pm$ 45  & 4.35 $\pm$ 0.12 & 0.65 $\pm$ 0.08 & $-$0&.20 $\pm$ 0.03 & $-$0&.06 $\pm$ 0.02 \\
\hline
\end{tabular}
\end{table*}
\begin{table*}[p]
\centering
  \caption[]{The same as Table~\ref{dwarfs_tab} but for 13 subgiant stars,
             from which 4 have planets.}
  \label{subgiants_tab}
\begin{tabular}{lccc r@{ }l cc r@{}l r@{}l}
\noalign{\smallskip}\hline\hline\noalign{\smallskip}
Star &
\parbox[c]{1.1cm}{\centering Spectral type} &
\parbox[c]{1.4cm}{\centering Population group} &
\parbox[c]{1.1cm}{\centering \Vbroad\ [\kms]} &
\multicolumn{2}{c}{\parbox[c]{1.0cm}{\centering \teff\ $ \pm\ \sigma$ [K]}} &
\parbox[c]{1.2cm}{\centering \logg\ $\pm\ \sigma$} &
\parbox[c]{1.1cm}{\centering $\xi$ $\pm\ \sigma$ [\kms]} &
\multicolumn{2}{c}{[Fe/H] $\pm\ \sigma$} &
\multicolumn{2}{c}{[C/Fe] $\pm\ \sigma$} \\
\noalign{\smallskip}\hline\noalign{\smallskip}
HD\,117176  & G5\,V       & thin       & 0.0  & 5562 & $\pm$ 43  & 4.01 $\pm$ 0.14 & 0.92 $\pm$ 0.05 & $-$0&.05 $\pm$ 0.03 & $-$0&.05 $\pm$ 0.02 \\
HD\,142091  & K1\,IVa     & thin       & 2.6  & 4839 & $\pm$ 163 & 3.16 $\pm$ 0.24 & 0.77 $\pm$ 0.16 &    0&.20 $\pm$ 0.07 & $-$0&.17 $\pm$ 0.03 \\
HD\,222404  & K1\,IV      & thin       & 2.4  & 4875 & $\pm$ 138 & 3.23 $\pm$ 0.27 & 1.17 $\pm$ 0.14 &    0&.05 $\pm$ 0.07 & $-$0&.01 $\pm$ 0.03 \\
HD\,38529   & G4\,V       & thin       & 2.9  & 5570 & $\pm$ 70  & 3.80 $\pm$ 0.14 & 1.11 $\pm$ 0.09 &    0&.30 $\pm$ 0.06 & $-$0&.01 $\pm$ 0.03 \\
\hline\noalign{\smallskip}
HD\,121370  & G0\,IV      & thin       & 6.4  & 6194 & $\pm$ 110 & 4.08 $\pm$ 0.32 & 2.29 $\pm$ 0.15 &    0&.17 $\pm$ 0.09 &    0&.08 $\pm$ 0.16 \\
HD\,161797A & G5\,IV      & thin       & 3.0  & 5583 & $\pm$ 78  & 3.99 $\pm$ 0.18 & 0.91 $\pm$ 0.11 &    0&.28 $\pm$ 0.05 &    0&.01 $\pm$ 0.02 \\
HD\,182572  & G8\,IV      & thin       & 2.4  & 5569 & $\pm$ 174 & 4.10 $\pm$ 0.20 & 0.67 $\pm$ 0.36 &    0&.40 $\pm$ 0.07 &    0&.01 $\pm$ 0.03 \\
HD\,185351  & G9\,IIIb    & thin       & 2.4  & 5086 & $\pm$ 85  & 3.45 $\pm$ 0.17 & 1.02 $\pm$ 0.10 &    0&.06 $\pm$ 0.07 & $-$0&.12 $\pm$ 0.03 \\
HD\,191026  & K0\,IV      & thin       & 1.6  & 5108 & $\pm$ 74  & 3.67 $\pm$ 0.18 & 0.97 $\pm$ 0.08 & $-$0&.03 $\pm$ 0.05 & $-$0&.08 $\pm$ 0.02 \\
HD\,198149  & K0\,IV      & thin/thick & 0.0  & 4920 & $\pm$ 61  & 3.29 $\pm$ 0.15 & 0.90 $\pm$ 0.06 & $-$0&.14 $\pm$ 0.05 & $-$0&.07 $\pm$ 0.02 \\
HD\,221585  & G8\,IV      & thin       & 2.2  & 5560 & $\pm$ 74  & 3.94 $\pm$ 0.17 & 0.87 $\pm$ 0.10 &    0&.30 $\pm$ 0.05 & $-$0&.02 $\pm$ 0.02 \\
HD\,57006   & F8\,V       & thin       & 7.6  & 6166 & $\pm$ 60  & 3.77 $\pm$ 0.24 & 1.86 $\pm$ 0.08 & $-$0&.05 $\pm$ 0.06 & $-$0&.01 $\pm$ 0.07 \\
HD\,9562    & G2\,IV      & thin       & 4.0  & 5895 & $\pm$ 49  & 4.10 $\pm$ 0.15 & 1.14 $\pm$ 0.06 &    0&.20 $\pm$ 0.04 & $-$0&.03 $\pm$ 0.03 \\
\hline
\end{tabular}
\end{table*}
\begin{table*}[p]
\centering
  \caption[]{The same as Table~\ref{dwarfs_tab} but for 96 giant stars,
             from which 7 have planets.}
  \label{giants_tab}
\begin{tabular}{lccc r@{ }l cc r@{}l r@{}l}
\noalign{\smallskip}\hline\hline\noalign{\smallskip}
Star &
\parbox[c]{1.1cm}{\centering Spectral type} &
\parbox[c]{1.4cm}{\centering Population group} &
\parbox[c]{1.1cm}{\centering \Vbroad\ [\kms]} &
\multicolumn{2}{c}{\parbox[c]{1.0cm}{\centering \teff\ $ \pm\ \sigma$ [K]}} &
\parbox[c]{1.2cm}{\centering \logg\ $\pm\ \sigma$} &
\parbox[c]{1.1cm}{\centering $\xi$ $\pm\ \sigma$ [\kms]} &
\multicolumn{2}{c}{[Fe/H] $\pm\ \sigma$} &
\multicolumn{2}{c}{[C/Fe] $\pm\ \sigma$} \\
\noalign{\smallskip}\hline\noalign{\smallskip}
HD\,137759  & K2\,III     & thin       & 0.0  & 4547 & $\pm$ 139 & 2.63 $\pm$ 0.18 & 1.27 $\pm$ 0.13 &    0&.07 $\pm$ 0.08 & $-$0&.12 $\pm$ 0.10 \\
HD\,16400   & G5\,III     & thin       & 1.3  & 4853 & $\pm$ 87  & 2.71 $\pm$ 0.24 & 1.38 $\pm$ 0.08 & $-$0&.02 $\pm$ 0.09 & $-$0&.20 $\pm$ 0.04 \\
HD\,170693  & K1.5\,III   & thin       & 2.3  & 4470 & $\pm$ 75  & 2.20 $\pm$ 0.28 & 1.37 $\pm$ 0.07 & $-$0&.37 $\pm$ 0.08 & $-$0&.09 $\pm$ 0.05 \\
HD\,221345  & G8\,III     & thick      & 2.4  & 4756 & $\pm$ 70  & 2.61 $\pm$ 0.20 & 1.43 $\pm$ 0.06 & $-$0&.29 $\pm$ 0.07 & $-$0&.03 $\pm$ 0.03 \\
HD\,28305   & G9.5\,III   & thin       & 4.5  & 4956 & $\pm$ 91  & 2.78 $\pm$ 0.30 & 1.73 $\pm$ 0.09 &    0&.04 $\pm$ 0.09 & $-$0&.13 $\pm$ 0.04 \\
HD\,62509   & K0\,III     & thin       & 2.3  & 4955 & $\pm$ 116 & 3.07 $\pm$ 0.22 & 1.15 $\pm$ 0.15 &    0&.16 $\pm$ 0.08 & $-$0&.25 $\pm$ 0.04 \\
HD\,81688   & K0\,III-IV  & thin       & 2.2  & 4895 & $\pm$ 60  & 2.72 $\pm$ 0.22 & 1.49 $\pm$ 0.04 & $-$0&.24 $\pm$ 0.06 & $-$0&.09 $\pm$ 0.03 \\
\hline\noalign{\smallskip}
HD\,101484  & K0\,III     & thin       & 2.3  & 4949 & $\pm$ 78  & 2.93 $\pm$ 0.21 & 1.29 $\pm$ 0.07 &    0&.06 $\pm$ 0.07 & $-$0&.27 $\pm$ 0.04 \\
HD\,102928  & K0\,III     & thin       & 0.5  & 4646 & $\pm$ 75  & 2.43 $\pm$ 0.29 & 1.26 $\pm$ 0.06 & $-$0&.20 $\pm$ 0.07 & $-$0&.15 $\pm$ 0.04 \\
HD\,104979  & G8\,IIIa    & thin       & 0.7  & 5045 & $\pm$ 46  & 2.96 $\pm$ 0.18 & 1.55 $\pm$ 0.04 & $-$0&.34 $\pm$ 0.05 & $-$0&.05 $\pm$ 0.03 \\
HD\,106714  & G8\,III     & thin       & 0.7  & 5017 & $\pm$ 68  & 2.88 $\pm$ 0.20 & 1.44 $\pm$ 0.05 & $-$0&.11 $\pm$ 0.07 & $-$0&.24 $\pm$ 0.04 \\
HD\,10975   & K0\,III     & thin       & 2.4  & 4943 & $\pm$ 63  & 2.78 $\pm$ 0.18 & 1.43 $\pm$ 0.05 & $-$0&.14 $\pm$ 0.06 & $-$0&.22 $\pm$ 0.03 \\
HD\,110024  & G9\,III     & thin       & 2.8  & 5003 & $\pm$ 83  & 3.03 $\pm$ 0.21 & 1.35 $\pm$ 0.08 &    0&.07 $\pm$ 0.08 & $-$0&.22 $\pm$ 0.04 \\
HD\,114357  & K3\,III     & thin       & 0.0  & 4498 & $\pm$ 111 & 2.46 $\pm$ 0.27 & 1.62 $\pm$ 0.10 & $-$0&.13 $\pm$ 0.08 & $-$0&.03 $\pm$ 0.06 \\
HD\,11559   & K0\,III     & thin       & 3.0  & 5064 & $\pm$ 98  & 3.18 $\pm$ 0.24 & 1.27 $\pm$ 0.09 &    0&.13 $\pm$ 0.08 & $-$0&.29 $\pm$ 0.05 \\
HD\,116292  & K0\,III     & thin       & 2.7  & 5036 & $\pm$ 58  & 3.00 $\pm$ 0.21 & 1.40 $\pm$ 0.05 & $-$0&.01 $\pm$ 0.06 & $-$0&.17 $\pm$ 0.03 \\
HD\,117304  & K0\,III     & thin       & 2.3  & 4723 & $\pm$ 75  & 2.66 $\pm$ 0.23 & 1.19 $\pm$ 0.08 & $-$0&.09 $\pm$ 0.07 & $-$0&.15 $\pm$ 0.04 \\
HD\,11749   & K0\,III     & thin       & 0.5  & 4740 & $\pm$ 66  & 2.55 $\pm$ 0.21 & 1.46 $\pm$ 0.06 & $-$0&.18 $\pm$ 0.07 & $-$0&.13 $\pm$ 0.03 \\
HD\,119126  & G9\,III     & thin       & 0.5  & 4890 & $\pm$ 68  & 2.74 $\pm$ 0.20 & 1.42 $\pm$ 0.06 & $-$0&.05 $\pm$ 0.07 & $-$0&.21 $\pm$ 0.04 \\
HD\,11949   & K0\,IV      & thin       & 1.5  & 4814 & $\pm$ 65  & 2.86 $\pm$ 0.18 & 1.02 $\pm$ 0.07 & $-$0&.09 $\pm$ 0.06 & $-$0&.20 $\pm$ 0.03 \\
HD\,120164  & K0\,III     & thin       & 0.5  & 4785 & $\pm$ 81  & 2.64 $\pm$ 0.21 & 1.36 $\pm$ 0.07 & $-$0&.07 $\pm$ 0.07 & $-$0&.17 $\pm$ 0.04 \\
HD\,120420  & K0\,III     & thin       & 0.0  & 4794 & $\pm$ 62  & 2.76 $\pm$ 0.20 & 1.25 $\pm$ 0.06 & $-$0&.19 $\pm$ 0.06 & $-$0&.21 $\pm$ 0.03 \\
HD\,12929   & K2\,III     & thin       & 1.0  & 4682 & $\pm$ 99  & 2.85 $\pm$ 0.29 & 1.50 $\pm$ 0.10 & $-$0&.30 $\pm$ 0.10 & $-$0&.01 $\pm$ 0.04 \\
HD\,133208  & G8\,IIIa    & thin       & 3.4  & 5121 & $\pm$ 62  & 2.76 $\pm$ 0.20 & 1.81 $\pm$ 0.06 & $-$0&.03 $\pm$ 0.07 & $-$0&.23 $\pm$ 0.04 \\
HD\,136138  & G8\,II-III  & thin       & 6.5  & 5022 & $\pm$ 80  & 2.86 $\pm$ 0.24 & 1.48 $\pm$ 0.07 & $-$0&.17 $\pm$ 0.09 & $-$0&.13 $\pm$ 0.04 \\
HD\,136512  & K0\,III     & thin       & 3.5  & 4830 & $\pm$ 58  & 2.69 $\pm$ 0.20 & 1.39 $\pm$ 0.05 & $-$0&.21 $\pm$ 0.06 & $-$0&.09 $\pm$ 0.02 \\
HD\,138852  & K0\,III-IV  & thin       & 0.8  & 4928 & $\pm$ 58  & 2.75 $\pm$ 0.20 & 1.47 $\pm$ 0.05 & $-$0&.21 $\pm$ 0.07 & $-$0&.20 $\pm$ 0.03 \\
HD\,148856  & G7\,IIIa    & thin       & 4.1  & 5116 & $\pm$ 62  & 2.91 $\pm$ 0.20 & 1.64 $\pm$ 0.06 & $-$0&.04 $\pm$ 0.07 & $-$0&.18 $\pm$ 0.03 \\
HD\,150997  & G7.5\,III   & thin       & 2.5  & 5069 & $\pm$ 59  & 2.99 $\pm$ 0.18 & 1.27 $\pm$ 0.05 & $-$0&.14 $\pm$ 0.06 & $-$0&.20 $\pm$ 0.03 \\
HD\,152224  & K0\,III     & thin       & 0.8  & 4780 & $\pm$ 68  & 2.83 $\pm$ 0.20 & 1.11 $\pm$ 0.06 & $-$0&.13 $\pm$ 0.06 & $-$0&.17 $\pm$ 0.04 \\
HD\,15596   & G5\,III-IV  & thick      & 0.5  & 4903 & $\pm$ 53  & 3.13 $\pm$ 0.18 & 1.07 $\pm$ 0.05 & $-$0&.56 $\pm$ 0.05 & $-$0&.09 $\pm$ 0.03 \\
HD\,15755   & K0\,III     & thin       & 0.8  & 4666 & $\pm$ 88  & 2.63 $\pm$ 0.22 & 1.11 $\pm$ 0.07 & $-$0&.02 $\pm$ 0.06 & $-$0&.18 $\pm$ 0.04 \\
HD\,15779   & G3\,III     & thin       & 0.5  & 4906 & $\pm$ 78  & 2.95 $\pm$ 0.18 & 1.28 $\pm$ 0.07 &    0&.01 $\pm$ 0.07 & $-$0&.18 $\pm$ 0.03 \\
HD\,159353  & K0\,III     & thin       & 0.8  & 4876 & $\pm$ 81  & 2.79 $\pm$ 0.20 & 1.36 $\pm$ 0.07 & $-$0&.06 $\pm$ 0.08 & $-$0&.18 $\pm$ 0.04 \\
HD\,161178  & G9\,III     & thin       & 0.8  & 4845 & $\pm$ 64  & 2.56 $\pm$ 0.20 & 1.39 $\pm$ 0.05 & $-$0&.14 $\pm$ 0.06 & $-$0&.17 $\pm$ 0.03 \\
HD\,162076  & G5\,IV      & thin       & 2.9  & 5160 & $\pm$ 84  & 3.39 $\pm$ 0.21 & 1.30 $\pm$ 0.08 &    0&.07 $\pm$ 0.08 & $-$0&.17 $\pm$ 0.04 \\
HD\,163993  & G8\,III     & thin       & 4.8  & 5168 & $\pm$ 86  & 3.21 $\pm$ 0.26 & 1.43 $\pm$ 0.08 &    0&.07 $\pm$ 0.09 & $-$0&.17 $\pm$ 0.04 \\
HD\,168653  & K1\,III     & thin       & 2.3  & 4800 & $\pm$ 80  & 2.96 $\pm$ 0.23 & 1.17 $\pm$ 0.07 & $-$0&.03 $\pm$ 0.06 & $-$0&.14 $\pm$ 0.03 \\
HD\,168723  & K0\,III-IV  & thin       & 0.0  & 4926 & $\pm$ 74  & 3.09 $\pm$ 0.17 & 1.08 $\pm$ 0.06 & $-$0&.19 $\pm$ 0.06 & $-$0&.16 $\pm$ 0.03 \\
HD\,17361   & K1.5\,III   & thin       & 0.8  & 4670 & $\pm$ 126 & 2.66 $\pm$ 0.27 & 1.51 $\pm$ 0.10 &    0&.00 $\pm$ 0.09 & $-$0&.08 $\pm$ 0.05 \\
HD\,180711  & G9\,III     & thin       & 1.6  & 4865 & $\pm$ 73  & 2.73 $\pm$ 0.22 & 1.38 $\pm$ 0.06 & $-$0&.13 $\pm$ 0.08 & $-$0&.19 $\pm$ 0.04 \\
HD\,185644  & K1\,III     & thin       & 0.8  & 4613 & $\pm$ 141 & 2.68 $\pm$ 0.28 & 1.41 $\pm$ 0.12 & $-$0&.02 $\pm$ 0.11 & $-$0&.14 $\pm$ 0.07 \\
HD\,19270   & K3\,III     & thin       & 2.3  & 4774 & $\pm$ 106 & 2.71 $\pm$ 0.28 & 1.51 $\pm$ 0.11 &    0&.00 $\pm$ 0.09 & $-$0&.03 $\pm$ 0.04 \\
HD\,192787  & K0\,III     & thin       & 3.1  & 5131 & $\pm$ 63  & 3.19 $\pm$ 0.20 & 1.29 $\pm$ 0.06 & $-$0&.02 $\pm$ 0.07 & $-$0&.21 $\pm$ 0.03 \\
\hline
\end{tabular}
\end{table*}
\addtocounter{table}{-1}
\begin{table*}[p]
\centering
  \caption[]{continued.}
\begin{tabular}{lccc r@{ }l cc r@{}l r@{}l}
\noalign{\smallskip}\hline\hline\noalign{\smallskip}
Star &
\parbox[c]{1.1cm}{\centering Spectral type} &
\parbox[c]{1.4cm}{\centering Population group} &
\parbox[c]{1.1cm}{\centering \Vbroad\ [\kms]} &
\multicolumn{2}{c}{\parbox[c]{1.0cm}{\centering \teff\ $ \pm\ \sigma$ [K]}} &
\parbox[c]{1.2cm}{\centering \logg\ $\pm\ \sigma$} &
\parbox[c]{1.1cm}{\centering $\xi$ $\pm\ \sigma$ [\kms]} &
\multicolumn{2}{c}{[Fe/H] $\pm\ \sigma$} &
\multicolumn{2}{c}{[C/Fe] $\pm\ \sigma$} \\
\noalign{\smallskip}\hline\noalign{\smallskip}
HD\,196134  & K0\,III-IV  & thin       & 0.8  & 4835 & $\pm$ 64  & 2.97 $\pm$ 0.20 & 1.10 $\pm$ 0.06 & $-$0&.10 $\pm$ 0.05 & $-$0&.21 $\pm$ 0.03 \\
HD\,19787   & K2\,III     & thin       & 2.3  & 4869 & $\pm$ 90  & 2.79 $\pm$ 0.25 & 1.41 $\pm$ 0.09 &    0&.06 $\pm$ 0.08 & $-$0&.16 $\pm$ 0.04 \\
HD\,197989  & K0\,III     & thin       & 1.9  & 4843 & $\pm$ 75  & 2.78 $\pm$ 0.17 & 1.34 $\pm$ 0.07 & $-$0&.11 $\pm$ 0.07 & $-$0&.17 $\pm$ 0.03 \\
HD\,19845   & G9\,III     & thin       & 2.0  & 5050 & $\pm$ 138 & 3.28 $\pm$ 0.26 & 1.45 $\pm$ 0.13 &    0&.14 $\pm$ 0.09 & $-$0&.17 $\pm$ 0.04 \\
HD\,199870  & K0\,IIIb    & thin       & 3.3  & 4968 & $\pm$ 85  & 3.03 $\pm$ 0.20 & 1.20 $\pm$ 0.09 &    0&.11 $\pm$ 0.08 & $-$0&.19 $\pm$ 0.04 \\
HD\,202109  & G8\,III     & thin       & 2.3  & 4998 & $\pm$ 93  & 2.78 $\pm$ 0.24 & 1.72 $\pm$ 0.07 & $-$0&.02 $\pm$ 0.10 & $-$0&.08 $\pm$ 0.04 \\
HD\,205435  & G5\,III     & thin       & 2.9  & 5180 & $\pm$ 63  & 3.24 $\pm$ 0.20 & 1.32 $\pm$ 0.06 & $-$0&.06 $\pm$ 0.06 & $-$0&.21 $\pm$ 0.03 \\
HD\,20791   & G8.5\,III   & thin       & 1.8  & 5046 & $\pm$ 82  & 2.94 $\pm$ 0.20 & 1.31 $\pm$ 0.08 &    0&.12 $\pm$ 0.08 & $-$0&.24 $\pm$ 0.04 \\
HD\,212496  & G8.5\,III   & thin       & 1.6  & 4760 & $\pm$ 60  & 2.72 $\pm$ 0.20 & 1.21 $\pm$ 0.05 & $-$0&.27 $\pm$ 0.06 & $-$0&.15 $\pm$ 0.03 \\
HD\,212943  & K0\,III     & thick      & 2.3  & 4683 & $\pm$ 73  & 2.77 $\pm$ 0.20 & 1.13 $\pm$ 0.06 & $-$0&.20 $\pm$ 0.06 & $-$0&.06 $\pm$ 0.03 \\
HD\,216131  & G8\,III     & thin       & 2.9  & 5087 & $\pm$ 68  & 3.05 $\pm$ 0.18 & 1.26 $\pm$ 0.06 &    0&.03 $\pm$ 0.07 & $-$0&.25 $\pm$ 0.03 \\
HD\,216228  & K0\,III     & thin       & 0.8  & 4811 & $\pm$ 81  & 2.75 $\pm$ 0.22 & 1.48 $\pm$ 0.07 & $-$0&.04 $\pm$ 0.08 & $-$0&.15 $\pm$ 0.04 \\
HD\,225216  & K1\,III     & thin       & 0.8  & 4734 & $\pm$ 91  & 2.53 $\pm$ 0.28 & 1.56 $\pm$ 0.07 & $-$0&.15 $\pm$ 0.08 & $-$0&.13 $\pm$ 0.04 \\
HD\,25602   & K0\,III-IV  & thin       & 0.5  & 4857 & $\pm$ 74  & 3.00 $\pm$ 0.21 & 1.05 $\pm$ 0.07 & $-$0&.16 $\pm$ 0.06 & $-$0&.12 $\pm$ 0.03 \\
HD\,25604   & K0\,III     & thin       & 1.1  & 4783 & $\pm$ 97  & 2.69 $\pm$ 0.18 & 1.38 $\pm$ 0.08 &    0&.07 $\pm$ 0.08 & $-$0&.17 $\pm$ 0.03 \\
HD\,26546   & K0\,III     & thin       & 0.8  & 4788 & $\pm$ 102 & 2.69 $\pm$ 0.32 & 1.48 $\pm$ 0.10 &    0&.00 $\pm$ 0.09 & $-$0&.18 $\pm$ 0.05 \\
HD\,26659   & G8\,III     & thin       & 5.3  & 5207 & $\pm$ 62  & 3.07 $\pm$ 0.18 & 1.38 $\pm$ 0.06 & $-$0&.14 $\pm$ 0.07 & $-$0&.28 $\pm$ 0.04 \\
HD\,26755   & K1\,III     & thin       & 0.0  & 4540 & $\pm$ 114 & 2.41 $\pm$ 0.29 & 1.43 $\pm$ 0.10 & $-$0&.09 $\pm$ 0.11 & $-$0&.12 $\pm$ 0.10 \\
HD\,27348   & G8\,III     & thin       & 3.5  & 5081 & $\pm$ 96  & 3.10 $\pm$ 0.25 & 1.36 $\pm$ 0.09 &    0&.09 $\pm$ 0.09 & $-$0&.21 $\pm$ 0.04 \\
HD\,27371   & K0\,III     & thin       & 3.7  & 5026 & $\pm$ 87  & 3.05 $\pm$ 0.20 & 1.46 $\pm$ 0.08 &    0&.10 $\pm$ 0.08 & $-$0&.15 $\pm$ 0.03 \\
HD\,27697   & K0\,III     & thin       & 4.5  & 4796 & $\pm$ 93  & 2.29 $\pm$ 0.31 & 1.64 $\pm$ 0.07 & $-$0&.12 $\pm$ 0.10 & $-$0&.14 $\pm$ 0.05 \\
HD\,28307   & K0\,III     & thin       & 3.7  & 5129 & $\pm$ 111 & 3.21 $\pm$ 0.26 & 1.32 $\pm$ 0.11 &    0&.18 $\pm$ 0.09 & $-$0&.18 $\pm$ 0.04 \\
HD\,2910    & K0\,III     & thin       & 1.0  & 4815 & $\pm$ 89  & 2.71 $\pm$ 0.26 & 1.35 $\pm$ 0.09 &    0&.03 $\pm$ 0.08 & $-$0&.08 $\pm$ 0.03 \\
HD\,30557   & G9\,III     & thin       & 1.3  & 4879 & $\pm$ 72  & 2.69 $\pm$ 0.23 & 1.43 $\pm$ 0.06 & $-$0&.07 $\pm$ 0.07 & $-$0&.23 $\pm$ 0.04 \\
HD\,33419   & K0\,III     & thin       & 0.5  & 4791 & $\pm$ 183 & 2.83 $\pm$ 0.34 & 1.40 $\pm$ 0.15 &    0&.15 $\pm$ 0.12 & $-$0&.16 $\pm$ 0.07 \\
HD\,34559   & G8\,III     & thin       & 3.7  & 5025 & $\pm$ 73  & 2.87 $\pm$ 0.20 & 1.23 $\pm$ 0.07 &    0&.07 $\pm$ 0.07 & $-$0&.24 $\pm$ 0.04 \\
HD\,35369   & G8\,III     & thin       & 1.2  & 4995 & $\pm$ 58  & 2.88 $\pm$ 0.16 & 1.46 $\pm$ 0.05 & $-$0&.14 $\pm$ 0.06 & $-$0&.21 $\pm$ 0.03 \\
HD\,3546    & G8\,III     & thin       & 3.5  & 5070 & $\pm$ 39  & 2.78 $\pm$ 0.16 & 1.64 $\pm$ 0.04 & $-$0&.54 $\pm$ 0.04 & $-$0&.08 $\pm$ 0.02 \\
HD\,37160   & K0\,III     & thin/thick & 0.0  & 4804 & $\pm$ 50  & 2.83 $\pm$ 0.15 & 1.18 $\pm$ 0.04 & $-$0&.54 $\pm$ 0.05 & $-$0&.02 $\pm$ 0.02 \\
HD\,37638   & G5\,III     & thin       & 2.9  & 5183 & $\pm$ 51  & 3.15 $\pm$ 0.18 & 1.34 $\pm$ 0.04 & $-$0&.01 $\pm$ 0.05 & $-$0&.27 $\pm$ 0.04 \\
HD\,40801   & K0\,III     & thin       & 0.5  & 4817 & $\pm$ 81  & 3.00 $\pm$ 0.23 & 1.10 $\pm$ 0.07 & $-$0&.17 $\pm$ 0.07 & $-$0&.06 $\pm$ 0.03 \\
HD\,45415   & G9\,III     & thin       & 1.5  & 4819 & $\pm$ 81  & 2.67 $\pm$ 0.26 & 1.34 $\pm$ 0.07 & $-$0&.04 $\pm$ 0.08 & $-$0&.21 $\pm$ 0.04 \\
HD\,46374   & K2\,III     & thin       & 0.8  & 4658 & $\pm$ 133 & 2.42 $\pm$ 0.43 & 1.69 $\pm$ 0.12 & $-$0&.17 $\pm$ 0.15 & $-$0&.07 $\pm$ 0.07 \\
HD\,47138   & G9\,III     & thin       & 3.5  & 5191 & $\pm$ 61  & 2.98 $\pm$ 0.20 & 1.35 $\pm$ 0.06 & $-$0&.17 $\pm$ 0.07 & $-$0&.24 $\pm$ 0.04 \\
HD\,47366   & K1\,III     & thin       & 1.5  & 4871 & $\pm$ 86  & 3.04 $\pm$ 0.21 & 1.05 $\pm$ 0.08 & $-$0&.01 $\pm$ 0.07 & $-$0&.20 $\pm$ 0.04 \\
HD\,48432   & K0\,III     & thin       & 0.5  & 4936 & $\pm$ 73  & 3.02 $\pm$ 0.19 & 1.11 $\pm$ 0.07 & $-$0&.07 $\pm$ 0.07 & $-$0&.26 $\pm$ 0.04 \\
HD\,5395    & G8\,III     & thin       & 2.8  & 4941 & $\pm$ 45  & 2.71 $\pm$ 0.16 & 1.52 $\pm$ 0.04 & $-$0&.34 $\pm$ 0.05 & $-$0&.10 $\pm$ 0.02 \\
HD\,58207   & G9\,III     & thin       & 1.8  & 4885 & $\pm$ 76  & 2.73 $\pm$ 0.24 & 1.44 $\pm$ 0.06 & $-$0&.08 $\pm$ 0.08 & $-$0&.14 $\pm$ 0.04 \\
HD\,60986   & K0\,III     & thin       & 3.5  & 5157 & $\pm$ 75  & 3.10 $\pm$ 0.20 & 1.41 $\pm$ 0.07 &    0&.09 $\pm$ 0.07 & $-$0&.30 $\pm$ 0.04 \\
HD\,61363   & K0\,III     & thin       & 1.3  & 4876 & $\pm$ 65  & 2.69 $\pm$ 0.25 & 1.48 $\pm$ 0.06 & $-$0&.23 $\pm$ 0.07 & $-$0&.17 $\pm$ 0.04 \\
HD\,61935   & G9\,III     & thin       & 0.8  & 4851 & $\pm$ 81  & 2.74 $\pm$ 0.26 & 1.41 $\pm$ 0.07 & $-$0&.02 $\pm$ 0.08 & $-$0&.17 $\pm$ 0.04 \\
HD\,65066   & K0\,III     & thin       & 1.2  & 4939 & $\pm$ 135 & 2.97 $\pm$ 0.33 & 1.46 $\pm$ 0.12 &    0&.05 $\pm$ 0.14 & $-$0&.17 $\pm$ 0.07 \\
HD\,65345   & K0\,III     & thin       & 1.2  & 5063 & $\pm$ 68  & 3.06 $\pm$ 0.18 & 1.27 $\pm$ 0.07 &    0&.02 $\pm$ 0.07 & $-$0&.34 $\pm$ 0.05 \\
HD\,68375   & G8\,III     & thin       & 0.6  & 5144 & $\pm$ 55  & 3.16 $\pm$ 0.16 & 1.29 $\pm$ 0.05 & $-$0&.02 $\pm$ 0.06 & $-$0&.27 $\pm$ 0.03 \\
HD\,70523   & K0\,III     & thin       & 0.8  & 4685 & $\pm$ 77  & 2.57 $\pm$ 0.21 & 1.38 $\pm$ 0.06 & $-$0&.20 $\pm$ 0.07 & $-$0&.01 $\pm$ 0.03 \\
HD\,73017   & G8\,IV      & thin       & 1.9  & 4842 & $\pm$ 55  & 2.80 $\pm$ 0.15 & 1.23 $\pm$ 0.04 & $-$0&.44 $\pm$ 0.06 & $-$0&.12 $\pm$ 0.03 \\
HD\,76291   & K1\,IV      & thin       & 0.5  & 4560 & $\pm$ 101 & 2.46 $\pm$ 0.33 & 1.29 $\pm$ 0.08 & $-$0&.13 $\pm$ 0.08 & $-$0&.06 $\pm$ 0.04 \\
HD\,76813   & G9\,III     & thin       & 2.1  & 5206 & $\pm$ 61  & 3.21 $\pm$ 0.18 & 1.44 $\pm$ 0.06 &    0&.03 $\pm$ 0.06 & $-$0&.29 $\pm$ 0.04 \\
HD\,78235   & G8\,III     & thin       & 3.4  & 5146 & $\pm$ 69  & 3.16 $\pm$ 0.21 & 1.32 $\pm$ 0.07 & $-$0&.06 $\pm$ 0.07 & $-$0&.24 $\pm$ 0.04 \\
HD\,83240   & K1\,III     & thin       & 0.5  & 4801 & $\pm$ 89  & 2.83 $\pm$ 0.23 & 1.26 $\pm$ 0.09 & $-$0&.03 $\pm$ 0.08 & $-$0&.18 $\pm$ 0.04 \\
HD\,9408    & G9\,III     & thin       & 1.0  & 4804 & $\pm$ 53  & 2.49 $\pm$ 0.17 & 1.43 $\pm$ 0.04 & $-$0&.28 $\pm$ 0.06 & $-$0&.10 $\pm$ 0.02 \\
HD\,95808   & G7\,III     & thin       & 1.2  & 5029 & $\pm$ 68  & 2.98 $\pm$ 0.22 & 1.35 $\pm$ 0.06 & $-$0&.02 $\pm$ 0.07 & $-$0&.35 $\pm$ 0.06 \\
\hline
\end{tabular}
\end{table*}

\end{document}